\documentclass[runningheads,openany]{svmult}

\usepackage{palatino}

\usepackage{euler}

\usepackage{helvet}

\usepackage{graphicx}  

\usepackage{physprbb}  

\usepackage{proc}  

\usepackage{epsfig}

\usepackage{amssymb}

\usepackage{axodraw}

\makeindex             

\usepackage{amsmath}   

\let\tocauthor\authorrunning

\begin{document}
\ifx\href\undefined\else\hypersetup{linktocpage=true}\fi

\frontmatter



\thispagestyle{empty}
\parindent=0pt

{\Large\sc Blejske delavnice iz fizike \hfill Letnik~5, \v{s}t. 2}

\smallskip

{\large\sc Bled Workshops in Physics \hfill Vol.~5, No.~2}

\smallskip

\hrule

\hrule

\hrule

\vspace{0.5mm}

\hrule

\medskip
{\sc ISSN 1580--4992}

\vfill

\bigskip\bigskip
\begin{center}

{\bfseries 
{\Large  Proceedings to the $7^\textrm{th}$ Workshop}\\
{\Huge What Comes Beyond the Standard Models\\}
\bigskip
{\Large Bled, July 19--31, 2004}}

\vspace{5mm}

\vfill

{\bfseries\large
Edited by

\vspace{5mm}
Norma Manko\v c Bor\v stnik\rlap{$^{1}$}

\smallskip

Holger Bech Nielsen\rlap{$^{2}$}

\smallskip

Colin D. Froggatt\rlap{$^{3}$}

\smallskip

Dragan Lukman\rlap{$^1$}

\bigskip

{\em\normalsize $^1$University of Ljubljana, $^2$ Niels Bohr Institute, %
$^3$ Glasgow University}

\vspace{12pt}

\vspace{3mm}

\vrule height 1pt depth 0pt width 54 mm}

\vspace*{3cm}

{\large {\sc  DMFA -- zalo\v{z}ni\v{s}tvo} \\[6pt]
{\sc Ljubljana, december 2004}}
\end{center}
\newpage

\thispagestyle{empty}
\parindent=0pt
\begin{flushright}
{\parskip 6pt
{\bfseries\large
                  The 7th Workshop \textit{What Comes Beyond  
                  the Standard Models}, 19.-- 31. July 2004, Bled}

\bigskip\bigskip

{\bfseries\large was organized by}

{\parindent8pt
\textit{Department of Physics, Faculty of Mathematics and Physics,
University of Ljubljana}

\textit{Primorska Institute of Natural Sciences and Technology, Koper}}

\bigskip

{\bfseries\large and sponsored by}

{\parindent8pt
\textit{Ministry of Education, Science and Sport of Slovenia}

\textit{Department of Physics, Faculty of Mathematics and Physics,
University of Ljubljana}

\textit{Primorska Institute of Natural Sciences and Technology, Koper}

\textit{Society of Mathematicians, Physicists and Astronomers
of Slovenia}}}
\bigskip
\medskip

{\bfseries\large Organizing Committee}

\medskip

{\parindent9pt
\textit{Norma Manko\v c Bor\v stnik}

\textit{Colin D. Froggatt}

\textit{Holger Bech Nielsen}}

\end{flushright}

\setcounter{tocdepth}{0}

\tableofcontents

\cleardoublepage

\chapter*{Preface}
\addcontentsline{toc}{chapter}{Preface}

The series of workshops on "What Comes Beyond the Standard Model?" started 
in 1998 with the idea of organizing a real workshop, in which participants 
would spend most of the time in discussions, confronting different 
approaches and ideas. The picturesque town of Bled by the lake of the 
same name, surrounded by beautiful mountains and offering pleasant walks, 
was chosen to stimulate the discussions.

The idea was successful and has developed into an annual workshop.  
Very open-minded and fruitful discussions have become the trade-mark of 
our workshop, producing several published works. It takes place in 
the house of Plemelj, which belongs to the Society of Mathematicians, 
Physicists and Astronomers of Slovenia.

In this seventh workshop, which took place from 19 to 31 of July 2004 at Bled, 
Slovenia, we have tried to answer some of the open questions which the 
Standard models leave unanswered, like:
 
\begin{itemize}
\item  Why has Nature made a choice of four (noticeable) dimensions? While all 
the others, if existing, are hidden?  And what are the properties of 
space-time in the hidden dimensions? 
\item  How could Nature make the decision about the breaking of symmetries 
down to the noticeable ones, if coming from some higher dimension $d$?
\item  Why is the metric of space-time Minkowskian and how is the choice 
of metric connected with the evolution of our universe(s)? 
\item  Where does the observed asymmetry between matter and antimatter 
originate from?
\item  Why do massless fields exist at all? Where does the weak scale 
come from?
\item  Why do only left-handed fermions carry the weak charge? Why does 
the weak charge break parity?
\item  What is the origin of Higgs fields? Where does the Higgs mass come from?
\item  Where does the small hierarchy come from? (Or why are some Yukawa 
couplings so small and where do they come from?) 
\item  Do Majorana-like particles exist?
\item  Where do the generations come from? 
\item  Can all known elementary particles be understood as different states of 
only one particle, with a unique internal space of spins and charges?
\item  How can all gauge fields (including gravity) be unified and quantized?
\item  Why do we have more matter than antimatter in our universe?
\item  What is our universe made out of (besides the baryonic matter)?
\item  What is the role of symmetries in Nature?
\item  What is the origin of the field which caused inflation?
\item  How are the randomness and the fundamental laws of Nature connected?
\end{itemize}

We have discussed these and other questions for ten days. Some results of 
this efforts appear in these Proceedings. Some of the ideas are treated 
in a very preliminary way.
Some ideas still wait to be discussed (maybe in the next workshop) 
and understood better before appearing in the next proceedings
of the Bled workshops.
The discussion will certainly continue next year, 
again at Bled, again in the house 
of Josip Plemelj.

The organizers are grateful to all the participants for the lively discussions 
and the good working atmosphere.\\[2 cm]

\parbox[b]{40mm}{%
                 \textit{Norma Manko\v c Bor\v stnik}\\
                 \textit{Holger Bech Nielsen}\\
                 \textit{Colin Froggatt}\\
                 \textit{Dragan Lukman} }
\qquad\qquad\qquad\qquad\qquad\qquad\quad
\textit{Ljubljana, December 2004}

\newpage


\cleardoublepage


\mainmatter

\parindent=20pt

\setcounter{page}{1}


\newcommand{\mjnSO}{\mathrm{SO}}
\newcommand{\mjnSU}{\mathrm{SU}}
\newcommand{\mjnunit}{\mathrm{U}}
\title{Predictions for Four Generations of Quarks Suggested by the Approach %
Unifying Spins and Charges}
\author{M. Breskvar, J. Mravlje, N.Manko\v c Bor\v stnik}
\institute{%
Department of Physics, University of
Ljubljana, Jadranska 19, 1111 Ljubljana}

\titlerunning{Predictions for Four Generations of Quarks \ldots}
\authorrunning{M. Breskvar, J. Mravlje, N.Manko\v c Bor\v stnik}
\maketitle

\begin{abstract} 
Ten years ago one of us (NSMB) has proposed the approach unifying all the internal degrees of freedom - that is the spin and 
all the char\-ges\cite{mjnnorma92,mjnnorma93,mjnnorma97,mjnnorma01} - within the group $SO(1,13)$,  a kind of Kaluza-Klein-like theories.
The approach predicts for families of quarks and leptons (in the low energy region), and suggets some approximate symmetries
of mass matrices. Following this  symmetries as far as possible, we study in this work properties of mixing matrices and
mass matrices of families of fermions, 
in particular of quarks, trying
to  limit the number of free parameters as far as possible. We 
present some very preliminary results, predicting not yet measured matrix elements for the mixing matrix for four generations of
quarks and the mass of the fourth $u$ quark.
\end{abstract}

\section{Introduction}
There is (not yet) any experimental evidence, which would be in disagreement 
with the Standard electroweak model (one should, of course, understand 
the neutrino nonzero masses just as a natural extension of 
the Standard model). 
But the Standard electroweak model has more than 20 parameters  and assumptions, 
the origin of which is not at all understood. There are also no theoretical approaches yet which
would be able to explain all these 
assumptions and parameters. 
We expect a lot from experiments on new extremely sophisticated and expensive 
accelerators and spectrometers, but measurements will first of all corroborate or not  predictions for 
several events calculated with models and theories.

The Standard electroweak model assumes the left handed weak charged doublets which are either colour triplets 
(quarks) or colour singlets (leptons) and the right handed weak chargeless singlets which are again 
either colour triplets
or colour singlets. 
It also assumes three families of quarks and leptons and the corresponding
''anti quarks'' and ''anti leptons'', without giving any explanation about the origin of  families and ''anti families''.
It assumes that the quarks and the leptons are massless - until gaining a (small) mass at low 
energies through the 
vacuum expectation value(s) of Higgs fields and Yukawa couplings, without giving any explanation, why is this so and
where does the weak scale come from.

The great advantage of the approach of (one of) us, which unifies spins and 
charges\cite{mjnnorma92,mjnnorma93,mjnnorma95,mjnnorma97,mjnnorma01,mjnholgernorma2002,mjnpikanormaproceedings1,mjnpikanormaproceedings2,mjnPortoroz03}, 
is, that it proposes possible answers to open
questions of the Standard electroweak model.
 We demonstrate that a left handed $SO(1,13)$ Weyl spinor multiplet includes, 
if the representation is interpreted
in terms of the subgroups $SO(1,3)$, $SU(2)$, $SU(3)$ and the sum of the two $U(1)$'s,  spinors and ``anti spinors`` of
the Standard model - that is the left handed $SU(2)$ doublets and the right handed  $SU(2)$ singlets of with the group 
$SU(3)$ charged quarks and  chargeless leptons, while the ``anti spinors`` are oppositely charged 
and have opposite handedness.
Right handed neutrinos and left handed anti neutrinos - both weak chargeless - are also included.

We demonstrate that, when starting with a spinor of one handedness only, although spinor representations of subgroups
 contain represetations of both handedness  with respect to each of the subgroups, yet 
each of the two representations might be distinguished
by the charges -  the gauge Kaluza-Klein charges - of subgroups  and accordingly, the choice of the 
representation of a particular handedness is still possible, and accordingly spinors couple chirally to
the corresponding gauge fields. 

Our gauge group is $SO(1,13)$ - the smallest complex Lorentz group with a left handed Weyl spinor
containing the needed representations of the Standard model.  The gauge fields of this group are spin 
connections and vielbeins\cite{mjnnorma93,mjnnorma01}, determining the gravitational field in (d = 14)-dimensional space.
Then a gauge gravitational field manifests in four dimensional subspace as all the gauge fields of the known charges, 
and (as alleady written) also as the Yukawa couplings.

It was demonstrated\cite{mjnpikanormaproceedings2,mjnPortoroz03,mjnastrinorma2003,mjnastridragannorma} that the approach offers a possible 
explanation for families of spinors and their masses, since a part of the gravitational gauge fields
originating in higher than four dimensions appears as terms manifesting the Yukawa couplings, postulated by 
the Standard electroweak model. The approach suggests (approximate) symmetries of mass matrices, 
requires that four - rather than three - 
families appear at low energies and suggests splitting of the four families into two by two families.
It also suggests that the proposed symmetries must be slightly broken, so that the mixing matrices (''CKM'' matrix in
the case of three families) would not be just unity.

In this work we start with the action, suggested by the approach, as well as with by the approach suggested 
approximate symmetries of the
matrix elements of the Yukawa couplings. We study, how the way of breaking the suggested symmetries influences the
mass matrices (expectation values of Yukawa couplings in the approach) and the mixing matrices.
Trying to keep suggesting symmetries as far as possible and minimizing the number of free parameters of the
model, we discuss the properties of the mixing matrices and by fitting the free parameters with the experimental data
(using the Monte-Carlo procedure), we were able to make some preliminary predictions for  matrix elements of the
mixing matrice for four 
families of quarks, as well as for the mass of the fourth $u$ quark. 

We shall repeat some properties of the approach as far as they are needed for our work.

\section{Spinor representation in terms of Clifford algebra objects}
\label{technique}

In this section we present some of properties of one Weyl spinor of the group $SO(1,13)$ in terms of the 
properties of subgroups $SO(1,7) \times SO(6)$ and of $SO(1,3) \times SU(3)\times U(1) \times U(1)$. We formulate spinor 
representations as products of binomials, which are either nilpotents or projectors - Clifford algebra odd and even binomials of
$\gamma^a$'s, respectively. For details we kindly ask the reader to see refs.\cite{mjnnormasuper94,mjnpikanormaproceedings2,%
mjnholgernorma2002,mjnPortoroz03}. 

We define two kinds of the Clifford algebra objects, with the properties
\begin{eqnarray}
\{\gamma^a,\gamma^b\}_{+} = 2\eta^{ab}, \quad \{\tilde{\gamma}^a,\tilde{\gamma}^b\}_{+} = 2\eta^{ab},
\quad \{\gamma^a,\tilde{\gamma}^b\}_{+} = 0.
\label{clifford}
\end{eqnarray}
The operators $\tilde{\gamma}^a$ are introduced formally as operating from the left hand side (as $\gamma^a$'s do)
on all the Clifford algebra  objects, but they are indeed define as operating from the right hand side as follows
\begin{eqnarray}
\tilde{\gamma}^a B : = (-)^{n_B} B \gamma^b,
\label{tildegclifford}
\end{eqnarray}
with $(-)^{n_B} = \pm 1$ for a Clifford even and odd object, respectively.

Accordingly two kinds of generators of the Lorentz transformations follow 
$S^{ab} = i/4 (\gamma^a \gamma^b - \gamma^b \gamma^a)$, as well  as
$\tilde{S}^{ab} = i/4 (\tilde{\gamma}^a \tilde{\gamma}^b - \tilde{\gamma}^b \tilde{\gamma}^a$, with the properties
$\{ S^{ab},\tilde{S}^{cd}\}_{-}=0$.

The generators of the subgroups $SO(1,3),SU(3), SU(2)$ and the two $U(1)$, needed to determine the spin, the
weak charge, the colour charge and the two hypercharges will be written in terms of $S^{ab}$
\begin{eqnarray}
\tau^{Ai} = \sum_{ab} c^{Ai}{}_{ab} S^{ab},
\quad \{\tau^{Ai},\tau^{Bj}\}_{-} = i \delta^{AB} f^{Aijk} \tau^{Ak},
\label{tau}
\end{eqnarray}
with $A$ representing the corresponding subgroup and $f^{Aijk}$ the corresponding structure constants.

We define spinors  as  
eigenstates of the chosen Cartan subalgebra of the Lo\-rentz algebra $SO(1,13)$,  with  the operators 
$S^{ab}$ and $\tilde{S}^{ab}$ in the two Cartan subalgebra sets, with the same indices in both cases.
By introducing the notation
\begin{eqnarray}
\stackrel{ab}{(\pm i)}: &=& \frac{1}{2}(\gamma^a \mp  \gamma^b),  \quad 
\stackrel{ab}{[\pm i]}: = \frac{1}{2}(1 \pm \gamma^a \gamma^b), \;{for} \; \eta^{aa} \eta^{bb} =-1, \nonumber\\
\stackrel{ab}{(\pm )}: &= &\frac{1}{2}(\gamma^a \pm i \gamma^b),  \quad 
\stackrel{ab}{[\pm ]}: = \frac{1}{2}(1 \pm i\gamma^a \gamma^b), \;{for} \; \eta^{aa} \eta^{bb} =1,
\label{eigensab}
\end{eqnarray}
it can be shown that  
\begin{eqnarray}
S^{ab} \stackrel{ab}{(\mp i)}: &=& \pm \frac{i}{2} \stackrel{ab}{(\mp i)}, \quad 
S^{ab} \stackrel{ab}{[\pm i]}:  =  \pm \frac{i}{2} \stackrel{ab}{[\pm i]}, \nonumber\\
S^{ab} \stackrel{ab}{(\pm )}:  &= &\pm \frac{1}{2} \stackrel{ab}{(\pm )},  \quad 
S^{ab} \stackrel{ab}{[\pm ]}:   =  \pm \frac{1}{2} \stackrel{ab}{[\pm ]}.
\label{mjneigensabev}
\end{eqnarray}
The above binomials are all ''eigenvectors''  of  the generators $S^{ab}$.
They are also eigenvectors of $\tilde{S}^{ab}$
\begin{eqnarray}
\tilde{S}^{ab} \stackrel{ab}{(\mp i)}: &=& \pm \frac{i}{2} \stackrel{ab}{(\pm i)}, \quad 
\tilde{S}^{ab} \stackrel{ab}{[\pm i]}:  =  \mp \frac{i}{2} \stackrel{ab}{[\pm i]}, \nonumber\\
\tilde{S}^{ab} \stackrel{ab}{(\pm )}:  &= &\pm \frac{1}{2} \stackrel{ab}{(\pm )},  \quad 
\tilde{S}^{ab} \stackrel{ab}{[\pm ]}:   =  \mp \frac{1}{2} \stackrel{ab}{[\pm ]}.
\label{eigensabev}
\end{eqnarray}
We further find 
\begin{eqnarray}
\gamma^a \stackrel{ab}{(k)}&=&\eta^{aa}\stackrel{ab}{[-k]},\nonumber\\
\gamma^b \stackrel{ab}{(k)}&=& -ik \stackrel{ab}{[-k]}, \nonumber\\
\gamma^a \stackrel{ab}{[k]}&=& \stackrel{ab}{(-k)},\nonumber\\
\gamma^b \stackrel{ab}{[k]}&=& -ik \eta^{aa} \stackrel{ab}{(-k)}
\label{graphgammaaction}
\end{eqnarray}
and
\begin{eqnarray}
\tilde{\gamma^a} \stackrel{ab}{(k)}: &=& - i\eta^{aa}\stackrel{ab}{[k]},\noindent\\
\tilde{\gamma^b} \stackrel{ab}{(k)}: &=&  - k \stackrel{ab}{[k]}, \nonumber\\
\tilde{\gamma^a} \stackrel{ab}{[k]}: &=& \;\; i \stackrel{ab}{[k]} \gamma^a = \;\;i\stackrel{ab}{(k)},\nonumber\\
\tilde{\gamma^b} \stackrel{ab}{[k]}: &=& \;\; i \stackrel{ab}{[k]} \gamma^b = -k \eta^{aa} \stackrel{ab}{(k)}.
\label{gammatilde}
\end{eqnarray}
The reader should notice that $\gamma^a$'s transform the binomial  $\stackrel{ab}{(k)}$ into the binomial $\stackrel{ab}{[-k]}$,
whose eigenvalue with respect to $S^{ab}$ change sign, while
$\tilde{\gamma}^a$'s transform the binomial $\stackrel{ab}{(k)}$ into the one ($\stackrel{ab}{[k]}$) with unchanged ''eigenvalue''
with respect to $S^{ab}$.

Let us select operators belonging to the Cartan subalgebra of $7$ elements of $SO(1,13)$ as follows
\begin{eqnarray}
S^{03}, S^{12}, S^{56}, \cdots, S^{13\; 14}.
\label{cartan}
\end{eqnarray}
One can find the operators of handedness for the Lorentz group $SO(1,13)$ and the subgroups $SO(1,3), SO(1,7),
SO(1,9), SO(6)$ and $SO(4)$ as follows 
\begin{eqnarray}
\Gamma^{(1,13)} &=&  2^{7}i \; S^{03} S^{12} S^{56} \cdots S^{13 \; 14},
\nonumber\\
\Gamma^{(1,3)}  &=&  - 4i  S^{03} S^{12},\nonumber\\
\Gamma^{(1,7)} &=&  - 2^{4} i S^{03} S^{12} S^{56} S^{78},\label{gammas}\\
\Gamma^{(1,9)} &=& 2^{5} i S^{03} S^{12} S^{9\;10} S^{11\;12} S^{13 \; 14},
\nonumber\\
\Gamma^{(6)} &=& -8 S^{9 \;10} S^{11\;12} S^{13 \; 14},
\nonumber\\
\Gamma^{(4)} &=& 4 S^{56} S^{78}.\nonumber
\end{eqnarray}

Let us make now a choice of a starting state  of one Weyl representation of the group $SO(1,13)$, which is the
eigenstate of all the members of the Cartan subalgebra (Eq.(\ref{cartan})) and is left handed
($\Gamma^{(1,13)} =-1$) 
\begin{eqnarray}
\stackrel{03}{(+i)}\stackrel{12}{(+)}|\stackrel{56}{(+)}\stackrel{78}{(+)}
||\stackrel{9 \;10}{(+)}\stackrel{11\;12}{(-)}\stackrel{13\;14}{(-)} |\psi \rangle &=&
\nonumber\\
 (\gamma^0 -\gamma^3)(\gamma^1 +i \gamma^2)| (\gamma^5 + i\gamma^6)(\gamma^7 +i \gamma^8) & & \nonumber\\
 ||(\gamma^9 +i\gamma^{10})(\gamma^{11} -i \gamma^{12})(\gamma^{13}-i\gamma^{14})|\psi \rangle & & .
\label{start}
\end{eqnarray}

The signs "$|$" and "$||$" are to point out the  $SO(1,3)$ (up to $|$), $SO(1,7)$ (up to $||$)
and $SO(6)$ (between $|$ and $||$) substructure of the starting state of the left handed multiplet of
$SO(1,13)$ which has $2^{14/2-1}= 64 $ vectors. Here $|\psi\rangle$ is any state, which is not transformed 
to zero and therefore we shall not write down $|\psi \rangle$ any longer.
One easily finds that the eigenvalues of the chosen
Cartan sub algebra elements of Eq.(\ref{cartan}) are $+i/2, 1/2, 1/2,1/2,1/2,-1/2,-1/2$, 
respectively. This state is a right handed spinor with respect to $SO(1,3)$ ($\Gamma^{(1,3)} =1$, 
Eq.(\ref{gammas})), with spin up 
($S^{12} =1/2$), it is $SU(2)$  
singlet ($\tau^{33} = 0$, Eq.(\ref{tau})), and it is the member 
of  the $SU(3)$ triplet (Eq.(\ref{tau})) with ($\tau^{53} =1/2, \tau^{58} = 1/(2 \sqrt{3})$),
it has $\tau^{43} = 1/2$ and $\tau^{6,1}= 1/2$. We further find
according to Eq.(\ref{gammas}) that $\Gamma^{(4)} =1, \Gamma^{(1,7)}= 1, \Gamma^{(6)} = -1$ and 
$\Gamma^{(1,9)} = -1$.

To obtain all the states of one Weyl spinor one only has to apply on the starting state of Eq.(\ref{start})
the generators $S^{ab}$. 
The generators $S^{01}, S^{02}, S^{31}, S^{32}$ transform spin up state (the 
$\stackrel{03}{(+i)}\stackrel{12}{(+)} $ part of the starting state (Eq.(\ref{start}) with
$S^{12}=1/2$ and $S^{03}=i/2$) into spin 
down state ( $\stackrel{03}{[-i]}\stackrel{12}{[-]}$, which has
$S^{12}=-1/2$ and $S^{03} = -i/2$), leaving all the other parts of the state and accordingly also all the other properties of 
this state unchanged. 
The generator $S^{08}$, for example, transforms one $SU(2)$ right handed singlet, if making a choice of the first row of Table~\ref{mjntable1}
this would be the right handed neutrino with spin up - $\nu_R $ - which is the $SU(3)$ singlet,
into a member of an $SU(2)$ doublet, that  is a left handed neutrino, again with spin up.   

\begin{table}
\begin{center}
\begin{tabular}{|r|c||c||c|c||c|c|c||c|c|c||r|r|}
\hline
i&$$&$|^a\psi_i>$&$\Gamma^{(1,3)}$&$ S^{12}$&$\Gamma^{(4)}$&
$\tau^{13}$&$\tau^{2}$&$\tau^{33}$&$\tau^{38}$&$\tau^{4}$&$Y$&$Y'$\\
\hline\hline
&& ${\rm Octet},\;\Gamma^{(1,7)} =1,\;\Gamma^{(6)} = -1,$&&&&&&&&&& \\
&& ${\rm of \; leptons}$&&&&&&&&&&\\
\hline\hline
1&$\nu_{R}$&$\stackrel{03}{(+i)}\stackrel{12}{(+)}|\stackrel{56}{(+)}\stackrel{78}{(+)}
||\stackrel{9 \;10}{(+)}\stackrel{11\;12}{[+]}\stackrel{13\;14}{[+]}$
&1&$\frac{1}{2}$&1&0&$\frac{1}{2}$&0&$0$&$-\frac{1}{2}$&0&-1\\
\hline 
2&$\nu_{R}$&$\stackrel{03}{[-i]}\stackrel{12}{[-]}|\stackrel{56}{(+)}\stackrel{78}{(+)}
||\stackrel{9 \;10}{(+)}\stackrel{11\;12}{[+]}\stackrel{13\;14}{[+]}$
&1&$-\frac{1}{2}$&1&0&$\frac{1}{2}$&0&$0$&$-\frac{1}{2}$&0&-1\\
\hline
3&$e_{R}$&$\stackrel{03}{(+i)}\stackrel{12}{(+)}|\stackrel{56}{[-]}\stackrel{78}{[-]}
||\stackrel{9 \;10}{(+)}\stackrel{11\;12}{[+]}\stackrel{13\;14}{[+]}$
&1&$\frac{1}{2}$&1&0&$-\frac{1}{2}$&0&$0$&$-\frac{1}{2}$&-1&0\\
\hline 
4&$e_{R}$&$\stackrel{03}{[-i]}\stackrel{12}{[-]}|\stackrel{56}{[-]}\stackrel{78}{[-]}
||\stackrel{9 \;10}{(+)}\stackrel{11\;12}{[+]}\stackrel{13\;14}{[+]}$
&1&$-\frac{1}{2}$&1&0&$-\frac{1}{2}$&0&$0$&$-\frac{1}{2}$&-1&0\\
\hline
5&$e_{L}$&$\stackrel{03}{[-i]}\stackrel{12}{(+)}|\stackrel{56}{[-]}\stackrel{78}{(+)}
||\stackrel{9 \;10}{(+)}\stackrel{11\;12}{[+]}\stackrel{13\;14}{[+]}$
&-1&$\frac{1}{2}$&-1&$-\frac{1}{2}$&0&0&$0$&$-\frac{1}{2}$&$-\frac{1}{2}$&$-\frac{1}{2}$\\
\hline
6&$e_{L}$&$\stackrel{03}{(+i)}\stackrel{12}{[-]}|\stackrel{56}{[-]}\stackrel{78}{(+)}
||\stackrel{9 \;10}{(+)}\stackrel{11\;12}{[+]}\stackrel{13\;14}{[+]}$
&-1&$-\frac{1}{2}$&-1&$-\frac{1}{2}$&0&0&$0$&$-\frac{1}{2}$&$-\frac{1}{2}$&$-\frac{1}{2}$\\
\hline
7&$\nu_{L}$&$\stackrel{03}{[-i]}\stackrel{12}{(+)}|\stackrel{56}{(+)}\stackrel{78}{[-]}
||\stackrel{9 \;10}{(+)}\stackrel{11\;12}{[+]}\stackrel{13\;14}{[+]}$
&-1&$\frac{1}{2}$&-1&$\frac{1}{2}$&0&0&$0$&$-\frac{1}{2}$&$-\frac{1}{2}$&$-\frac{1}{2}$\\
\hline
8&$\nu_{L}$&$\stackrel{03}{(+i)}\stackrel{12}{[-]}|\stackrel{56}{(+)}\stackrel{78}{[-]}
||\stackrel{9 \;10}{(+)}\stackrel{11\;12}{[+]}\stackrel{13\;14}{[+]}$
&-1&$-\frac{1}{2}$&-1&$\frac{1}{2}$&0&0&$0$&$-\frac{1}{2}$&$-\frac{1}{2}$&$-\frac{1}{2}$\\
\hline\hline
\end{tabular}
\end{center}
\caption{\label{mjntable1}%
The 8-plet of leptons, belonging to one Weyl spinor of $SO(1,13)$ (64-plet), is presented. 
The whole multiplet contains  spinors - all the quarks 
and the leptons - and the corresponding ``anti spinors''- anti quarks  and anti leptons 
of the Standard model, that is left handed weak charged quarks and leptons and right handed 
weak chargeless anti quarks and anti leptons, as well as left handed weak chargeless anti quarks and anti leptons
and weak charged right handed anti quarks and anti leptons. The reader can find the whole
spinor in the ref.\cite{mjnPortoroz03}. The lepton part  contains also right handed weak chargeless 
neutrinos and left handed weak chargeless antineutrinos. $\Gamma$'s stay for operator of handedness, while 
$\tau^{13}$ is the third component of the weak charge, $\tau^{2}$ is one $U(1)$ charge, $\tau^{33}$ and $\tau^{38}$
are the two colour charges and $\tau^{4}$ is the second $U(1)$ charge, so that $Y= \tau^2 + \tau^{4}$,$Y'= - \tau^2
+ \tau^4$.}
\end{table}
One Weyl spinor in $d=(1+13)$-dimensional space appears as all the quarks and all the leptons and
all the anti quarks and all the anti leptons in the ''physical'' part
of space, provided that the symmetries properly break from $SO(1,13)$ to the observable symmetries of 
the Standard electroweak model.

\subsection{Operators transforming one family into another}
\label{operators}

We would like to point out that while the generators of the Lorentz group $S^{ab}$, with  a pair of $a,b$, which does not belong
to the Cartan subalgebra (Eq.(\ref{cartan})), transform one vector of the representation into another, 
transform the generators $\tilde{S}^{ab}$ (again the pair $a,b$ should not belong to the Cartan set) 
a member of one family into the same member of another family, leaving all the other quantum numbers (determined by $S^{ab}$)
unchanged\cite{mjnnorma92,mjnnorma93,mjnnorma95,mjnnorma01,mjnholgernorma00,mjnpikanormaproceedings2,mjnPortoroz03}. 
This is happening since  the multiplication by $\gamma^a$ from the left  changes the operator $\stackrel{ab}{(+)}$ or
the operator $\stackrel{ab}{(+i)}$
into the operator $\stackrel{ab}{[-]}$ or the operator $\stackrel{ab}{[-i]}$, respectively, while the operator
$\tilde{\gamma}^a$ changes $\stackrel{ab}{(+)}$ or $\stackrel{ab}{(+i)}$ 
 into $\stackrel{ab}{[+]}$
or to $\stackrel{ab}{[+i]}$, respectively, whithout changing the ''eigen value'' of $S^{ab}$.

\section{ Lagrange function } 
\label{sec:lagrange}

Refereeing to the work\cite{mjnnorma92,mjnnorma93,mjnnorma95,mjnnorma01,mjnholgernorma00,mjnpikanormaproceedings2,mjnPortoroz03} we write the action 
for a Weyl (massless) spinor and the gauge field
in $d(=1+13)$ - dimensional space as follows
\begin{eqnarray}
S &=& \int \; d^dx \;E {\mathcal L} +  \int \; d^dx \;E R, \; {\rm with} \nonumber\\
{\mathcal L} &=& \bar{\psi}\gamma^a p_{0a} \psi = \bar{\psi} \gamma^a f^{\alpha}_a p_{0\alpha} \; {\rm and}\;
R= f^{\alpha}{}_{[a} f^{\beta}{}_{b]} (\omega^{ab}{}_{\alpha,\beta} + \omega^a{}_{c \alpha} \omega^{c b}{}_{\beta}),
\nonumber\\
p_{0\alpha} &=& p_{\alpha} - \frac{1}{2}S^{ab} \omega_{ab\alpha} - \frac{1}{2}\tilde{S}^{ab} \tilde{\omega}_{ab\alpha}.
\label{lagrange}
\end{eqnarray}
Here $f^{\alpha}_a$ are vielbeins, while 
 $\omega_{ab\alpha}$ and $\tilde{\omega}_{ab\alpha} $ are spin connections, the gauge fields of $S^{ab}$ and
$\tilde{S}^{ab}$, respectively. We point out that there are two kinds of the Clifford algebra objects\cite{mjnPortoroz03}.( Besides
the usual $\gamma^a$ operators, there are also the operators $\tilde{\gamma}^a$, the first connected with 
the left multiplication, the second with the right multiplication. The two types of Clifford algebra objects
anti commute ($\{\gamma^a,\tilde{\gamma}^b\}_+ =0$),
while the two corresponding types of the generators of the Lo\-rentz transformations commute ($\{ S^{ab}, \tilde{S}^{cd}\}_-=0$.))
Indices $a,b$ are flat indices, while $\alpha, \beta $ are Einstein indices, $m,n,..$ and $\mu,\nu,..$ will be used to
describe the  coordinates in $d=(1+3)$-dimensional space, while $s,t,..$ and $\sigma,\tau$ belong to higher than four-
dimensional space.

While  one Weyl spinor in $d=(1+13)$ with the spin as the only internal degree of freedom, should manifest  in
four-dimensional part of space  as the ordinary ($SO(1,3)$) spinor with all the known charges 
of one family of the Standard model, the gravitational field presented with spin connections and vielbeins
should accordingly in four 
dimensions  manifest as all the known gauge fields as
well as the Yukawa couplings, if the break of symmetries occurs in an appropriate way.

To see that the Yukawa-like  couplings are the part of the Lagrangean of Eq.(\ref{lagrange}), we rewrite
the spinor part of the Lagrangean of 
Eq.(\ref{lagrange}) as\cite{mjnPortoroz03} 
\begin{eqnarray}
{\mathcal L} &=& \bar{\psi}\gamma^{\alpha} (p_{\alpha}- \sum_{A,i}\; g^{A}\tau^{Ai} A^{Ai}_{\alpha} \psi) 
+ i\psi^+ S^{0h} S^{k k'} f^{\sigma}_h \omega_{k k' \sigma} \psi \nonumber\\
 & &\qquad {} +  
i \psi^+ S^{0h} \tilde{S}^{k k'} f^{\sigma}_h \tilde{\omega}_{kk' \sigma} \psi,
\label{yukawa}
\end{eqnarray}
with  $\psi$, which is (for low energy solution) assumed not to depend on coordinates $x^{\sigma}, \sigma=\{5,6, \cdots ,14 \}$.
The second and the third term look like a mass term, since 
$f^{\sigma}_h \omega_{kk' \sigma}$ and $f^{\sigma}_h \tilde{\omega}_{kk' \sigma}$ behaves in $d(=1+3)-$
dimensional part of space like a scalar field, while the operator $S^{0h}, h=7,8$, for example, 
transforms a right handed
weak chargeless spinor (above we looked at $\nu_R$) into a left handed weak charged spinor ($\nu_L$), 
without changing the spin in $d=1+3$ - just
what the Yukawa couplings with the Higgs doublet included, do in the Standard model formulation.
The reader should note, that no Higgs weak charge doublet is needed here, as $S^{0h}, h=7,8$ does his job.

\section{Break of symmetries}
\label{break}

There are several ways of breaking  the group $SO(1,13),$ down to subgroups of the Standard model.
(One) of the most probable breaks, suggested by the approach unifying spins and charges, would 
be the following one
\[
 \begin{array}{c}
 \begin{array}{c}
 \underbrace{%
 \begin{array}{rrcll}
  & & \mjnSO(1,13) \\
  & & \downarrow \\
  & & \mjnSO(1,7) \otimes \mjnSO(6) \\
  & \swarrow & &  \searrow \\
  & \mjnSO(1,7) & & \mjnSU(4)\\
  \swarrow\qquad & & & \qquad\downarrow \\
  \mjnSO(1,3)\otimes \mjnSO(4) & & & \mjnSU(3)\otimes\mjnunit(1) \\
  \downarrow\qquad & & & \qquad\downarrow \\
 \mjnSO(1,3)\otimes\mjnSU(2)\otimes\mjnunit(1) & & & \mjnSU(3)\otimes\mjnunit(1)\\
 & & & \\
 \end{array}} \\
 \mjnSO(1,3)\otimes\mjnSU(2)\otimes\mjnunit(1)\otimes\mjnSU(3)\otimes \mjnunit(1)\\
 \end{array}\\
 \downarrow \\
 \mjnSO(1,3)\otimes\mjnSU(2)\otimes\mjnunit(1)\otimes\mjnSU(3)\\
 \end{array}
 \]
This break of symmetries leads to the scheme of the running coupling constants\cite{mjnholgernormaren}, presented on the
figure.
\begin{figure} 
 \centering
 \includegraphics[trim= 125 255 55 85 ,clip ,width=10cm]{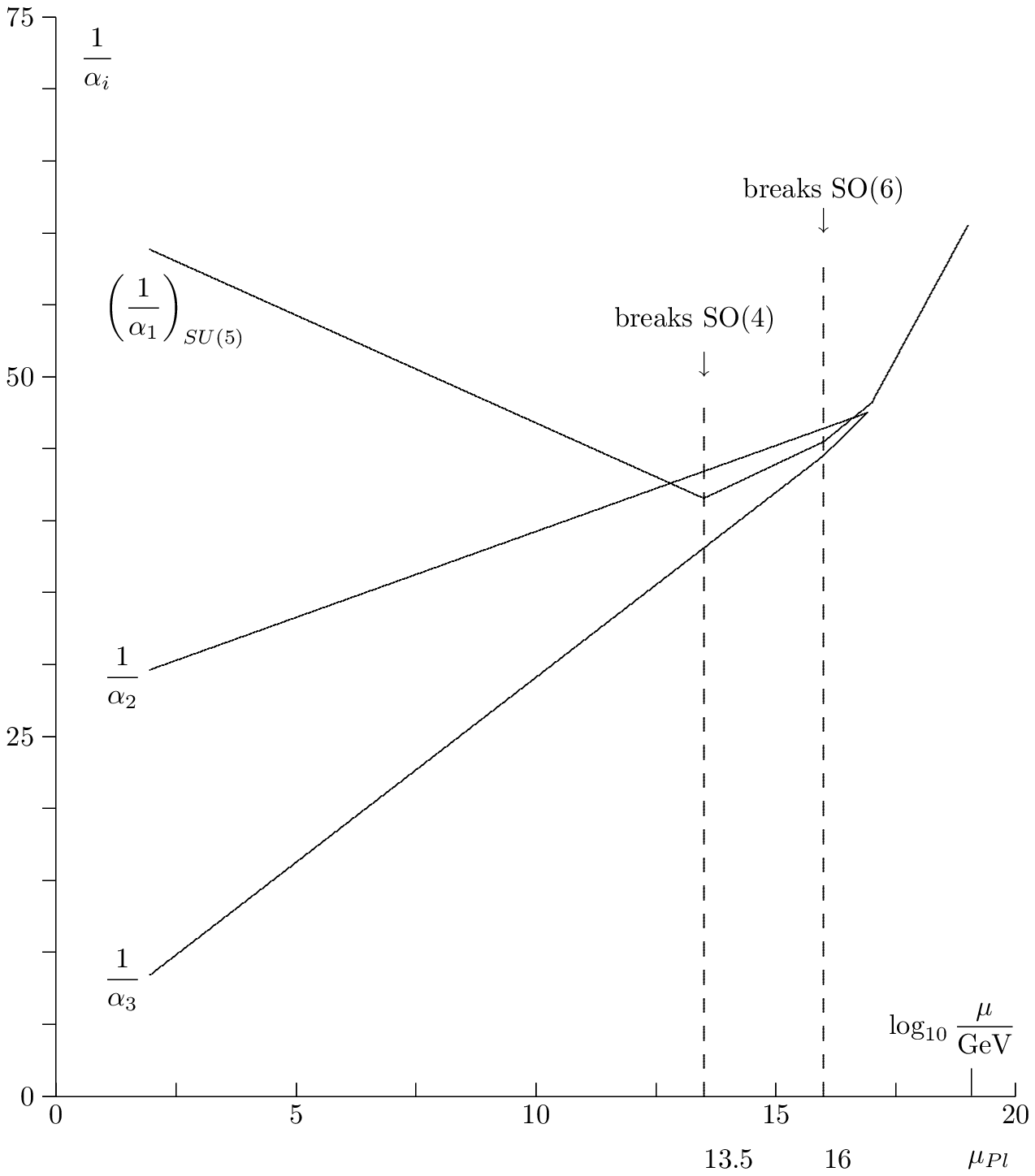}
\caption{\label{snmbfig1}
We found the running coupling constants, extrapolated from 
the experimental values by the assumption that the gauge 
group $SO(4)$ breaks at much lower  scale (at around $10^{13}$ GeV)
than the gauge group $SO(6)$ (which breaks at around $10^{17}$ GeV). 
The $SU(2)$ gauge group coupling constant
does not change when running together with $(\alpha_{U(1),SO(4)})^{-1}$.  
The three $\alpha$'s meet and then run together as $SO(1,13)$ (rather SO(10)).}
\end{figure}
It is worthwhile to notice that all the running coupling constants meet and run then together without additional
degrees of freedom.

\section{Mass matrices predicted by the approach unifying spins and charges}
\label{massapproach}

Let us assume that the break of symmetries appeared in a way that only terms like $\psi^+ S^{08} (S^{st} 
f^{\sigma}_8 \omega_{st \sigma} + \tilde{S}^{mn} 
f^{\sigma}_8 \tilde{\omega}_{mn \sigma} +\tilde{S}^{st} 
f^{\sigma}_8 \tilde{\omega}_{st \sigma} )\psi $
appear in the Lagrange density, with $s,t $ and $\sigma \in 5,6,7,8$ and   $m,n$ equal 
$0,1,2,3$. Then there are only four families which are measurable at low energies,
namely
\begin{eqnarray}
\stackrel{03}{(+i)} \stackrel{12}{(+)} \stackrel{56}{(+)} \stackrel{78}{(+)} \nonumber\\
\stackrel{03}{[+i]} \stackrel{12}{[+]} \stackrel{56}{(+)} \stackrel{78}{(+)} \nonumber\\
\stackrel{03}{(+i)} \stackrel{12}{(+)} \stackrel{56}{[+]} \stackrel{78}{[+]} \nonumber\\
\stackrel{03}{[+i]} \stackrel{12}{[+]} \stackrel{56}{[+]} \stackrel{78}{[+]}.
\label{threefamilies}
\end{eqnarray}
while they all have  the same $SO(6)$ segment (like $\stackrel{9\;10}{(+)]} \stackrel{11\;12}{(-)}
\stackrel{13\;14}{(-)} $ for the right han\-ded quark of the $SU(3)$ charge $(1/2),(1/2\sqrt{3})$, or 
$\stackrel{9\;10}{(+)]} \stackrel{11\;12}{[+]} \stackrel{13\;14}{[+]} $ for the above mentioned 
right handed neutrino).

Let us write down formally the matrix ellements determining the mass matrices
as follows
\begin{eqnarray}
M^{\alpha ij} = < \psi^{\alpha i}{}_L| Y_{\alpha}{}^{ij}|\psi^{\alpha j}{}_R>,
\label{fij}
\end{eqnarray}
where  $Y_{\alpha}^{ij}$ stays for all the operators and the corresponding gauge fields, which are in Eq.(\ref{yukawa})
responsible for the transitions from  the right  handed  $SU(2)$ chargeless spinor of a type $\alpha$  to a left handed
weak charged  spinor
of a type $ \alpha$ and $i,j$ denote a family. 
Let $S^{\alpha}$ be the unitary matrix which diagonalizes the mass matrix of the type $\alpha$.

To manifest the symmetry, suggested by the approach, let us  denote by $A_{\alpha}$ the matrix element for the 
transition from  a right handed weak chargeless 
 spinor of type $\alpha = u,d,e,\nu$ to the left handed weak charged spinor (these transitions occur within 
 one family and are caused by the second term $- \gamma^0 \gamma^8
\tau^{Ai} A^{Ai}_{\sigma} f^{\sigma}_8$) of Eq.(\ref{yukawa} ), by $B_{\alpha}$ the matrix element 
causing the transition,
 in which $\stackrel{03}{(+i)} \stackrel{12}{(+)}$ changes to $\stackrel{03}{[+i]} \stackrel{12}{[+]}$
 or opposite (such are transitions between the first and the second family or transitions between the 
 third family and the fourth family of Eq.(\ref{threefamilies}) caused by 
 $\tilde{S}^{mm'}f^{\sigma}_8 \tilde{\omega}_{mm' \sigma}$ with $m,m'=0,1,2,3$ ), 
 by $C_{\alpha}$  the matrix element causing the transition 
 in which $\stackrel{56}{(i)} \stackrel{78}{(+)}$ changes to $\stackrel{56}{[+]} \stackrel{78}{[+]}$
 or opposite (such are transitions between the first and the third family or transitions between the 
 second and the fourth family of Eq.(\ref{threefamilies}) caused by $ \tilde{S}^{st'}f^{\sigma}_8
 \tilde{\omega}_{st \sigma}$ with $s,t=5$,$6$,$7$,$8$) and by $D_{\alpha}$  transitions in which all four 
 factors change, that is the transitions,
 in which $\stackrel{03}{(+i)} \stackrel{12}{(+)}$ changes to $\stackrel{03}{[+i]} \stackrel{12}{[+]}$
 or opposite and  $\stackrel{56}{(+i)} \stackrel{78}{(+)}$ changes to $\stackrel{56}{[+]} \stackrel{78}{[+]}$ or
 opposite
 (such are transitions between the first and the fourth family or transitions between the 
 second  and the third family of Eq.(\ref{threefamilies}))  and if we further assume that 
 the elements are real numbers, we find 
 the following mass matrix $M$
\begin{displaymath}
\left( \begin{array}{cccc}
A_{\alpha}&B_{\alpha}&C_{\alpha}&D_{\alpha}\\
B_{\alpha}&A_{\alpha}&D_{\alpha}&C_{\alpha}\\
C_{\alpha}&D_{\alpha}&A_{\alpha}&B_{\alpha}\\
D_{\alpha}&C_{\alpha}&B_{\alpha}&A_{\alpha}
\end{array} \right).
\label{mass}
\end{displaymath}
Since $u,d,\nu,e$  have different values 
for the quantum numbers $\tau^{Ai}$, one expects even on a ''tree level'' different values for matrix elements
$A_{\alpha},B_{\alpha},C_{\alpha},D_{\alpha}$. 
To evaluate the matrix elements $A_{\alpha},B_{\alpha},C_{\alpha},D_{\alpha}$  one should make a precise model, 
which would take into account 
not only that  matrix elements within one family depend on quantum numbers of the members
of the family on ''a tree level'', but  also on next orders , as well as on the way how symmetries are broken.

One notices, that  the mass matrices have the symmetry
\begin{displaymath}
\left( \begin{array}{cc}
X&Y\\
Y&X
\end{array} \right),
\label{xy}
\end{displaymath}
which makes the diagonalization of the mass matrix of Eq.(\ref{mass})
simple. We find immediatelly that
\begin{eqnarray}
\lambda_{\alpha 1} &=& A_{\alpha}-B_{\alpha}-C_{\alpha}+D_{\alpha},\nonumber\\
\lambda_{\alpha 2} &=& A_{\alpha}-B_{\alpha}+C_{\alpha}-D_{\alpha},\nonumber\\
\lambda_{\alpha 3} &=& A_{\alpha}+B_{\alpha}-C_{\alpha}-D_{\alpha},\nonumber\\
\lambda_{\alpha 4} &=& A_{\alpha}+B_{\alpha}+C_{\alpha}+D_{\alpha}. 
\label{formalabcd}
\end{eqnarray}
We see that a ''democratic'' matrix with $A_{\alpha}=B_{\alpha}=C_{\alpha}=D_{\alpha}$  (ref.\cite{mjnfritsch})
leads to $\lambda_{\alpha 1}=\lambda_{\alpha 2}=\lambda_{\alpha 3}= 0, \lambda_{\alpha 4} = 4 A_{\alpha}$. 
The diagonal matrix leads to
four equal values $\lambda_{\alpha i} = A_{\alpha}. $ One could expect that the  break of symmetries of 
the group $SO(1,13)$ down
to $SO(1,3), SU(3)$ and $U(1)$ will lead to something in between.
If we fit $\lambda_{\alpha i}$ with the masses of families $m_{\alpha i}$, with $\alpha= u,d,\nu,e$ and $i$ is
the number of family,
we find
\begin{eqnarray}
A_{\alpha} &=& \{(m_{\alpha 4}+ m_{\alpha 3}) + (m_{\alpha 2} + m_{ \alpha 1})\}/4,\nonumber\\
B_{\alpha} &=& \{(m_{\alpha 4}+ m_{\alpha 3}) - (m_{\alpha 2} - m_{ \alpha 1})\}/4,\nonumber\\
C_{\alpha} &=& \{(m_{\alpha 4}- m_{\alpha 3}) + (m_{\alpha 2} - m_{ \alpha 1})\}/4,\nonumber\\
D_{\alpha} &=& \{(m_{\alpha 4}- m_{\alpha 3}) - (m_{\alpha 2} - m_{ \alpha 1})\}/4.
\label{formalabcdwithm}
\end{eqnarray}
For the masses of quarks and leptons to agree with the experimentally
determined  
$m_{u_i}/\mathrm{GeV} = 0.0004$, $1.4$, 
$180$, $285 (215)$ and 
$m_{d_i}/\mathrm{GeV} = 0.009$, $0.2$, $6.3$, $215 (285)$ for quarks, and
$m_{e_i}/GeV = 0.0005, 0.105, 1.78, 100$ and $m_{\nu_i}/\mathrm{GeV}$ let say $1.10^{-11}, 2.10^{-11}, 6.10{-11}$
and $50$ for leptons, which would agree also with what Okun and coauthors\cite{mjnokun} have found as possible values 
for masses of the fourth family, 
we find 
\begin{equation}
\begin{array}{cccc}
A_u = 116.601 & B_u = 115.899 & C_u = 26.599 & D_u = 25.901 \\
(A_u' = 99.101 & B_u' = 98.399 & C_u' = 9.099 & D_u' = 8.401)  \\
A_d = 55.377 & B_d = 55.2728 & C_d = 52.223 & D_d = 52.127 \\
(A_d' = 72.877 & B_d' = 72.773 & C_d' = 69.723 & D_d' = 69.627) \\
A_e = 25.471 & B_e = 25.419 & C_e = 24.581 & D_e = 24.529 \\
A_\nu = 12.5 & B_\nu = 12.5 & C_\nu = 12.5  & D_\nu = 12.5.  
\end{array}
\end{equation}
Values for matrix elements in  the parentheses correspond to the values of masses of quarks in the parentheses.
The mass matrices are for leptons and even for $d$ quarks very close to a ''democratic''
one. One could also notice that for 
quarks $A_{\alpha}$ are roughly proportional to the charge $Y$. 

The unitary transformations $S^{\alpha}$ diagonalizing the mass matrices $M^{\alpha}$ 
\begin{eqnarray}
S^{\alpha +} M^{\alpha} S^{\alpha} = \lambda^{\alpha},
\label{diagalpha}
\end{eqnarray}
where $\lambda^{\alpha}$ is a diagonal matrix, are due to the symmetry of Eq.(\ref{xy}), four  two by two matrices
of the type
\begin{eqnarray}
 \begin{pmatrix}I \cos \varphi & -I \sin \varphi \\
I\sin \varphi & \cos \varphi I,\;
 \end{pmatrix}
\label{sxy}
\end{eqnarray}
with $\varphi$ equal to $\pi/4$ and lead to  two two by two off diagonal matrices, which again 
can be diagonalized with the same type of matrices of  Eq.{\ref{sxy}}, again with the  $\varphi$ equal to $\pi/4$,
independently of  of matrix elements. Accordingly, the mixing matrix (the ''CKM'' matrix for four 
families)
is the identity
\begin{eqnarray}
V_{\alpha \beta} = S^{\alpha +} S^{\beta} = I.
\label{ckm}
\end{eqnarray}

We can conclude, that the symmetries, suggested by the approach, advise that 
i) the four by four mass matrices first split into two by two off diagonal matrices (splitting four families  into
two families), ii) and that the symmetry of Eq.(\ref{xy}) should be only approximate, since  the deviation from this
symmetry makes the mixing matrix different from the identity.

\section{Study of approximate symmetries of mass matrices}
\label{MJN}

We start with a mass matrix with symmetries, suggested by the approach, unifying spins and charges.
We do not treat discrete symmetries, like
the parity (P) and the charge conjugation (P) and will not accordingly pay attention on non conservation of
the (CP) symmetry, assuming that the break of (CP) symmetry is a small effect and can accordingly 
be neglected in this study, which concerns the masses and the mixing angles only.
Since the symmetries, presented in Eq.(\ref{xy}), lead to the mixing matrix which is unity, it is obvious that 
the mass matrices, determined by the operators of Eq.(\ref{yukawa}), should  in the realistic case deviate from the type
suggested by the Eq.(\ref{xy}). We expect, that the Yukawa terms from Eq.(\ref{yukawa}),
if gauge fields would appropriately be chosen, will demonstrate a weak break of this symmetry.
We shall not in this work proceed this way. We shall rather respect the suggested symmetries as much as possible, 
trying to extract out of it, when taking into account the experimental data, as much as possible with the smallest number
of fitted parameters.

We therefore assume 

i) four families of spinors, 

ii) symmetries, which suggest that four families manifest the natural
break into two by two families, 

iii) the mass matrices to be closed to the
''democratic'' ones, 

iv) small deviations from the above symmetries to determine both: the mixing matrix and teh (nonzero)
masses of the first three generations and

v) the mass matrices to be real.

Accordingly  the mass matrices should be diagonalizable in a two step process as follows
\begin{eqnarray}
\Lambda & = & S^{ 2 T}S^{ 1 T} M^{} S^{ 1}S^{
2},
 \label{twobytwoS}
 \end{eqnarray}
with 
\begin{eqnarray}
S^{1}\left( \varphi \right) & = & \left(\begin{array}{cc} 
I_{2x2}\cos \varphi & I_{2x2} \sin \varphi\\
-I_{2x2} \sin\varphi & I_{2x2}\cos\varphi 
\nonumber
\end{array}\right)\\
S^{2}\left( \varphi^{a},\varphi^{b}\right) & = & \left(\begin{array}{cc}
s^{2}\left(\varphi^{a}\right) & 0\\
0 & s^{2}\left(\varphi^{b} \right)
\nonumber
\end{array}\right)\\
s^{2}\left(\varphi\right) & = & \left(\begin{array}{cc}
\cos \varphi & \sin \varphi\\
-\sin\phi & \cos\phi
\nonumber
\end{array}\right)\\
I_{2x2} & = & \left(\begin{array}{cc}
1 & 0\\
0 & 1\end{array}\right).
\label{twobytwoSexplicit}
\end{eqnarray}
The above requirement leads to the following mass matrix
\begin{equation}
M=\left(\begin{array}{cc}
X & Y\\
Y & X+\alpha Y\end{array}\right),
\label{xyext}
\end{equation}
with the two matrices $X$ and $Y$ which are symmetric and real and with $\alpha$, which is a small real number.
We expect again that the two matrices $X$ and $Y$ are very close to each other and  that
each of them are of the same  shape as the matrix of Eq.(\ref{xyext})
\begin{equation}
X=\left(\begin{array}{cc}
a & b\\
b & a+\beta b \end{array}\right),
\label{xyextpart}
\end{equation}
where $\beta$ again is a small real number. One can then easily show, that the suggested shape of the matrices 
easily lead to two families of spinors, which have the desired properties with respect to measured masses of 
quarks and leptons. The question then arises, how many different parameters are enough to reproduce also the
up to now measured mixing matrices of three families and which predictions then comes out for masses of the fourth families
and for the not yet measured (or badly measured) matrix elements of the mixing matrices.

\subsection{Properties of mass matrices and mixing matrices of four families of spinors}
\label{predictionsq}

We treat properties of the mixing matrices and masses which follow from our choice of 
symmetries, discussed above, in details.
We may write the mixing matrix as follows 
\begin{eqnarray}
V_{\alpha \beta} & = & S_{\alpha}^{\dagger}S_{\beta}=\\
 & = & S_{\alpha}^{2T}S_{\alpha}^{1T}S_{\beta}^{1}S_{\beta}^{2}=\\
 & = & \left(\begin{array}{cc}
U_{\alpha \beta} & V_{\alpha \beta}\\
W_{\alpha \beta} & Z_{\alpha \beta}\end{array}\right)\label{Vckmsplosno}\end{eqnarray}
with  $U_{\alpha \beta}$, $V_{\alpha \beta}$, $W_{\alpha \beta}$ and $Z_{\alpha \beta}$ expressed 
with the angles as follows 
\begin{eqnarray}
U_{\alpha \beta} & = & \cos\left(\varphi_{\alpha\beta}\right)\left(\begin{array}{cc}
\cos\left(\varphi_{\alpha\beta}^{aa}\right) & -\sin\left(\varphi_{\alpha\beta}^{aa}\right)\\
\sin\left(\varphi_{\alpha\beta}^{aa}\right) & \cos\left(\varphi_{\alpha\beta}^{aa}\right)\end{array}\right)\\
V_{\alpha \beta} & = & \sin\left(\varphi_{\alpha\beta}\right)\left(\begin{array}{cc}
-\cos\left(\varphi_{\alpha\beta}^{ab}\right) & \sin\left(\varphi_{\alpha\beta}^{ab}\right)\\
-\sin\left(\varphi_{\alpha\beta}^{ab}\right) & -\cos\left(\varphi_{\alpha\beta}^{ab}\right)\end{array}\right)\\
W_{\alpha \beta} & = & \sin\left(\varphi_{\alpha\beta}\right)\left(\begin{array}{cc}
\cos\left(\varphi_{\beta\alpha}^{ab}\right) & \sin\left(\varphi_{\beta\alpha}^{ab}\right)\\
-\sin\left(\varphi_{\beta\alpha}^{ab}\right) & \cos\left(\varphi_{\beta\alpha}^{ab}\right)\end{array}\right)\\
Z_{\alpha \beta} & = & \cos\left(\varphi_{\alpha\beta}\right)\left(\begin{array}{cc}
\cos\left(\varphi_{\alpha\beta}^{bb}\right) & -\sin\left(\varphi_{\alpha\beta}^{bb}\right)\\
\sin\left(\varphi_{\alpha\beta}^{bb}\right) & \cos\left(\varphi_{\alpha\beta}^{bb}\right)\end{array}\right)\label{VckmUVWZ}\end{eqnarray}

where $\varphi_{\alpha \beta}^{ab} = \varphi_{\alpha}^{a}-\varphi_{\beta}^{b},$  since in
the mixing matrix only the differences of the angles, which are responsible for the diagonalization of the mass matrices enter.
It correspondingly follows
\begin{eqnarray}
& & V_{\alpha \beta} = \qquad\qquad\qquad\qquad\qquad\qquad \nonumber\\
& &\left(\begin{array}{cccc}
c\left(\varphi_{\alpha\beta}\right)c\left(\varphi_{\alpha\beta}^{aa}\right) & -c\left(\varphi_{\alpha\beta}\right)s\left(\varphi_{\alpha\beta}^{aa}\right) & -s\left(\varphi_{\alpha\beta}\right)c\left(\varphi_{\alpha\beta}^{ab}\right) & s\left(\varphi_{\alpha\beta}\right)s\left(\varphi_{\alpha\beta}^{ab}\right)\\
c\left(\varphi_{\alpha\beta}\right)s\left(\varphi_{\alpha\beta}^{aa}\right) & c\left(\varphi_{\alpha\beta}\right)c\left(\varphi_{\alpha\beta}^{aa}\right) & -s\left(\varphi_{\alpha\beta}\right)s\left(\varphi_{\alpha\beta}^{ab}\right) & -s\left(\varphi_{\alpha\beta}\right)c\left(\varphi_{\alpha\beta}^{ab}\right)\\
s\left(\varphi_{\alpha\beta}\right)c\left(\varphi_{\beta\alpha}^{ab}\right) & s\left(\varphi_{\alpha\beta}\right)s\left(\varphi_{\beta\alpha}^{ab}\right) & c\left(\varphi_{\alpha\beta}\right)c\left(\varphi_{\alpha\beta}^{bb}\right) & -c\left(\varphi_{\alpha\beta}\right)s\left(\varphi_{\alpha\beta}^{bb}\right)\\
-s\left(\varphi_{\alpha\beta}\right)s\left(\varphi_{\beta\alpha}^{ab}\right) & s\left(\varphi_{\alpha\beta}\right)c\left(\varphi_{\beta\alpha}^{ab}\right) & c\left(\varphi_{\alpha\beta}\right)s\left(\varphi_{\alpha\beta}^{bb}\right) & c\left(\varphi_{\alpha\beta}\right)c\left(\varphi_{\alpha\beta}^{bb}\right)\end{array}\right),\nonumber\\
  \label{eq: Vckm, nas model, splosno, explicitno}
\end{eqnarray}

One notices that when $\varphi_{\alpha\beta}$ is close to 
 $ n \pi$, with $n$ an integer, the matrix elements, which  connect the third and the fourth family with  the first two
 (or opposite) are much smaller then the matrix elements, connecting the members within  the first two or within
 the last two families. While if $\varphi_{\alpha\beta}$ is close to $n \pi/2$ the situation is reversed.
The assumption, that the symmetries, suggested by the approach unifying spins and charges, should be respected as 
much as possible (Eq.(\ref{xyext})), limits the number of free parameters in the mass matrices.
Accordingly, the model allows at most $5$ mixing angles and $2\times 4$ masses, all together $13$  free parameters for 
quarks and the same number of parameters for leptons, 
with which it should then be possible to describe 
$2\times$ four masses of quarks and $2\times $ four masses of leptons together with $2\times $ $\frac{4(4-1)}{2}=
2\times6$ members of the 
mixing matrices, that is $14$ measurable (but not yet measured) quantities for quarks and the same number for leptons. 

The mixing matrix, which corresponds to two step diagonalization of 
$\alpha $ and $\beta $ mass matrices, has five angles as free parameters, as we have seen above. To reduce further
the number of free parameters, let us remind the reader that the (approximate) symmetry suggested by the approach 
unifying spins and charges, lead to the angles $\varphi_{\alpha}, \varphi_{\alpha}^a $ and $\varphi_{\alpha}^b$ which all are
equal to $\pi/4$. It seems straightforward to assume in the first approximation that each of the transformation 
$S^{\alpha}$ ($\alpha = u,d$ in this particular case) contribute the same ammount to the mixing angles. 
This assumtion then further reduces the number of angles of the mixing matrix to only three, we call them
$\varphi, \varphi^a$ and $\varphi^b$.
Then we have antisymmetric mixing matrix of the following form
\begin{equation}
V_{ud}=\left(\begin{array}{cccc}
c\left(\varphi\right)c\left(\varphi^a\right) & -c\left(\varphi\right)s\left(\varphi\right) & -s\left(\varphi\right)c
\left(\varphi^{ab}\right) & s\left(\varphi\right)s\left(\varphi^{ab}\right)\\
c\left(\varphi\right)s\left(\varphi^a\right) & c\left(\varphi\right)c\left(\varphi^a\right) & -s\left(\varphi\right)s
\left(\varphi^{ab}\right) & -s\left(\varphi\right)c\left(\varphi^{ab}\right)\\
s\left(\varphi\right)c\left(\varphi^{ab}\right) & s\left(\varphi\right)s\left(\varphi^{ab}\right) & c\left(\varphi\right)c
\left(\varphi^b\right) & -c\left(\varphi\right)s\left(\varphi^b\right)\\
-s\left(\varphi\right)s\left(\varphi^{ab}\right) & s\left(\varphi\right)c\left(\varphi^{ab}\right) & c\left(\varphi\right)s
\left(\varphi^b\right) & c\left(\varphi\right)c\left(\varphi^b\right)\end{array}\right),\label{Vckmexp1}\end{equation}
with $\varphi^{ab} = (\varphi^a + \varphi^b)/2 = -  \varphi^{ba}$.

If we fit these tree angles by the measured mixing matrix elements, the rest of the matrix elements for the four families
are within our model uniquely (within measuring errors) determined.

It rests then to determine the two quark masses of the fourth family. At the same time the whole mass matrices for $u$
and $d $ follow. Since the mass matrices are only measured in the diagonal form, the number of free parameters is still too
large, so that the model would predict the two masses. 

One further assumption, which requires that the matrix elements of the mass matrix are equal
(which would in the simplified model of Eq.(\ref{formalabcdwithm}) mean that $B=D$), reduces the whole number of
free parameters
of the model to 10 and accordingly the fourth quarks masses can be predicted.

Since the three mixing angles can only approximately be determined, due to experimental errors, we used the 
Monte-Carlo method to find the best fit.

\section{Results and conclusions}
\label{conclusions}

In this work we study the propreties of the mixing matrices and the mass matrices for four generations of 
quarks and leptons. We followed the suggestions, which the approach unifying spins and charges offers for
these properties. This approach namely offers the mechanism for generating families, as well as the operators
accompanied by the corresponding gauge fields, which cause  transitions from right handed weak chargeless quarks and leptons 
to left handed weak charged quarks and leptons, just as the Standard electroweak model postulates and makes the
realization of by ''dressing'' the right handed fermions with the Higgs field.

The most symmetric mass matrices - the ''democratic'' ones or even the less 
symmetric ones from Eq.(\ref{xy}) lead to the
mixing matrix, which is unity. It means accordingly, that the non unity mixing matrices between the 
fermions of different flavour may be due to small deviations from the (by the approach unifying spins and charges)
suggested approximate symmetry of Eq.(\ref{xy}). To keep the symmetry that the four by four mass matrices first split
into
the two by two off diagonal matrices (demonstrating the split of four families into two families), we have to require
the mass matrix of Eq.(\ref{xyext}). 

We neglect small effects as it is the nonconservation of the
(CP) symmetry and accordingly work with only real matrix elements of mass matrices. 
We treat in this work only quarks. Since we assume that both $u$ and $d$ quarks are equally ''responsible'' for
the effect, that the mixing matrix is not just the unity, we end up with $3$ free parameters for the antisymmetric mixing matrix, which then, taking into 
account the measured values of the matrix elements,
predicts the whole  mixing matrix
\[
\left(\begin{array}{cccc}
0.9730-0.9746 & 0.2174-0.2241 & 0.0030-0.0044 & \mathbf{0.039-0.044}\\
0.213-0.226 & 0.968-0.975 & 0.039-0.044 & \mathbf{0.0030-0.0044}\\
\;\;\;\;\,0-0.08 & \;\;\;\;\,0-0.11 & \mathbf{0.900-0.935} & \mathbf{0.350-0.434}\\
\mathbf{0.039-0.044} & \mathbf{0.0030-0.0044} & \mathbf{0.350-0.434} & \mathbf{0.900-0.935}\end{array}\right)\]

We predict in addition the $u$ quark  mass of the fourth family.We find 
(preliminarily) for the mass of the fourth $u$ quark the value 
$m_{u_{4}}  = (210.5\pm11.5)\; \mathrm{GeV}$.
The work is in progress.

\section*{Acknowledgments} We would like to express many thanks to Ministry of education, Science and sport for the
grant.
It is a pleasure to thank all the participants of the seventh  workshop What comes beyond the Standard model at Bled, July
2004 for fruitful discussions.

\title{No--scale Supergravity and the Multiple Point Principle}
\author{ C.Froggatt${}^{1}$, L.Laperashvili${}^{2}$, R.Nevzorov${}^{3,2}$, H.B.Nielsen${}^{4}$}
\institute{%
${}^{1}$ Department of Physics and Astronomy, Glasgow University, Scotland\\
${}^{2}$ Theory Department, ITEP, Moscow, Russia\\ 
${}^{3}$ School of Physics and Astronomy, University of Southampton, UK\\
${}^{4}$ The Niels Bohr Institute, Copenhagen, Denmark}
\titlerunning{No--scale Supergravity and the Multiple Point Principle}
\authorrunning{C.Froggatt, L.Laperashvili, R.Nevzorov, H.B.Nielsen}
\maketitle

\begin{abstract}
\noindent We review symmetries protecting a zero value for the
cosmological constant in no--scale supergravity and reveal the
connection between the Multiple Point Principle, no--scale and
superstring inspired models.

\end{abstract}

\section{Introduction}

Nowadays the existence of a tiny energy density spread all over
the Universe (the cosmological constant), which is responsible for
its accelerated expansion, provides the most challenging problem
for modern particle physics. A fit to the recent data shows that
$\Lambda \sim 10^{-123}M_{Pl}^4 \sim 10^{-55} M_Z^4$ \cite{CDF0}. At
the same time the presence of a gluon condensate in the vacuum is
expected to contribute an energy density of order
$\Lambda_{QCD}^4\simeq 10^{-74}M_{Pl}^4$. On the other hand if we
believe in the Standard Model (SM) then a much larger contribution
$\sim v^4\simeq 10^{-62}M_{Pl}^4$ must come from the electroweak
symmetry breaking. The contribution of zero--modes is expected to
push the vacuum energy density even higher up to $\sim M_{Pl}^4$.
Thus, in order to reproduce the observed value of the cosmological
constant, an enormous cancellation between the various
contributions is required. Therefore the smallness of the
cosmological constant should be considered as a fine-tuning
problem. For its solution new theoretical ideas must be employed.

Unfortunately the cosmological constant problem can not be
resolved in any available generalization of the SM. An exact
global supersymmetry (SUSY) ensures zero value for the vacuum
energy density. But in the exact SUSY limit bosons and fermions
from one chiral multiplet get the same mass. Soft supersymmetry
breaking, which guarantees the absence of superpartners of
observable fermions in the $100\,\mbox{GeV}$ range, does not
protect the cosmological constant from an electroweak scale mass
and the fine-tuning problem is re-introduced.

It was argued many years ago that soft breaking of global
supersymmetry at low energies could be consistent with a zero
value for the cosmological constant in the framework of $N=1$
supergravity (SUGRA) models \cite{CDF01}. Moreover there is a class
of models (so called no--scale supergravity) where the vacuum
energy density vanishes automatically \cite{CDF34}. It happens
because no--scale models possess an enlarged global symmetry. Even
after breaking, this symmetry still protects zero vacuum energy
density at the tree level. All vacua in the no--scale models are
degenerate, which provides a link between no--scale supergravity
and the Multiple Point Principle (MPP) \cite{CDF02}. MPP postulates
that in Nature as many phases as possible, which are allowed by
the underlying theory, should coexist. On the phase diagram of the
theory it corresponds to the special point -- the multiple point
-- where many phases meet. According to the MPP, the vacuum energy
densities of these different phases are degenerate at the multiple
point.

This article is organized as follows: in section 2 we describe the
structure of $(N=1)$ SUGRA models; in section 3 we study
symmetries protecting the zero value of the cosmological constant
in the no-scale models ignoring the superpotential; the no--scale
models with a non--trivial superpotential are considered in
section 4. The connection between the MPP, no--scale and
superstring inspired models is discussed in section 5.

\section{$N=1$ supergravity}

The full $N=1$ SUGRA Lagrangian (see \cite{CDF34},\cite{CDF21})
is specified in terms of an analytic gauge kinetic function
$f_a(\phi_{M})$ and a real gauge-invariant K\"{a}hler function
$G(\phi_{M},\bar{\phi}_{M})$, which depend on the chiral superfields
$\phi_M$. The function $f_{a}(\phi_M)$ determines the kinetic
terms for the fields in the vector supermultiplets and the gauge
coupling constants $Re f_a(\phi_M)=1/g_a^2$, where the index $a$
designates different gauge groups. The K\"{a}hler function is a
combination of two functions
\begin{equation}
G(\phi_{M},\bar{\phi}_{M})=K(\phi_{M},\bar{\phi}_{M})+
\ln|W(\phi_M)|^2\, ,
\label{CDF22}
\end{equation}
where $K(\phi_{M},\bar{\phi}_{M})$ is the K\"{a}hler potential whose
second derivatives define the kinetic terms for the fields in the
chiral supermultiplets. $W(\phi_M)$ is the complete analytic
superpotential of the considered SUSY model. In this article
standard supergravity mass units are used:
$\displaystyle\frac{M_{Pl}}{\sqrt{8\pi}}=1$.

The SUGRA scalar potential can be presented as a sum of $F$-- and D--terms
$V=V_{F}+V_{D}$, where the F--part is given by \cite{CDF34},\cite{CDF21}
\begin{equation}
\begin{array}{c}
V_{F}(\phi_M,\bar{\phi}_M)=e^{G}\left(\displaystyle\sum_{M,\,\bar{N}} G_{M}G^{M\bar{N}}
G_{\bar{N}}-3\right)\, ,\\[3mm]
G_M \equiv\partial_{M} G\equiv\partial G/\partial \phi_M,
\qquad G_{\bar{M}}\equiv
\partial_{\bar{M}}G\equiv\partial G/ \partial \phi^{*}_M\, , \\[3mm]
G_{\bar{N}M}\equiv
\partial_{\bar{N}}\partial_{M}G=\partial_{\bar{N}}\partial_{M}K\equiv
K_{\bar{N}M}\, . \end{array} \label{CDF6} \end{equation} The matrix $G^{M\bar{N}}$ is
the inverse of the K\"{a}hler metric $K_{\bar{N}M}$. In order to
break supersymmetry in $(N=1)$ SUGRA models, a hidden sector is
introduced. It contains superfields $(h_m)$, which are singlets
under the SM $SU(3)\times SU(2)\times U(1)$ gauge group. If, at
the minimum of the scalar potential (\ref{CDF6}), hidden sector
fields acquire vacuum expectation values so that at least one of
their auxiliary fields 
\begin{equation}
F^{M}=e^{G/2}\sum_{\bar{P}}G^{M\bar{P}}G_{\bar{P}} \label{CDF61} 
\end{equation}
is non-vanishing, then local SUSY is spontaneously broken. At the
same time a massless fermion with spin $1/2$ -- the goldstino --
is swallowed by the gravitino which becomes massive
$m_{3/2}=<e^{G/2}>$. This phenomenon is called the super-Higgs
effect.

It is assumed that the superfields of the hidden sector interact
with the observable ones only by means of gravity. Therefore they
are decoupled from the low energy theory; the only signal they
produce is a set of terms that break the global supersymmetry of
the low-energy effective Lagrangian of the observable sector in a
soft way. The size of all soft SUSY breaking terms is
characterized by the gravitino mass scale $m_{3/2}$.

In principle the cosmological constant in SUGRA models tends to be
huge and negative. To show this, let us suppose that, the
K\"{a}hler function has a stationary point, where all
derivatives $G_M=0$. Then it is easy to check that this point is
also an extremum of the SUGRA scalar potential. In the vicinity of
this point local supersymmetry remains intact while the energy
density is $-3<e^{G}>$, which implies the vacuum energy density
must be less than or equal to this value. In general enormous
fine-tuning is required to keep the cosmological constant around
its observed value in supergravity theories.
\vspace{10mm}

\section{$SU(1,1)$ and $SU(n,1)$ symmetries in the no--scale models}

We know that the smallness of the parameters in a physical theory
can usually be related to an almost exact symmetry. Since the
cosmological constant is extremely tiny, one naturally looks for a
symmetry reason to guarantee its smallness in supergravity. In the
simple case when there is only one singlet chiral multiplet
$\hat{z}$, the scalar potential can be written as 
\begin{equation}
V(z,\bar{z})=9e^{4G/3}G_{z\bar{z}}
\left(\partial_z\partial_{\bar{z}}e^{-G/3}\right)\,. 
\label{CDF31}
\end{equation} 
In order that the vacuum energy density of $V(z,\bar{z})$
should vanish, we must either choose some parameters inside $G$ to
be fine--tuned or, alternatively, demand that the K\"{a}hler
function $G$ satisfies the differential equation
$\partial_z\partial_{\bar{z}}e^{-G/3}=0$, whose solution is
\cite{CDF31}: 
\begin{equation} 
G=-3\ln\left(f(z)+f^{*}(\bar{z})\right)\,.
\label{CDF32} 
\end{equation} 
For the K\"{a}hler function given by
Eq.~(\ref{CDF32}), fine--tuning is no longer needed for the vanishing
of the vacuum energy, since the scalar potential is flat and
vanishes at any point $z$. The kinetic term for the field $z$ is
then given by \begin{equation} L_{kin}=\displaystyle\frac{3\partial_z
f(z)\partial_{\bar{z}}f^{*}(\bar{z})}{(f(z)
+f^{*}(\bar{z}))^2}\left|\partial_{\mu}z\right|^2=
\displaystyle\frac{3\left|\partial_{\mu}f(z)\right|^2}{(f(z)+f^{*}(\bar{z}))^2}\,.
\label{CDF33} 
\end{equation} 
As follows from Eq.~(\ref{CDF33}), $L_{kin}$ can be
rewritten so that only the field $T=f(z)$ appears in the kinetic
term. Actually this holds for the whole Lagrangian. The considered
theory depends only on the field $T$ and all theories obtained by
the replacement $T=f(z)$ are equivalent.

One expects that such a theory with a completely flat potential
possesses an enlarged symmetry. For the case
$\mbox{$T=(z+1)/(z-1)$}$ the scalar kinetic term becomes
$$
L_{kin}=\displaystyle\frac{3|\partial_{\mu}z|^2}{(|z|^2-1)^2}
$$
which is evidently invariant under the following set of
transformations: \begin{equation} z\to\displaystyle\frac{az+b}{b^{*}z+a^{*}}\,.
\label{CDF34} \end{equation} The set of transformations (\ref{CDF34}) forms the
group $SU(1,1)$, which is non--compact and characterized by the
parameters $a$ and $b$ which obey $|a|^2-|b|^2=1$. Hence $SU(1,1)$
is a three--dimensional group. Transformations of $SU(1,1)$ are
defined by $2\times 2$ matrices
$$
U=\left(
\begin{array}{cc}
a & b\\
b^{*}& a^{*}
\end{array}
\right)\,,
$$
which can also be written in the form \cite{CDF32} 
\begin{equation}
U=\exp\biggl\{i\frac{\omega_0}{2}\sigma_3+i\frac{\omega^{*}}{2}\sigma_{-}
-i\frac{\omega}{2}\sigma_{+}\biggr\}\,, \qquad
\sigma_{\pm}=(\sigma_1\pm i\sigma_2)/2 
\label{CDF35} 
\end{equation} 
Here $\omega_0$ is a real parameter and $\sigma_{1,2,3}$ are the
conventional Pauli matrices. The matrices $U$ acting on the space
$ \displaystyle \left( \begin{array}{c} x\\ y \end{array} \right) $ leave the element $|x|^2-|y|^2$ invariant, in
contrast with the $SU(2)$ group where we have invariance of the
element $|x|^2+|y|^2$\,. The $SU(1,1)$ transformations of the
field variable $T$ are
$$
T\to\frac{(\alpha T+i\beta)}{(i\gamma T +\delta)}\,
\qquad \alpha\delta+\beta\gamma=1\,,
$$
where $\alpha$, $\beta$, $\gamma$ and $\delta$ are real parameters.

The group $SU(1,1)$ contains the following subgroups \cite{CDF33}:
\begin{equation}
\begin{array}{rll}
i)& \mbox{Imaginary translations:}&\qquad  T\to T+i\beta;\\[3mm]
ii)& \mbox{Dilatations:} & \qquad T\to\alpha^2 T ;\\[1mm]
iii)&\mbox{Conformal transformations:}&\qquad
T\to\displaystyle\frac{\cos\theta T+i\sin\theta}{i\sin\theta T+\cos\theta}\,. 
\end{array} 
\label{CDF36} 
\end{equation} 
The K\"{a}hler function
(\ref{CDF32}) is invariant under the first set of transformations,
but not under dilatations and conformal transformations. The
gravitino mass term in the SUGRA Lagrangian, which appears when
SUSY is broken, results in the breaking of $SU(1,1)\to U_a(1)$,
where $U_a(1)$ is a subgroup of imaginary translations. One can
wonder whether $SU(1,1)$ invariance implies a flat potential. The
invariance of the scalar potential with respect to imaginary
translations implies that $V(z,\bar{z})$ is a function of the sum
$z+\bar{z}$. At the same time the invariance under dilatation
forces $V(z,\bar{z})$ to depend only on the ratio $z/\bar{z}$.
These two conditions are incompatible unless $V(z,\bar{z})$ is a
constant. Moreover the $SU(1,1)$ invariance requires this constant
to be zero \cite{CDF33}. In order to get a flat non--zero potential
in SUGRA models, one should break $SU(1,1)$. The $SU(1,1)$
structure of the $N=1$ SUGRA Lagrangian can have its roots in
supergravity theories with extended supersymmetry ($N=4$ or $N=8$)
\cite{CDF34}.

Let us consider a SUGRA model in which there are $n$ chiral
multiplets $z$ and $\varphi_i$, $i=1,2,...n-1$, where $z$ is a
singlet field while $\varphi_i$ are non--singlets under the gauge
group. If the K\"{a}hler function has the form 
\begin{equation}
G=-3\ln\left(f(z)+f^{*}(\bar{z})+g(\varphi_i,\bar{\varphi}_i)\right)\,,
\label{CDF37} 
\end{equation} 
then the F--part of the scalar potential vanishes
and only D--terms give a non--zero contribution, so that 
\begin{equation}
V=\displaystyle\frac{1}{2}\sum_{a}(D^{a})^2\,,\qquad
D^{a}=g_{a}\sum_{i,\,j}\left(G_i
T^a_{ij}\varphi_j\right)\,, 
\label{CDF38} 
\end{equation} 
where $g^a$ is the gauge coupling constant associated with the 
generator $T^a$ of the gauge transformations. Owing to the particular
form of the K\"{a}hler function (\ref{CDF37}), the scalar potential
(\ref{CDF38}) is positive definite. Its minimum is attained at the
points for which $<D^{a}>=0$ and the vacuum energy density
vanishes \cite{CDF35}.

In the case when $g(\varphi_i,\bar{\varphi}_i)=-\sum_i\varphi_i
\bar{\varphi}_i$, the kinetic terms of the scalar fields are
invariant under the isometric transformations of the non--compact
$SU(n,1)$ group \cite{CDF35}. The manifestation of the extended
global symmetry of $L_{kin}$ can be clearly seen, if one uses new
field variables $y_i$, i=0,1,...n-1, related to $f(z)$ and
$\varphi_i$ by
$$
f(z)=\frac{1-y_0}{2(1+y_0)}\,,\qquad \varphi_i=\frac{y_i}{1+y_0}\,.
$$
Then the K\"{a}hler function takes the form
$$
G=-3\ln\left(1-\sum_{i=0}^{n-1}y_i\bar{y}_i\right)
+3\ln|1+y_0|^2\,,
$$
from which it follows that the kinetic terms of the scalar fields
are 
\begin{equation}
L_{kin}=\displaystyle\sum_{j}\frac{3\partial_{\mu}y_j\partial_{\mu}\bar{y}_j}
{\left(1-\sum_{i}y_i\bar{y}_i \right)^2}\,. 
\label{CDF39} 
\end{equation} 
In particular the kinetic terms (\ref{CDF39}) remain intact if 
\begin{equation}
y_i\to \displaystyle\frac{a_i y_i+b_i}{b_i^{*}y_i+a_i^{*}}\,;\qquad
y_j\to\displaystyle\frac{y_j}{b^{*}_iy_i+a^{*}_i}\,\quad 
\mbox{for}\quad i\ne j\,, 
\label{CDF310} 
\end{equation} 
where $|a_i|^2-|b_i|^2=1$. The $SU(n,1)$ symmetry implies a zero 
contribution of the $F$--terms to the potential, which protects the 
vacuum energy density.

The $SU(n,1)$ symmetry can be derived from an extended ($N\ge 5$)
supergravity theory \cite{CDF36}. This symmetry is broken by the
gauge interactions (D--terms) in $N=1$ supergravity models,
leaving only an $SU(1,1)$ symmetry. In terms of the symmetry
transformations (\ref{CDF310}), the kinetic terms and scalar
potential are still invariant with respect to the replacement 
\begin{equation}
y_0\to \displaystyle\frac{a_0 y_0+b_0}{b_0^{*}y_0+a_0^{*}}\,;\qquad
y_i\to\displaystyle\frac{y_i}{b^{*}_0y_0+a^{*}_0}\,\quad \mbox{for}\quad
i\ne 0\,. 
\label{CDF311} 
\end{equation} 
The gravitino mass breaks $SU(1,1)$ further to $U_a(1)$, since 
the K\"{a}hler function (\ref{CDF37}) is not invariant under 
the dilatation subgroup.

\section{No--scale models with nontrivial superpotential and MPP}

The introduction of the superpotential complicates the analysis.
Suppose that the K\"{a}hler potential $K$ of the model is given
by Eq.~(\ref{CDF37}) and the superpotential does not depend on the
singlet superfield $z$. Then one can define the vector $\alpha_i$
\begin{equation} 
\alpha_i=e^{-K/3}\biggl[\displaystyle\frac{1}{3}F_i(\varphi_{\alpha})-
\displaystyle\frac{3+\sum_j
g_{\bar{j}}(\varphi_{\alpha},\bar{\varphi}_{\alpha})
F_{j}(\varphi_{\alpha})}{3|\partial_z f(z)|^2}f_i(z)
\biggr]\,, 
\label{CDF312} 
\end{equation} 
where $F(\varphi_{\alpha})=\ln W(\varphi_{\alpha})$ and the indices i and j on 
the functions $f(z)$, $g(\varphi_{\alpha},\bar{\varphi}_{\alpha})$ and 
$F(\varphi_{\alpha})$ denote the derivatives with respect to $z$ and $\varphi_{\alpha}$. 
The vector $\alpha_i$ satisfies the following property
$$
\sum_{j}G_{i\bar{j}}\alpha_j=G_i
$$
from which one deduces that 
\begin{equation}
\sum_{i,\,k}G_{i}G^{i\bar{k}}G_{\bar{k}}=\sum_{k}\alpha_{k}G_{\bar{k}}\,.
\label{CDF313} 
\end{equation} 
As a result the scalar potential takes the form
\begin{equation} 
V=\frac{1}{3}e^{2K/3}\sum_{\alpha}\biggl|\displaystyle\frac{\partial
W(\varphi_{\alpha})}{\partial\varphi_{\alpha}}
\biggr|^2+\displaystyle\frac{1}{2}\sum_{a}(D^{a})^2\,. 
\label{CDF314}
\end{equation} 
The potential (\ref{CDF314}) leads to a supersymmetric particle
spectrum at low energies. It is positive definite and its minimum
is reached when $\biggl<\displaystyle\frac{\partial
W(\varphi_{\alpha})}{\partial \varphi_{\alpha}}\biggr>=\,
<D^{a}>=0$, so that the cosmological constant goes to zero.

It is interesting to investigate what kind of symmetries protect
the cosmological constant when $W(z,\varphi_{\alpha})\ne const$.
As discussed above, it is natural to seek such symmetries within
the subgroups of $SU(1,1)$. The invariance of the K\"{a}hler
function under the imaginary translations of the hidden sector
superfields 
\begin{equation} 
z_i\to z_i+i\beta_i\,;\qquad
\varphi_{\alpha}\to\varphi_{\alpha} 
\label{CDF315} 
\end{equation} 
implies that the K\"{a}hler potential depends only on 
$z_i+\bar{z}_i$, while the superpotential is given by 
\begin{equation}
W(z_i,\varphi_{\alpha})=\exp\left\{\sum_{i=1}^{m} a_i
z_i\right\}\tilde{W}(\varphi_{\alpha})\,, 
\label{CDF316} 
\end{equation} 
where $a_i$ are real. Here we assume that the hidden sector involves $m$
singlet superfields. Since $G(\phi_M,\bar{\phi}_M)$ does not
change if
$$
\left\{
\begin{array}{l}
K(\phi_M,\bar{\phi}_M)\to K(\phi_M,\bar{\phi}_M)-g(\phi_M)
-g^{*}(\bar{\phi}_M)\,,\\[2mm]
W(\phi_M)\to \displaystyle e^{g(\phi_M)}W(\phi_M)
\end{array}
\right.\,.
$$
the most general K\"{a}hler function can be written as
\begin{equation}
G(\phi_M,\bar{\phi}_M)=K(z_i+\bar{z}_i,\varphi_{\alpha},
\bar{\varphi}_{\alpha})+\ln|W(\varphi_{\alpha})|\,,
\label{CDF318}
\end{equation}
where $W(\varphi_{\alpha})=\tilde{W}(\varphi_{\alpha})$.

The dilatation invariance constrains the K\"{a}hler potential
and superpotential further. Suppose that hidden and observable
superfields transform differently 
\begin{equation} 
z_i\to\alpha^k z_i\,,\qquad
\varphi_{\sigma}\to\alpha\varphi_{\sigma}\,. 
\label{CDF319} 
\end{equation} 
Then the superpotential $W(\varphi_{\alpha})$ may contain either bilinear or
trilinear terms involving the chiral superfields $\varphi_{\alpha}$ 
but not both. Because in phenomenologically acceptable
theories the masses of the observable fermions are generated by
trilinear terms, all others should be omitted. If there is only
one field $T$ in the hidden sector, then the K\"{a}hler function
is fixed uniquely by the gauge invariance and symmetry transformations 
(\ref{CDF315}) and (\ref{CDF319}): 
\begin{equation}
\begin{array}{c}
K(T+\bar{T},\varphi_{\sigma},\bar{\varphi}_{\sigma})=-\displaystyle\frac{6}{k}\ln(T+\bar{T})+
\sum_{\sigma} C_{\sigma}\frac{|\varphi_{\sigma}|^2}{(T+\bar{T})^{2/k}}\\[2mm]
W(\varphi_{\alpha})=\displaystyle\sum_{\sigma,\beta,\gamma}\displaystyle\frac{1}{6}
Y_{\sigma\beta\gamma}\varphi_{\sigma}\varphi_{\beta}\varphi_{\gamma}\,, 
\end{array} 
\label{CDF320}
\end{equation} 
where $C_{\sigma}$ and $Y_{\sigma\beta\gamma}$ are constants.
The scalar potential of the hidden sector induced by the
K\"{a}hler function, with $K$ and $W$ given by Eq.~(\ref{CDF320}),
is
$$
V(T+\bar{T})=\displaystyle\frac{3}{(T+\bar{T})^{6/k}}\biggl[\displaystyle\frac{2}{k}-1\biggr]
$$
and vanishes when $k=2$. In this case the subgroups of $SU(1,1)$
--- imaginary translations and dilatations ($T\to\alpha^2 T$,
$\varphi_{\sigma}\to \alpha\varphi_{\sigma}$) --- keep the value
of the cosmological constant equal to zero.

The invariance of the K\"{a}hler function with respect to
imaginary translations and dilatations prevents the breaking of
supersymmetry. In order to demonstrate this, let us consider the
SU(5) SUSY model with one field in the adjoint representation
$\Phi$ and with one singlet field $S$. The superpotential that
preserves gauge and global symmetries has the form 
\begin{equation}
W(S,\Phi)=\displaystyle\frac{\varkappa}{3}S^3+\lambda \mbox{Tr}\Phi^3+\sigma
S \mbox{Tr}\Phi^2\,. 
\label{CDF321} 
\end{equation} 
In the general case the
minimum of the scalar potential, which is induced by the
superpotential (\ref{CDF321}), is attained when $<S>=<\Phi>=0$ and
does not lead to the breaking of local supersymmetry or of gauge
symmetry. But if $\varkappa=-40\sigma^3/(3\lambda^2)$ there is a
vacuum configuration \begin{equation} <\Phi>=\displaystyle\frac{\Phi_0}{\sqrt{15}}\left(
\begin{array}{ccccc}
1 & 0 & 0 & 0 & 0 \\[0mm]
0 & 1 & 0 & 0 & 0 \\[0mm]
0 & 0 & 1 & 0 & 0 \\[0mm]
0 & 0 & 0 & -3/2 & 0 \\[0mm]
0 & 0 & 0 & 0 & -3/2
\end{array}
\right)\,,\qquad
\begin{array}{l}
<S>=S_0\,, \\
\\
\Phi_0=\displaystyle\frac{4\sqrt{15}\sigma}{3\lambda}S_0\,, 
\end{array} 
\label{CDF322}
\end{equation} 
which breaks SU(5) down to $SU(3)\times SU(2)\times U(1)$.
However, along the valley (\ref{CDF322}), the superpotential and all
auxiliary fields $F_i$ vanish preserving supersymmetry and the
zero value of the vacuum energy density.

In order to get a vacuum where local supersymmetry is broken, one
should violate dilatation invariance, allowing the appearance of
the bilinear terms in the superpotential of SUGRA models.
Eliminating the singlet field from the considered SU(5) model and
introducing a mass term for the adjoint representation, we get 
\begin{equation}
W(\Phi)=M_X\mbox{Tr}\Phi^2+\lambda \mbox{Tr}\Phi^3\,. 
\label{CDF323}
\end{equation} 
In the resulting model, there are a few degenerate vacua with
vanishing vacuum energy density. For example, in the scalar
potential there exists a minimum where $<\Phi>=0$ and another
vacuum, which has a configuration similar to Eq.~(\ref{CDF322}) but
with $\Phi_0=\displaystyle\frac{4\sqrt{15}}{3\lambda}M_X$. In the first
vacuum the SU(5) symmetry and local supersymmetry remain intact,
while in the second one the auxiliary field $F_{T}$ acquires a
vacuum expectation value and a non-zero gravitino mass is
generated: 
\begin{equation} 
\begin{array}{rclcl}
<|F_T|>&\simeq&\left<\displaystyle\frac{|W(\Phi)|}{(T+\bar{T})^{1/2}}\right>&
=&m_{3/2}(T+\bar{T})\,,\\[3mm]
m_{3/2}&=&\left<\displaystyle\frac{|W(\Phi)|}{(T+\bar{T})^{3/2}}\right>&
=&\displaystyle\frac{40}{9}\frac{M_X^3}{\lambda^2(T+\bar{T})^{3/2}}\,. 
\end{array}
\label{CDF324} 
\end{equation} 
As a result, local supersymmetry and gauge symmetry are broken in the second vacuum. 
However it does not break global supersymmetry in the observable sector 
at low energies (see Eq.(\ref{CDF314})). When $M_X$ goes to zero
the dilatation invariance, SU(5) symmetry and local supersymmetry
are restored.

A simple model with the superpotential (\ref{CDF323}) can serve as a
basis for the Multiple Point Principle (MPP) assumption in  SUGRA
models, which was formulated recently in \cite{CDF37}. When applied
to supergravity, MPP implies that the scalar potential contains at
least two degenerate minima. In one of them local supersymmetry is
broken in the hidden sector, inducing a set of soft SUSY breaking
terms for the observable fields. In the other vacuum the low
energy limit of the considered theory is described by a pure
supersymmetric model in flat Minkowski space. Since the vacuum
energy density of supersymmetric states in flat Minkowski space is
just zero, the cosmological constant problem is thereby solved to
first approximation by the MPP assumption. An important point is
that the vacua with broken and unbroken local supersymmetry are
degenerate and have zero energy density in the model considered
above. However, in the vacuum where local supersymmetry is broken,
all soft SUSY breaking terms vanish making this model irrelevant
for phenomenological studies.

\section{No--scale models and the superstring}

The K\"{a}hler function and the structure of the hidden sector
should be fixed by an underlying renormalizable or even finite
theory. Nowadays the best candidate for the ultimate theory is
$E_8\times E_8$ (ten dimensional) heterotic superstring theory
\cite{CDF48}. The minimal possible SUSY--breaking sector in string
models involves dilaton ($S$) and moduli ($T_m$) superfields. The
number of moduli varies from one string model to another. But
dilaton and moduli fields are always present in four--dimensional
heterotic superstrings, because $S$ is related with the
gravitational sector while vacuum expectation values of $T_m$
determine the size and shape of the compactified space. Amongst
the moduli $T_m$ we concentrate here on the overall modulus $T$.
In this case Calabi--Yau and orbifold compactifications lead to
rather similar results for the K\"{a}hler potential,
superpotential and gauge kinetic functions at the tree level: 
\begin{equation}
\begin{array}{c}
K=-\ln(S+\bar{S})-3\ln(T+\bar{T})+\sum_{\alpha}(T+\bar{T})^{n_{\alpha}}
\varphi_{\alpha}\bar{\varphi}_{\alpha}\,,\\[2mm]
W=W^{(ind)}(S,\,T,\,\varphi_\alpha)+\displaystyle\sum_{\sigma,\beta,\gamma}
\displaystyle\frac{1}{6} Y_{\sigma\beta\gamma}\varphi_{\sigma}
\varphi_{\beta}\varphi_{\gamma}\,,\qquad f_a=k_a S\,, 
\end{array}
\label{CDF41} 
\end{equation} 
where $k_a$ is the Kac--Moody level of the gauge
factor ($k_3=k_2=\displaystyle\frac{3}{5}k_1=1$). In the case of orbifold
compactifications, the $n_{\alpha}$ are negative integers
sometimes called modular weights of the matter fields. Orbifold
models have a symmetry (``target--space duality'') which is either
the modular group $SL(2,\bf{Z})$ or a subgroup of it. Under
$SL(2,\bf{Z})$, the fields transform like 
\begin{equation} 
\begin{array}{c}
T\to\displaystyle\frac{aT-ib}{icT+d}\,,\qquad ad-bc=1\,
\quad a,\,b,\,c,\,d\,\in\, \bf{Z};\\[2mm]
S\to S\,;\qquad\varphi_{\alpha}\to(icT+
d)^{n_{\alpha}}\varphi_{\alpha}\,. 
\end{array} 
\label{CDF42} 
\end{equation} 
In the large $T$ limit of the Calabi--Yau compactifications, 
$n_{\alpha}=-1$ and the Lagrangian of the effective SUGRA models 
is also invariant with respect to the field transformations (\ref{CDF42}) 
if $n_{\alpha}=-1$. So one can see that the form of the
K$\Ddot{a}$hler function is very close to the no--scale structure
discussed in the previous sections.

In the classical limit $W^{(ind)}(S,\,T,\,\varphi_\alpha)$ is
absent. The superpotential of the hidden sector and supersymmetric
mass terms of the observable superfields may be induced by
non-perturbative corrections, which violate the invariance under
$SL(2,\bf{Z})$ symmetry. In the gaugino condensation scenario for
SUSY breaking, the superpotential of the hidden sector takes the
form: \begin{equation} W(S,\,T)\sim \exp\left\{-3S/2b_Q\right\}\,, \label{CDF43}
\end{equation} where $b_{Q}$ is the beta--function of the hidden sector gauge
group. For an $SU(N)$ model without matter superfields
$b_{Q}=3N/(16\pi^2)$. Assuming that the superpotential does not
depend on $T$, we get 
\begin{equation}
V(S,\,T)=\displaystyle\frac{1}{(S+\bar{S})(T+\bar{T})^3}\biggl|\frac{\partial
W(S)}{\partial S}-\frac{W(S)}{S+\bar{S}}\biggr|^2\,. 
\label{CDF44}
\end{equation} 
The scalar potential (\ref{CDF44}) of the hidden sector is
positive definite. All its vacua are degenerate and have zero
energy density. Among them there can be a minimum where the vacuum
expectation value of the hidden sector superpotential vanishes. It
is easy to check that, in this vacuum, local supersymmetry remains
intact. In other vacua where $<W(S)>\ne 0$ local supersymmetry is
broken, since $F_{T}\ne 0$. Thus the MPP conditions can be
realized in superstring inspired models as well.

But at low energies the SUGRA Lagrangian, corresponding to the
K\"{a}hler function given by Eq.~(\ref{CDF41}) with $n_{\alpha}=-1$
and a superpotential that does not depend on the overall modulus
$T$, exhibits structure inherent in global supersymmetry. In order
to destroy the degeneracy between bosons and fermions, the
$SL(2,\bf{Z})$ symmetry should be broken further. Non-zero gaugino
masses $M_a$ are generated when the gauge kinetic function gets a
dependence on $T$, i.e. $f_a=k_a (S-\sigma T)$. The soft scalar
masses $m_{\alpha}^2$ and trilinear couplings
$A_{\alpha\beta\gamma}$ arise for the minimal choice of the
K\"{a}hler metric of the observable superfields, when the
K\"{a}hler potential is given by 
\begin{equation}
K=-\ln(S+\bar{S})-3\ln(T+\bar{T})+
\sum_{\alpha}\varphi_{\alpha}\bar{\varphi}_{\alpha}\,. 
\label{CDF45}
\end{equation} 
In this case we have 
\begin{equation} 
A_{\alpha\beta\gamma}=3m_{3/2}\,,\qquad
m_{\alpha}^2=m_{3/2}^2, 
\label{CDF46} 
\end{equation} 
It is worth emphasizing that
the energy densities of vacua still vanish in models with the
modified gauge kinetic function and K\"{a}hler potential
(\ref{CDF45}). It clears the way to the construction of realistic
SUGRA models based on the MPP assumption.

\section*{Acknowledgements}
The authors are grateful to E.I.~Guendelman and
N.S.~Mankoc-Borstnik for stimulating questions and comments, and
S.F.~King, O.V.Kancheli, S.~Moretti, M.~Sher and
M.I.~Vysotsky for fruitful discussions. The work of RN was
supported by the Russian Foundation for Basic Research (projects
00-15-96562 and 02-02-17379) and by a Grant of the President of
Russia for young scientists (MK--3702.2004.2).

\title{The Two-Higgs Doublet Model and the Multiple Point Principle}
\author{C.Froggatt${}^{1}$, L.Laperashvili${}^{2}$,
R.Nevzorov${}^{3,2}$,
H.B.Nielsen${}^{4}$, M.Sher${}^{5}$}
\institute{%
${}^{1}$ Department of
Physics and Astronomy, Glasgow University, Scotland\\
${}^{2}$ Theory Department, ITEP, Moscow, Russia\\
${}^{3}$ School of Physics and Astronomy,
University of Southampton, UK\\
${}^{4}$ The Niels Bohr Institute, Copenhagen, Denmark\\
${}^{5}$ Physics Department, College of William and Mary,
Williamsburg, Virginia, USA 23187}
\titlerunning{The Two-Higgs Doublet Model and the Multiple Point Principle}
\authorrunning{C.Froggatt, L.Laperashvili, R.Nevzorov,
H.B.Nielsen, M.Sher}
\maketitle

\begin{abstract}
According to the multiple point principle, Nature
adjusts coupling parameters so that many vacuum states exist and
each has approximately zero vacuum energy density. We apply this
principle to the general two-Higgs doublet extension of the
Standard Model, by requiring the existence of a large set of
degenerate vacua at an energy scale much higher than the presently
realized electroweak scale vacuum. It turns out that two scenarios
are allowed. In the first scenario, a CP conserving Higgs
potential and the absence of flavour changing neutral currents are
obtained without fine-tuning. In the second scenario, the photon
becomes massive in the high scale vacua. We briefly discuss the
resulting phenomenology.
\end{abstract}

\section{Introduction}
\label{cdf2Hintroduction}

The success of the Standard Model (SM) strongly supports the concept of
spontaneous $SU(2)\times U(1)$ symmetry breaking. The mechanism of
electroweak symmetry breaking, in its minimal version, requires the
introduction of a single doublet of scalar complex Higgs fields
and leads to the existence of a neutral massive particle --- the Higgs
boson. Over the past two decades the upper \cite{cdf2H1} and lower
\cite{cdf2H1}-\cite{cdf2H2} theoretical bounds on its mass have been
established. Although the Higgs boson still remains elusive, the
combined analysis of electroweak data indicates that its mass lies
below 251 GeV with 95\% confidence level \cite{cdf2H4A}.

Recently the experimental lower limit on the Higgs mass of 115.3
GeV was set by the unsuccessful search at LEPII \cite{cdf2H4}. The
upgraded Tevatron, LHC and LC have a good chance to discover the
Higgs boson in the near future.

There is, of course, no strong argument for the existence of just
a single Higgs doublet, apart from simplicity. Indeed the
symmetries of many models for physics beyond the SM, such as
supersymmetry or the Peccei-Quinn symmetry \cite{cdf2H21}, naturally
introduce extra Higgs doublets with unit weak hypercharge. In this
paper, we consider the application of the Multiple Point
Principle to the general two Higgs doublet model, without any
symmetries imposed beyond those of the SM gauge group.

The Multiple Point Principle (MPP) \cite{cdf2H11} postulates the
co-existence in Nature of many phases, which are allowed by a
given theory. It corresponds to the special (multiple) point on
the phase diagram of the considered theory where these phases
meet. At the multiple point the vacuum energy densities (the
cosmological constants) of the neighbouring phases are degenerate.
Thus, according to MPP, Nature fine-tunes the couplings to their
values at the multiple point. We have not identified the physical
mechanism underlying MPP, but it seems likely \cite{cdf2H11} that a
mild form of non-locality is required, due to baby universes say
\cite{cdf2H14}, as in quantum gravity.

When applied to the pure SM, the MPP exhibits a remarkable agreement
\cite{cdf2H12} with the top quark mass measurements. According to MPP, the
renormalization group improved SM Higgs effective potential
\begin{equation}
V_{eff}(\phi)=-m^2(\phi)\phi^2+\displaystyle\frac{\lambda(\phi)}{2}\phi^4\,,
\label{cdf2H1}
\end{equation}
has two rings of minima with the same vacuum energy density \cite{cdf2H12}.
The radius of the little ring is equal to the electroweak vacuum expectation
value of the Higgs field $|\phi| = v = 246$ GeV. The second vacuum was
assumed to be near the fundamental scale of the theory\footnote{Here
we assume the existence of the hierarchy $v/M_{Pl} \sim 10^{-17}$.
However some of us have speculated \cite{cdf2H13} that this huge scale ratio
could be derived from MPP, as a consequence of the existence of yet
another SM vacuum at the electroweak scale, formed by the condensation
of a strongly bound S-wave state of 6 top quarks and 6 anti-top quarks.},
identified as the Planck scale $|\phi| \approx M_{Pl}$.
The mass parameter $m$ in the effective potential (\ref{cdf2H1}) has to be of the
order of the electroweak scale $v$ and is negligible compared to $M_{Pl}$.
The conditions for a second degenerate minimum of $V_{eff}$ at the
Planck scale then become
\begin{equation}
\beta_{\lambda}(\lambda(M_{Pl}),g_t(M_{Pl}),g_i(M_{Pl}))
= \displaystyle\frac{d\lambda}{d\ln\phi}(M_{Pl}) = \lambda(M_{Pl}) = 0
\label{cdf2Hsm}
\end{equation}
where $g_i(\phi)$ and $g_t(\phi)$ denote the gauge and top quark Yukawa
couplings respectively. Hence, by virtue of MPP, $\lambda(M_{Pl})$ and
$g_t(M_{Pl})$ are determined and one can compute quite precisely the
predicted top quark (pole) and Higgs boson masses using the renormalization
group flow \cite{cdf2H12}:
\begin{equation}
M_t=173\pm 5\ \mbox{GeV}\, ,\qquad M_H=135\pm 9\ \mbox{GeV}\, .
\label{cdf2H3}
\end{equation}

Here we study the MPP predictions for the general two Higgs
doublet extension of the SM \cite{cdf2H211},\cite{cdf2H16}. The structure of
the general two Higgs doublet model is outlined in the next
section. The MPP conditions are then formulated in section 3.
Section 4 contains our conclusions.

\section{Two Higgs doublet extension of the SM}
\label{cdf2H2HDM}

The most general renormalizable $SU(2)\times U(1)$ gauge invariant potential
of the model involving two Higgs doublets is given by
\begin{equation}
\begin{array}{c}
V_{eff}(H_1, H_2) = m_1^2(\Phi)H_1^{\dagger}H_1 +
m_2^2(\Phi)H_2^{\dagger}H_2
- \biggl[m_3^2(\Phi) H_1^{\dagger}H_2+h.c.\biggr]+\\[3mm]
\displaystyle\frac{\lambda_1(\Phi)}{2}(H_1^{\dagger}H_1)^2 +
\frac{\lambda_2(\Phi)}{2}(H_2^{\dagger}H_2)^2 +
\lambda_3(\Phi)(H_1^{\dagger}H_1)(H_2^{\dagger}H_2) +
\lambda_4(\Phi)|H_1^{\dagger}H_2|^2\\[3mm]
\displaystyle + \biggl[\frac{\lambda_5(\Phi)}{2}(H_1^{\dagger}H_2)^2 +
\lambda_6(\Phi)(H_1^{\dagger}H_1)(H_1^{\dagger}H_2)+
\lambda_7(\Phi)(H_2^{\dagger}H_2)(H_1^{\dagger}H_2)+h.c. \biggr]
\end{array}
\label{cdf2H4}
\end{equation}
where
$$
H_n=\left(
\begin{array}{c}
\chi^+_n\\[2mm]
(H_n^0+iA_n^0)/\sqrt{2}
\end{array}
\right) \qquad n=1,2\,.
$$
It is easy to see that the number of couplings in the two Higgs
doublet model (2HDM) compared with the SM grows from two to ten.
Furthermore, four of them $m_3^2$, $\lambda_5$, $\lambda_6$ and
$\lambda_7$ can be complex, inducing CP--violation in the Higgs
sector. In what follows we suppose that the mass parameters $m_i^2$
and Higgs self--couplings $\lambda_i$ of the effective potential
(\ref{cdf2H4}) only depend on the overall sum of the squared norms of
the Higgs doublets, i.e.
$$
\Phi^2=\Phi_1^2+\Phi_2^2\,,\qquad \Phi_n^2=H_n^{\dagger}H_n =
\frac{1}{2}\biggl[(H_n^0)^2+(A_n^0)^2\biggr]+|\chi_n^+|^2\,.
$$
The running of these couplings is described by the 2HDM renormalization group
equations \cite{cdf2H171}--\cite{cdf2H17}, where the renormalization scale is
replaced by $\Phi$.

At the physical minimum of the scalar potential (\ref{cdf2H4}) the
Higgs fields develop vacuum expectation values
\begin{equation}
<\Phi_1>=\displaystyle\frac{v_1}{\sqrt{2}}\,,\qquad\qquad
<\Phi_2>=\displaystyle\frac{v_2}{\sqrt{2}}
\label{cdf2H41}
\end{equation}
breaking the $SU(2)\times U(1)$ gauge symmetry and
generating masses for the bosons and fermions. Here the overall
Higgs norm $<\Phi>=\sqrt{v_1^2+v_2^2}=v=246\,\mbox{GeV}$ is fixed
by the electroweak scale. At the same time the
ratio of the Higgs vacuum expectation values remains arbitrary.
Hence it is convenient to introduce $\tan\beta=v_2/v_1$.

In general the Yukawa couplings of the quarks to the Higgs fields
$H_1$ and $H_2$ generate phenomenologically unwanted flavour
changing neutral currents, unless there is a protecting custodial
symmetry \cite{cdf2H19}. Such a custodial symmetry requires the
vanishing of the Higgs couplings $\lambda_6$ and $\lambda_7$. It
also requires the down-type quarks to couple to just one Higgs
doublet, $H_1$ say, while the up-type quarks couple either to
the same Higgs doublet $H_1$ (Model I) or to the second Higgs doublet
$H_2$ (Model II) but not both\footnote{Similarly the leptons are
required to only couple to one Higgs doublet, usually chosen to be the
same as the down-type quarks. However there are variations of Models I and
II, in which the leptons couple to $H_2$ rather than to $H_1$.}. If,
in addition, the Higgs coupling $\lambda_5$ vanishes, as in supersymmetric
and Peccei-Quinn models, there is no CP-violation in the Higgs sector.

We emphasize that, in this paper, we do not impose any custodial
symmetry but rather consider the general Higgs potential
(\ref{cdf2H4}). Instead we require that at some high energy scale
($M_Z<<\Lambda\lesssim M_{Pl}$), which we shall refer to as the
MPP scale $\Lambda$, a large set of degenerate vacua allowed by
the 2HDM is realized. In compliance with the MPP, these vacua and
the physical one must have the same energy density. Thus the MPP
implies that the couplings $\lambda_i(\Lambda)$ should be adjusted
with an accuracy of order $v^2/\Lambda^2$, in order to arrange an
appropriate cancellation among the quartic terms in the effective
potential (\ref{cdf2H4}).

\section{Implementation of the MPP in the 2HDM}

In this section, we aim to determine a large set of minima of the
2HDM scalar potential with almost vanishing energy density, which
may exist at the MPP high energy scale $\Lambda$ where the mass
terms in the potential can be neglected. The most general vacuum
configuration takes the form: \begin{equation} <H_1>=\Phi_1\left( \begin{array}{c}
0\\[2mm]
1
\end{array}
\right)\,,\qquad
<H_2>=\Phi_2\left(
\begin{array}{c}
\sin\theta\\[2mm]
\cos\theta\, e^{i\omega}
\end{array}
\right)\,,\\[2mm]
\label{cdf2H6}
\end{equation}
where $\Phi_1^2+\Phi_2^2=\Lambda^2$. Here, the gauge is fixed so
that only the real part of the lower component of $H_1$ gets a vacuum
expectation value.

We now consider the conditions that must be satisfied in order
that minima of $V_{eff}$ should exist for all possible values
of the phase $\omega$. The $\omega$-dependent part of the
potential takes the form:
\begin{equation}
V_{\omega} =
\frac{\lambda_5(\Phi)}{2} \Phi_1^2 \Phi_2^2 \cos^2\theta
e^{2i\omega} +\left[ \lambda_6(\Phi) \Phi_1^3 \Phi_2
+\lambda_7(\Phi) \Phi_1 \Phi_2^3  \right]\cos\theta e^{i\omega}
+h.c.
\label{cdf2HVomega}
\end{equation}
In order that $V_{\omega}$ should become
independent of $\omega$ at the MPP scale minima, we require that
the coefficients of $e^{i\omega}$ and $e^{2i\omega}$ in
(\ref{cdf2HVomega}) both vanish at $\Phi = \Lambda$. Similarly for
minima to exist for all values of $\omega$, we require the
derivatives:
\begin{eqnarray}
\lefteqn{\frac{\partial V_{\omega}}{\partial \Phi_1} =
\left[ \lambda_5(\Phi) \Phi_1 \Phi_2^2  +
\beta_{\lambda_5}(\Phi) \frac{\Phi_1^3 \Phi_2^2}{2\Phi^2}
\right] \cos^2\theta e^{2i\omega}  +
}  \nonumber \\
& & \left[3\lambda_6 \Phi_1^2 \Phi_2   +
\beta_{\lambda_6} \frac{\Phi_1^4 \Phi_2}{\Phi^2}  +
\lambda_7 \Phi_2^3  +
\beta_{\lambda_7} \frac{\Phi_1^2 \Phi_2^3}{\Phi^2}
\right]\cos\theta e^{i\omega}  +h.c.
\label{cdf2HdV1}
\end{eqnarray}
and
\begin{eqnarray}
\lefteqn{\frac{\partial V_{\omega}}{\partial \Phi_2} =
\left[ \lambda_5(\Phi) \Phi_1^2 \Phi_2  +
\beta_{\lambda_5}(\Phi) \frac{\Phi_1^2 \Phi_2^3}{2\Phi^2}
\right]\cos^2\theta e^{2i\omega}  + } \nonumber \\
& & \left[\lambda_6 \Phi_1^3   +
\beta_{\lambda_6} \frac{\Phi_1^3 \Phi_2^2}{\Phi^2}  +
3\lambda_7 \Phi_2^2 \Phi_1  +
\beta_{\lambda_7} \frac{\Phi_1 \Phi_2^4}{\Phi^2}
\right]\cos\theta e^{i\omega} +h.c.
\label{cdf2HdV2}
\end{eqnarray}
to be independent of $\omega$ at $\Phi = \Lambda$. Here
$\beta_{\lambda_i}(\Phi) = \frac{d \lambda_i}{d \ln \Phi}(\Phi)$
is the renormalisation group beta function for the Higgs self-coupling
$\lambda_i(\Phi)$.

It is readily verified (unless $\cos\theta$ = 0) that the
vanishing of the coefficients of $e^{i\omega}$ and $e^{2i\omega}$
in Eqs.~(\ref{cdf2HVomega}) - (\ref{cdf2HdV2}) leads to the conditions:
\begin{equation}
\lambda_5(\Lambda)=\lambda_6(\Lambda) = \lambda_7(\Lambda) = 0
\label{cdf2Hlambda0}
\end{equation}
and
\begin{equation}
 \beta_{\lambda_5}(\Lambda) = \beta_{\lambda_6}\Phi_1^2 +
 \beta_{\lambda_7}\Phi_2^2 = 0. \label{cdf2Hbetalambda0}
 \end{equation}
 When
$\lambda_5 = \lambda_6 = \lambda_7 = 0$, the Higgs potential
manifests an extra Peccei-Quinn-like $U(1)$ symmetry, and the only
non-vanishing contributions to the beta functions
$\beta_{\lambda_5}$, $\beta_{\lambda_6}$ and $\beta_{\lambda_7}$
arise from the Yukawa couplings to the fermion sector. We shall
consider just the third generation fermions here and neglect the
smaller Yukawa couplings from the first two generations. An
obvious method of ensuring that $\beta_{\lambda_5}$,
$\beta_{\lambda_6}$ and $\beta_{\lambda_7}$ also vanish, and
thereby satisfy Eq.~(\ref{cdf2Hbetalambda0}), is to extend the $U(1)$
symmetry to the fermion sector at the MPP scale. In other words
the Yukawa couplings at the MPP scale can be taken to be of the
2HDM Model I or Model II form discussed in section \ref{cdf2H2HDM} This
is illustrated by the explicit expression for $\beta_{\lambda_5}$
(in a notation where we have re-defined the Higgs doublets so that
the top quark only couples to $H_2$ at the MPP scale): \begin{equation}
 \beta_{\lambda_5}\biggl( \lambda_5=\lambda_6=\lambda_7=0, \Lambda
\biggr)= -\frac{1}{(4\pi)^2}\biggl[12h_b^2(\Lambda)g_b^2(\Lambda)+
4h_{\tau}^2(\Lambda) g^{\ast 2}_{\tau}(\Lambda)\biggr]\,.
\label{cdf2H34}
\end{equation}
Here $h_b$ and $g_b$ are the couplings of $H_1$ and
$H_2$ to the $b$--quark, while $h_{\tau}$ and $g_{\tau}$ are the
corresponding couplings of the Higgs doublets to the
$\tau$--lepton. For definiteness, we have chosen a phase
convention in which $h_t$, $h_b$, $h_{\tau}$ and $g_b$ are real
and $g_{\tau}$ is complex. The beta function (\ref{cdf2H34}) vanishes
when
\begin{equation}
\begin{array}{clcl} (I)\quad&
h_b(\Lambda)=h_{\tau}(\Lambda)=0\,;&\qquad\qquad (II)\quad &
g_b(\Lambda)=g_{\tau}(\Lambda)=0\,;\\
(III)\quad& h_b(\Lambda)=g_{\tau}(\Lambda)=0\,;&\qquad\qquad
(IV)\quad & g_b(\Lambda)=h_{\tau}(\Lambda)=0\,.
\end{array} \label{cdf2H35}
\end{equation}
corresponding to the 2HDM Model I and Model II Yukawa couplings
and their leptonic variations.

An alternative method of solving the MPP conditions
(\ref{cdf2Hlambda0}, \ref{cdf2Hbetalambda0}), without a
Peccei-Quinn-like $U(1)$ symmetry, is for the $b$ and $\tau$
contributions to cancel in Eq.~(\ref{cdf2H34})
with $g_{\tau}$ being imaginary. However the
manifold of such MPP solutions in the space of coupling constants
is of the same dimension as that of the Peccei-Quinn-like solutions.
Hence no fine-tuning is required to obtain one of the
Peccei-Quinn-like MPP solutions, as they are just as abundant as
MPP solutions without such a $U(1)$ symmetry. We shall therefore
concentrate on the phenomenologically favoured MPP solutions,
having the 2HDM Model I or Model II Yukawa couplings.

The Peccei-Quinn-like $U(1)$ custodial symmetry of the Higgs and
Yukawa sector implies that
\begin{equation}
\beta_{\lambda_5}(\Lambda)
= \beta_{\lambda_6}(\Lambda) = \beta_{\lambda_7}(\Lambda) = 0.
\label{cdf2Hbeta0}
\end{equation}
It then follows from Eqs.~(\ref{cdf2Hlambda0}) and
(\ref{cdf2Hbeta0}) that the renormalization group evolution does not
generate any $U(1)$ custodial symmetry breaking couplings below
the MPP scale, where we thus have:
\begin{equation}
\lambda_5(\Phi) =
\lambda_6(\Phi) = \lambda_7(\Phi)=0.
\label{cdf2Hlambda567}
\end{equation}
In this
way we have naturally obtained the absence of flavour changing neutral
currents and a CP conserving Higgs potential from the MPP
requirement that vacua at the MPP scale should be degenerate with
respect to the phase $\omega$.

We now consider whether we can impose further MPP conditions on
the couplings. The simplest way to ensure that the quartic part of
the effective potential vanishes for any vacuum configuration
(\ref{cdf2H6}) at the MPP scale is to impose the condition that all the
self-couplings should vanish there:
\begin{equation}
\lambda_1(\Lambda)=\lambda_2(\Lambda)=\lambda_3(\Lambda)=
\lambda_4(\Lambda)=\lambda_5(\Lambda)=\lambda_6(\Lambda)=
\lambda_7(\Lambda)=0\,.
\label{cdf2H5}
\end{equation}
However, further
investigation reveals that the configurations (\ref{cdf2H6}) do not
correspond to minima of the effective potential in this case. This
can be shown by consideration of the 2HDM renormalization group
equations for the quartic couplings \cite{cdf2H171}--\cite{cdf2H17}. The
detailed results depend on the choice of Model I or Model II
Yukawa couplings, but they are qualitatively similar. So, for
convenience, we shall concentrate on the Model II couplings here.

\begin{figure}[t]
  \centering
  \includegraphics[clip, trim=10 220 10 220, width=12cm]{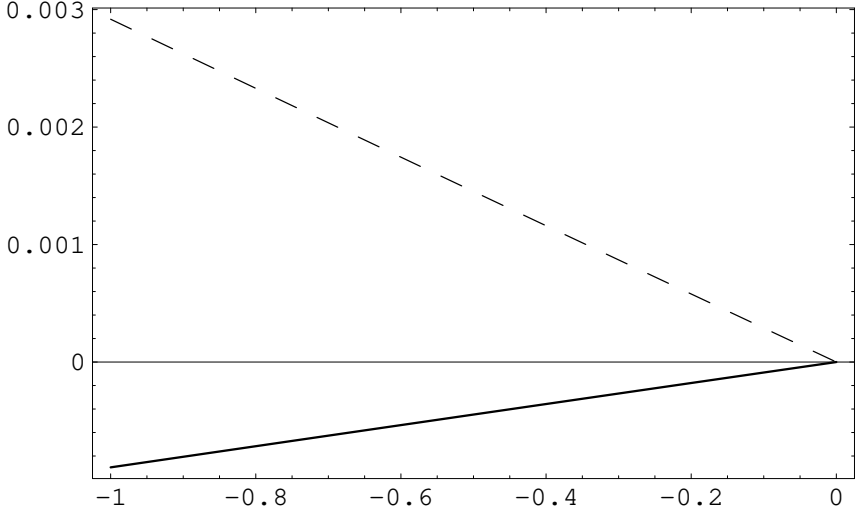}
\caption{The running of $\lambda_1$ and $\lambda_2$,
as a function of $\log[\Phi^2/M_{Pl}^2]$,
below $M_{Pl}$ for $\lambda_i(M_{Pl})=0$,
$m_t(M_t)=165\,\mbox{GeV}$ and $\alpha_3(M_Z)=0.117$. The
renormalization group flow is plotted for $\tan\beta=2$.
The solid and dashed lines correspond to $\lambda_1$
and $\lambda_2$ respectively.} \label{cdf2Hfig:higgs1}
\end{figure}

\begin{figure}[t]
  \centering
  \includegraphics[clip, trim=10 220 10 220, width=12cm]{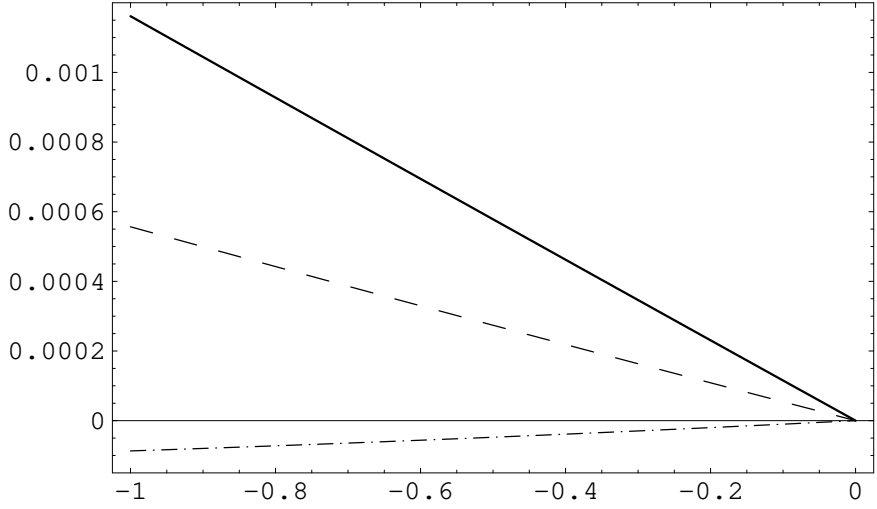}
\caption{The running of $\lambda_1$, $\lambda_2$ and
$\tilde{\lambda}$, as a function of $\log[\Phi^2/M_{Pl}^2]$,
below $M_{Pl}$ for $\lambda_i(M_{Pl})=0$,
$m_t(M_t)=165\,\mbox{GeV}$ and $\alpha_3(M_Z)=0.117$. The
renormalization group flow is plotted for $\tan\beta=50$.
The solid, dashed and dash--dotted lines correspond to
$\lambda_1$, $\lambda_2$ and $\tilde{\lambda}$
respectively.} \label{cdf2Hfig:higgs2}
\end{figure}

For moderate values of $\tan\beta$ the Higgs self--coupling
$\lambda_1$ becomes negative just below
the MPP scale (see Fig.~\ref{cdf2Hfig:higgs1}).
The renormalization group running of
$\lambda_2$ exhibits the opposite behaviour, because of the large
and negative top quark contribution to the corresponding
beta--function. This means that $V_{eff}$ does not have a
minimum at the
MPP scale and, just below it, there is a  huge negative energy density
$(V_{eff}\sim -\Lambda^4)$ where $ <\Phi_2>=0$ and
$<\Phi_1>\lesssim \Lambda$.

The renormalization group flow of $\lambda_1$
changes at very large $\tan\beta$ (see Fig.~\ref{cdf2Hfig:higgs2}).
The absolute
value of the $b$--quark and $\tau$--lepton contribution to
$\beta_{\lambda_1}$, although negligible at moderate values of
$\tan\beta$, grows with increasing $\tan\beta$. At $\tan\beta\sim
m_t(M_t)/m_b(M_t)$ their negative contribution to the
beta function of $\lambda_1$ prevails over the positive contributions
coming from loops containing Higgs and gauge bosons. The
negative sign of $\beta_{\lambda_1}$ results in
$\lambda_1(\Phi)>0$ if the overall Higgs norm $\Phi$ is less than
$\Lambda$.
However the positive sign of $\lambda_1$ does not ensure the
stability of the vacua (\ref{cdf2H6}). Substituting the vacuum
configuration (\ref{cdf2H6}) into the quartic part of the 2HDM scalar
potential, and using Eq.~(\ref{cdf2Hlambda567}), one
finds for any $\Phi$ below the MPP scale:
\begin{equation}
\begin{array}{rcl}
V(H_1,H_2)&\approx&\displaystyle\frac{1}{2}\biggl(\sqrt{\lambda_1(\Phi)}\Phi_1^2-
\sqrt{\lambda_2(\Phi)}\Phi_2^2\biggr)^2+\\[3mm]
&&+\left(\sqrt{\lambda_1(\Phi)\lambda_2(\Phi)}+\lambda_3(\Phi)+
\lambda_4(\Phi)\cos^2\theta\right) \Phi_1^2\Phi_2^2\, .
\end{array}
\label{cdf2H8}
\end{equation}
The Higgs scalar potential (\ref{cdf2H8})
attains its minimal value for $\cos\theta=0$ if $\lambda_4>0$ or
$\cos\theta=\pm 1$ when $\lambda_4<0$. For these values of $\cos \theta$,
the scalar potential can be written as
\begin{equation}
V_{eff}(H_1,H_2)\approx\displaystyle\frac{1}{2}\biggl(\sqrt{\lambda_1(\Phi)}\Phi_1^2-
\sqrt{\lambda_2(\Phi)}\Phi_2^2\biggr)^2+
\tilde{\lambda}(\Phi)\Phi_1^2\Phi_2^2\,,
\label{cdf2H9}
\end{equation}
where
$$
\tilde{\lambda}(\Phi)=\sqrt{\lambda_1(\Phi)\lambda_2(\Phi)}+\lambda_3(\Phi)+
\min\{0,\lambda_4(\Phi)\}\,.
$$
If at some intermediate scale the combination of the Higgs
self--couplings $\tilde{\lambda}(\Phi)$ is
less than zero, then there exists a minimum with negative
energy density that causes the instability of
the vacua at the electroweak and MPP scales. Otherwise
the Higgs effective potential is positive definite
and the considered vacua are stable.

In Fig.~\ref{cdf2Hfig:higgs2} the Higgs
self--couplings $\lambda_1(\Phi)$ and
$\lambda_2(\Phi)$, as well as $\tilde{\lambda}(\Phi)$,
are plotted as a function of $\Phi$ for a very large value of
$\tan\beta$. It is clear
that the vacuum stability conditions, i.e.
\begin{equation}
\lambda_1(\Phi)\gtrsim 0\,,\qquad\qquad\lambda_2(\Phi)\gtrsim 0\,,
\qquad\qquad\tilde{\lambda}(\Phi)\gtrsim 0
\label{cdf2H10}
\end{equation}
are not fulfilled simultaneously. The value of $\tilde{\lambda}(\Phi)$
becomes negative for $\Phi<\Lambda$. So we conclude that indeed the
conditions (\ref{cdf2H5}) can not provide
a self--consistent realization of the MPP in the 2HDM.

At the next stage it is worth relaxing the conditions (\ref{cdf2H5}),
by permitting $\lambda_1(\Lambda)$, $\lambda_2(\Lambda)$
and $\lambda_3(\Lambda)$ to take on non--zero values. In order to
avoid a huge and negative vacuum energy density in the
global minimum of the 2HDM effective potential that precludes
the implementation of MPP, the vacuum stability conditions
(\ref{cdf2H10}) should be satisfied for any $\Phi$ in the interval:
$v\lesssim \Phi\lesssim \Lambda$. In this case
both terms in the quartic part of the scalar potential (\ref{cdf2H8})
are positive. In order to achieve degeneracy of the vacua
at the electroweak and MPP scales, they must go to zero
separately at the scale $\Lambda$. Since $\lambda_4(\Lambda)$ is still
taken to be zero, the second term in Eq.~(\ref{cdf2H8}) vanishes when
\begin{equation}
\lambda_3(\Lambda)\simeq-\sqrt{\lambda_1(\Lambda)\lambda_2(\Lambda)}\,.
\label{cdf2H11}
\end{equation}
For finite values of $\lambda_1(\Lambda)$ and $\lambda_2(\Lambda)$
the first term in the quartic part of
the scalar potential can also be eliminated by the appropriate choice of
Higgs vacuum expectation values:
\begin{equation}
\Phi_1=\Lambda\cos\gamma\,,\qquad \Phi_2=\Lambda\sin\gamma\,,\qquad
\tan\gamma=\Biggl(\displaystyle\frac{\lambda_1(\Lambda)}{\lambda_2(\Lambda)}
\Biggr)^{1/4}\,.
\label{cdf2H12}
\end{equation}
The sum of the quartic terms in $V_{eff}(H_1,H_2)$ then tend to
zero at the MPP scale independently of the angle $\theta$ and the
phase $\omega$.

Nevertheless the situation is not as promising as it first appears,
since again we can show it does not correspond to a local minimum of
$V_{eff}$ at $\Phi = \Lambda$, in which all partial derivatives of the 2HDM
scalar potential go to zero. The degeneracy of the vacua, parameterized
by Eqs.~(\ref{cdf2H6}) and (\ref{cdf2H12}), implies that the following derivatives
\begin{equation}
\displaystyle\frac{\partial V_{eff}(H_1,H_2)}{\partial \Phi_i}\propto
\displaystyle\frac{1}{2}\beta_{\lambda_1}\tan^{-2}\gamma+
\frac{1}{2}\beta_{\lambda_2}\tan^2\gamma+\beta_{\lambda_3}+
\beta_{\lambda_4}\cos^2\theta\,.
\label{cdf2H13}
\end{equation}
should vanish at the MPP scale for any choice of $\theta$
and $\omega$. In order for these derivatives to be independent of
$\theta$, we require $\beta_{\lambda_4}(\Lambda)=0$. However,
for $\lambda_4(\Lambda)=0$, this requirement is in conflict with the
form of the beta function:
\begin{equation}
\beta_{\lambda_4}(\Lambda) =
\displaystyle\frac{1}{(4\pi)^2}\biggl[3g_2^2(\Lambda)g_1^2(\Lambda)
+12h_t^2(\Lambda)h_b^2(\Lambda)\biggr]
\label{cdf2H14}
\end{equation}
which is strictly positive. Thus our attempt to
adapt the MPP idea to the 2HDM with $\lambda_4(\Lambda)=0$ fails.
Then we have two MPP  scenarios.

When $\lambda_4(\Lambda) < 0$ (the first scenario), a
self--consistent implementation of the MPP can only be obtained if
$\lambda_1(\Lambda)$, $\lambda_2(\Lambda)$ and
$\lambda_3(\Lambda)$ have non--zero values. Then $\cos \theta =
\pm1$ near the MPP scale minima, where the Higgs effective
potential takes the form (\ref{cdf2H9}). In order to ensure the
vanishing of $V_{eff}$ at the MPP scale with an accuracy of order
$v^2/\Lambda^2$, the combination of Higgs self--couplings
$\tilde{\lambda}(\Lambda)$ must go to zero. Furthermore, if the
2HDM effective potential is to possess a set of local minima at
the MPP scale, the derivative of $\tilde{\lambda}(\Phi)$ must
vanish when $\Phi = \Lambda$ and hence
$\beta_{\tilde{\lambda}}(\Lambda) = 0$. In this first scenario,
the following set of MPP scale vacua \begin{equation}<H_1>=\left(
\begin{array}{c}
0\\ \Phi_1
\end{array}
\right)\; , \qquad <H_2>=\left(
\begin{array}{c}
0\\ \Phi_2\, e^{i\omega}
\end{array}
\right) \label{cdf2H16} \end{equation} have the same energy density for any
$\omega$. The ratio of the Higgs field norms $\Phi_1$ and $\Phi_2$
in Eq.~(\ref{cdf2H16}) is defined by the equations for the extrema of
the 2HDM scalar potential, whose solution is given by
Eq.~(\ref{cdf2H12}). In the minima (\ref{cdf2H16}) the photon remains
massless and electric charge is conserved.

In the second scenario $\lambda_4(\Lambda)>0$ the parameter
$\cos\theta$ tends to zero. We note that our general derivation of
the MPP conditions (\ref{cdf2Hlambda0}, \ref{cdf2Hbetalambda0}), and the
consequent $U(1)$ custodial symmetry without fine-tuning, breaks
down in this case, since the Higgs potential does not depend on
the phase $\omega$ near its minimum (where $\cos \theta =0$). If
$\lambda_4(\Lambda)-|\lambda_5(\Lambda)|>0$ and
$$
\lambda_1(\Lambda)=\lambda_2(\Lambda)=\lambda_3(\Lambda)=
\lambda_6(\Lambda)=\lambda_7(\Lambda)=0
$$
the following set of vacua
\begin{equation}
<H_1>=\left(
\begin{array}{c}
0\\ \Phi_1
\end{array}
\right)\; , \qquad <H_2>=\left(
\begin{array}{c}
\Phi_2\\ 0
\end{array}
\right) \label{cdf2H15} \end{equation} are degenerate for any $\Phi_1$ and
$\Phi_2$ satisfying $\Phi_1^2+\Phi_2^2=\Lambda^2$. In order to
ensure the existence of the minima given by Eq.~(\ref{cdf2H15}), the
conditions for extrema must be fulfilled which lead to
$\beta_{\lambda_1}(\Lambda)=\beta_{\lambda_2}(\Lambda)
=\beta_{\lambda_3}(\Lambda)=0$. The cancellation of different
contributions to these $\beta$-functions only becomes possible for
large values of the Yukawa couplings at the MPP scale
(corresponding to large $\tan\beta$ in Model II). The resulting
vacuum energy density vanishes because $\cos\theta$ goes to zero
in these vacua. At the set of minima (\ref{cdf2H15}) the $SU(2)\times
U(1)$ gauge symmetry is broken completely and the photon gains a
mass of the order of $\Lambda$. This is not in conflict with
phenomenology, since an MPP scale minimum is not presently
realised in Nature. However, on phenomenological grounds, we
prefer the first scenario, although it is consistent to impose an
{\it ad hoc} $Z_2$ custodial symmetry on the second scenario which
we will discuss separately elsewhere.

The conditions
\begin{equation}
\left\{
\begin{array}{l}
\lambda_5(\Lambda)=\lambda_6(\Lambda)=\lambda_7(\Lambda)
=\beta_{\lambda_5}(\Lambda) =\beta_{\lambda_6}(\Lambda)
=\beta_{\lambda_7}(\Lambda) = 0 \,,\\[2mm]
\tilde{\lambda}(\Lambda)= \beta_{\tilde{\lambda}}(\Lambda) = 0\,,
\end{array}
\right.
\label{cdf2H17}
\end{equation}
leading to the appearance of the degenerate vacua (\ref{cdf2H16}) in our
preferred scenario, should be identified
with the MPP conditions analogous to those of Eq.~(\ref{cdf2Hsm}).
The conditions (\ref{cdf2H17}) have to be supplemented
by the vacuum stability requirements (\ref{cdf2H10}), which must be valid
everywhere from the electroweak to the MPP scale. Any failure of either
the conditions (\ref{cdf2H17}) or the inequalities (\ref{cdf2H10}) prevents a
consistent realization of the MPP in the 2HDM.

We have made a detailed numerical analysis of these MPP constraints on the
Higgs spectrum in the 2HDM for high energy scales ranging from
$\Lambda = M_{Pl}$ down to $\Lambda = 10$ TeV, which we shall report
elsewhere. In the large $\tan\beta$ limit, the allowed range of the
Higgs self--couplings is severely constrained by the MPP conditions
(\ref{cdf2H17}) and vacuum stability requirements (\ref{cdf2H10}).
As a consequence, using the Model II Yukawa couplings and the
lower limit on the charged scalar mass \cite{cdf2H24} deduced from the
non-observation of $B\to X_s\gamma$ decay, the Higgs spectrum exhibits a
hierarchical structure for most of the large $\tan\beta$
($\tan\beta\gtrsim 2$) region. While the heavy scalar, pseudoscalar and
charged Higgs particles are nearly degenerate with a mass greater than
$300\,\mbox{GeV}$, the mass of the SM--like Higgs boson $m_h$ does not
exceed $180\,\mbox{GeV}$ for any scale $\Lambda\gtrsim 10\,\mbox{TeV}$.
The bounds on $m_h$ become stronger as the MPP scale is increased. For
$\Lambda=M_{Pl}$ and large values of $\tan\beta$ we find:
$m_h=137\pm 12\,\,\mbox{GeV}$. However, for very large
$\tan\beta\simeq m_t(M_t)/m_b(M_t)$ the MPP restrictions on the Higgs
self--couplings and the lightest Higgs scalar mass turn out to be
substantially relaxed, due to a loosening of the allowed upper limit
on $\lambda_2(\Lambda)$. In particular the upper bound on $m_h$ is
increased to 180 GeV for $\Lambda = M_{Pl}$ and very large $\tan\beta$.

\section{Conclusion}
We have studied the constraints imposed by the Multiple Point
Principle on the general two Higgs doublet model, by requiring the
existence of a large number of vacua at a high energy scale
$\Lambda$ which are degenerate with the electroweak scale vacuum.
The MPP conditions at the scale $\Lambda$, derived in our
preferred scenario with the vacua (\ref{cdf2H16}), are summarized in
Eq.~(\ref{cdf2H17}). In addition the vacuum stability conditions
(\ref{cdf2H10}) must be satisfied. The MPP conditions in the first line
of Eq.~(\ref{cdf2H17}) give CP invariance of the Higgs potential
and the presence of a softly broken (by the
$m^2_3H_1^{\dagger}H_2$ term in $V_{eff}$) $Z_2$ symmetry of the
usual type responsible for the absence of flavour changing neutral
currents without fine-tuning. The $Z_2$ invariance of the 2HDM
Lagrangian is not
spoiled by the renormalization group flow. This means that the MPP
provides an alternative mechanism for the suppression of flavour
changing neutral currents in the 2HDM.

In addition the MPP conditions in the second line of Eq.~(\ref{cdf2H17})
provide two relationships between the non-zero Higgs self-couplings,
at the scale $\Lambda$, which can in principle be checked when the
masses and couplings of the Higgs bosons are measured at future
colliders. It is interesting to remark that these relationships
are satisfied identically in the minimal supersymmetric standard model
at all high energy scales, where the soft SUSY breaking terms can be
neglected.

In conclusion we have constructed a new simple MPP inspired
non-super\-symmetric two Higgs doublet extension of the SM.

\section*{Acknowledgements}

The authors are grateful to E.Boos, I.Ginzburg, S.King,
M.Krawczyk, L.Okun and S.Moretti for fruitful discussions.
RN and HN are indebted to
the DESY Theory Group for hospitality extended to them in
2002 when this project was started. RN is also grateful to
Alfred Toepfer Stiftung for the scholarship and
for the favour disposition during his stay in Hamburg (2001--2002).

LL and RN would like to acknowledge the support of the
Russian Foundation for Basic Research (RFBR),
projects 00-15-96786, 00-15-96562 and 02-02-17379.
RN was also partly supported by a Grant of the President
of Russia for young scientists (MK--3702.2004.2).
The work of CF was
supported by PPARC grant PPA/G/0/2002/00463 and that of MS by the
National Science Foundation grant PHY0243400.

\title{New Physics From a Dynamical Volume Element}
\author{E. Guendelman${}^1$ and A. Kaganovich${}^1$\thanks{guendel@bgumail.bgu.ac.il, alexk@bgumail.bgu.ac.il},%
E. Nissimov${}^2$ and S. Pacheva${}^2$\thanks{nissimov@inrne.bas.bg, svetlana@inrne.bas.bg}}
\institute{%
${}^1$Department of Physics, Ben-Gurion University, Beer-Sheva, Israel\\
${}^2$Institute for Nuclear Research and Nuclear Energy,\\
Bulgarian Academy of Sciences, Sofia, Bulgaria}
\titlerunning{New Physics From a Dynamical Volume Element}
\authorrunning{E. Guendelman and A. Kaganovich,E. Nissimov and S. Pacheva}
\maketitle

\begin{abstract}
The use in the action integral of a volume element of the form
$\Phi d^{D}x$ where $\Phi$ is a metric independent measure can
give new interesting results in all types of known generally
coordinate invariant theories: (1) 4-D theories of gravity plus
matter fields; (2) Parametrization invariant theories of extended
objects; (3) Higher dimensional theories. In the case (1), a large
number of new effects appears: under normal particle physics
conditions (primordial) fermions split into three families; when
matter is highly diluted, neutrinos increase their mass and can
contribute both to dark energy and to dark matter. In the case
(2), it leads to dynamically induced tension; to string models of
non abelian confinement; to the possibility of new Weyl invariant
light-like branes which dynamically adjust themselves to sit at
black hole horizons; in the context of higher dimensional theories
it can provide examples of massless 4-D particles with nontrivial
Kaluza Klein quantum numbers. In the case  (3), i.e. in brane and
Kaluza Klein scenarios, the use of a metric independent measure
makes it possible to construct naturally models where only the
extra dimensions get curved and the 4-D remain flat.
\end{abstract}

\section{Introduction}

 We have studied models of the new class of
theories\cite{G1}-\cite{G20} based on the idea that the action
integral may contain the new metric-independent measure of
integration. For example, in four dimensions the new measure can
be built of four scalar fields $\varphi_{a}$ \, ($a=1,2,3,4$)
\begin{equation}
\Phi
=\varepsilon^{\mu\nu\alpha\beta}\varepsilon_{abcd}\partial_{\mu}\varphi_{a}
\partial_{\nu}\varphi_{b}\partial_{\alpha}\varphi_{c}
\partial_{\beta}\varphi_{d}.
\label{Phi}
\end{equation}
$\Phi$ is the scalar density under general coordinate
transformations and the action can be chosen in the form $S = \int
L\Phi d^{4}x$. This has been applied to three different
directions:

\medskip
I. Investigation of the four-dimensional gravity and matter fields
models containing the new measure of integration that appears to
be promising for resolution of the dark energy and dark matter
problems, fermion families problem, the fifth force problem, etc..

\medskip
II. Studying new type of string and brane models based on the use
of a modified world-sheet/world-volume integration measure. It
allows new types of objects and effects like for example:
spontaneously induced string tension; classical mechanism for a
charge confinement; Weyl-invariant light-like (WILL) brane having
the promising results for black hole physics.

\medskip
III. Studying higher dimensional realization of the idea of the
modified measure in the context of the Kaluza-Klein and brane
scenarios with the aim to solve the cosmological constant problem.

\section{Gravity, Particle Physics and Cosmology}.

 Since $\Phi$ is a
total derivative, a shift of $L$ by a constant has no effect on
the equations of motion. Similar shift of $L$ in usual theories,
i.e. with the action $S = \int L\sqrt{-g} d^{4}x$, would lead to
the shift of the constant part of the Lagrangian which in the
Einstein's GR is the cosmological constant. The exploitation of
this circumstance for a resolution of the "old" cosmological
constant problem was the initial motivation\cite{G1} for using the
measure $\Phi$ instead of $\sqrt{-g}$. It turns out that working
with the volume element $\Phi d^{4}x$ {\em instead of} $\sqrt{-g}
d^{4}x$ it is impossible to construct realistic models, e.g. with
a nontrivial scalar field dynamics.

However the situation is dramatically changed if one to apply the
action principle to the action of the general form
\begin{equation}
    S = \int L_{1}\Phi d^{4}x +\int L_{2}\sqrt{-g}d^{4}x,
\label{S}
\end{equation}
 including two Lagrangians $ L_{1}$ and $L_{2}$ and two
measures of the volume elements ($\Phi d^{4}x$ and
$\sqrt{-g}d^{4}x$ respectively). To provide parity conservation,
one can  choose for example one of $\varphi_{a}$'s to be
pseudoscalar. Constructing the field theory with the action
(\ref{S}), we make only two basic additional  assumptions:

(A) $L_{1}$ and $L_{2}$ are independent of the measure fields
$\varphi_{a}$. Then the action (\ref{S}) is invariant under volume
preserving diffeomorphisms\cite{G2},\cite{G6}. Besides, it is
invariant (up to an integral of a total divergence) under the
infinite dimensional group of shifts of the measure fields
$\varphi_{a}$: $\varphi_{a}\rightarrow\varphi_{a}+f_{a}(L_{1})$,
where $f_{a}(L_{1})$ is an arbitrary differentiable function of
the Lagrangian density $L_{1}$. This symmetry prevents the
appearance of terms of the form $h(\Phi/\sqrt{-g})\Phi$ in the
effective action (where quantum corrections are taken into
account) with single possible exception when the function
$h(\Phi/\sqrt{-g})$ is of the form
$h(\Phi/\sqrt{-g})=c\sqrt{-g}/\Phi$; here $c$ is $\Phi/\sqrt{-g}$
independent but may be a function of all other fields. This makes
possible that quantum corrections generate an additive
contribution to the cosmological constant term which may present
also in the second term of the action $(\ref{S})$. Moreover, one
can think of a theory where we start from the action $\int L\Phi
d^{4}x$ but quantum effects modify $L$ to some new Lagrangian
density $L_{1}$ and they also generate a term
$h(\Phi/\sqrt{-g})\Phi$ with
$h(\Phi/\sqrt{-g})=L_{2}\sqrt{-g}/\Phi$. In other words, the
action of the form $(\ref{S})$ may be an effective quantum action
corresponding to the classical action $\int L\Phi d^{4}x$. The
structure of the action (\ref{S}) may be motivated also in the
brane scenario\cite{G12}.

(B) We proceed in the first order formalism where all fields,
including metric $g_{\mu\nu}$ (or vierbeins ${e}_{a\mu}$),
connection coefficients (or spin-connection $\omega_{\mu}^{ab}$)
and the measure fields $\varphi_{a}$ are independent dynamical
variables. All the relations between them follow from equations of
motion. The  field theory based on the listed assumptions we call
"Two Measures Theory" (TMT).

It turns out that the measure fields $\varphi_{a}$ affect the
theory only via the ratio of the two measures
\begin{equation}
\zeta\equiv \Phi /\sqrt{-g} \label{zeta}
\end{equation}
 which is the scalar field. It is determined by a constraint in
the form of an algebraic equation which is exactly a consistency
condition of equations of motion.  {\it The constraint determines
$\zeta$ in terms of the fermion density and scalar fields}.

Applying the Palatini formalism in TMT one can show (see for
example\cite{G6} or Appendix C of Ref.\cite{G12})  that the
resulting relation between metric and  connection includes also
the gradient of $\zeta$. This means that with the original set of
variables we are not in a Riemannian (or Riemann-Cartan)
space-time. Gradient of $\zeta$ presents also in all equations of
motion. By an appropriate change of the dynamical variables which
consists of a conformal transformation of the metric and a
multiplicative redefinitions of the fermion fields, one can
formulate the theory as that in a Riemannian (or Riemann-Cartan)
space-time. The corresponding conformal frame we call "the
Einstein frame". The big advantage of TMT is that in the very wide
class of models, {\em the equations of motion in the Einstein
frame take the canonical general form of those of GR}, including
the field theory models in curved space-time. All the novelty
consists in the structure of the scalar fields effective
potential, masses of fermions and their interactions to scalar
fields as well as the structure of fermion contributions to the
energy-momentum tensor: all these now depend via $\zeta$ on the
scalar fields (e.g., dilaton, Higgs) and the fermion energy
densities. In addition to the canonical fermion contribution to
the energy-momentum tensor there appears the non-canonical one
proportional to $g_{\mu\nu}m(\zeta)\bar{\Psi}\Psi$, where
$m(\zeta)$ is the effective $\zeta$ depending "mass" of the
primordial fermion.

The surprising feature of the theory is that although  the
gravitational equations are used for obtaining the constraint,
neither Newton constant nor curvature appears in the constraint.
This means that the {\it geometrical scalar field} $\zeta (x)$ is
determined by the matter fields configuration locally and
straightforward (that is without gravitational interaction). As a
result of this, $\zeta (x)$ has a decisive influence in the
determination of the effective (that is appearing in the Einstein
frame) interactions and particle masses, and due to this, in the
gravity and particle physics, cosmology and astrophysics.

In Ref.\cite{G9}-\cite{G11} we have started to study the models
with the most general form for $L_1$ and $L_2$ (without higher
derivatives) such that the action (\ref{S}) possesses  both a
non-Abelian gauge symmetry and a special type of scale symmetry
(the latter includes the shift symmetry\cite{Carroll} of the
dilaton $\phi\rightarrow\phi +const$). For short, in a schematic
form $L_1$ can be represented as
\begin{equation}
L_1=e^{\alpha\phi /M_{p}}\left[-\frac{1}{\kappa}R(\omega ,e)
-\frac{1}{2}g^{\mu\nu}\phi_{,\mu}\phi_{,\nu}+(Higgs)+(gauge)+(fermions)\right]
 \label{L1}
\end{equation}
and similarly for $L_2$ (with different choice of the
normalization factors in front of each of the terms). Here
$R(\omega ,e)$ is the scalar curvature in the first order
formalism where the spin-connection $\omega_{\mu}^{ab}$ and the
vierbein $e^{a}_{\mu }$ are independent; $M_P$ is the Planck mass;
$\alpha$ is the dimensionless parameter. Varying the measure
fields $\varphi_{a}$ and assuming $\Phi\neq 0$, we get equations
that yield
\begin{equation}
 L_{1}=sM^{4} =const
\label{varphi}
\end{equation}
where  $s=\pm 1$ and $M$ is a constant of integration with the
dimension of mass. The appearance of a nonzero integration
constant $sM^{4}$ {\em spontaneously breaks the scale invariance}.

In TMT there is no need\cite{G9}-\cite{G12}  to postulate the
existence of three species for each type of fermions (like three
neutrinos, three charged leptons, etc.). Instead of this we start
from one "primordial" fermion field for each type of leptons and
quarks: one primordial neutral lepton $N$, one primordial charged
lepton $E$, etc.. Splitting of each of them into three generations
occurs as a dynamical effect of TMT in normal particle physics
conditions, that is when fermions are localized (in nuclei, atoms,
etc.) and constitute the regular (visible) matter with energy
density tens orders of magnitude larger then the vacuum energy
density. The crucial role in this effect belongs to the
above-mentioned constraint which dictates the balance (in orders
of magnitude) between the vacuum energy density  and the fermion
energy density. In normal particle physics conditions this balance
may be satisfied if $\zeta$ gets the set of pairs of constant
values $\zeta^{(i)}_{1,2}$ that correspond to two different states
of the each type of primordial fermions ($i=N,E,...$) with {\it
different constant masses}. It turns out that with those constant
values of $\zeta$, the non-canonical fermion contribution to the
energy-momentum tensor disappears and the gravitational equations
of our TMT model are reduced exactly to the Einstein equations in
the corresponding field theory model (i.e. when the scalar field
and massive fermions are sources of gravity). Since the classical
tests of GR deal with matter built of the fermions of the first
generation (with a small touch  of the second generation), one
should identify the states of the primordial fermions obtained as
$\zeta =\zeta_{1,2}^{(i)}$ with the first two generations of the
regular fermions. One can show that the model allows to quantize
the matter fields and provides right flavor properties of the
electroweak interactions, at least for the first two lepton
generations.

It turns out that besides the discussed two solutions for $\zeta$
there is only one more additional possibility to satisfy the
constraint when primordial fermion is in the normal particle
physics conditions and to provide that the non-canonical fermion
contribution to the energy-momentum tensor is much less than the
canonical one. We associate this solution with the third
generation of fermions. (for details see \cite{G9}-\cite{G11}).
The described effect of splitting of the primordial fermions into
three generations in the normal particle physics conditions can be
called "{\em fermion families birth effect}".

Fermion families birth effect (at the normal particle physics
conditions) and reproduction of Einstein equations (as the
fermionic matter source of gravity built of the fermions of the
first two generations) do not exhaust the remarkable features of
the theory. Simultaneously with this the theory automatically
provides an extremely strong suppression of the Yukawa coupling of
the scalar field $\phi$ to the fermions observable in
gravitational experiments. The mechanism by means of which the
model solves the long-range scalar force problem is very unusual:
primordial fermions interact with quintessence-like scalar field
$\phi$, but this interaction practically disappears when
primordial fermions are in the states of the regular fermions
observed in gravitational experiments with visible matter. The
fact that {\em the same condition provides simultaneously both
reproduction of GR and the first two families birth effect} seems
very surprising because we did not make any special assumptions
intended for obtaining these results.

In the fermion vacuum the constraint determines $\zeta$ as the
function of the dilaton $\phi$ (and of the Higgs field if it is
included in the model). If the integration constant is chosen to
be negative ($s=-1$ in Eq.(\ref{varphi})) then the effective
potential of the scalar sector implies a scenario\cite{G7} where
zero vacuum energy is achieved without any fine tuning. This
allows to suggest a new way for {\em resolution of the old
cosmological constant problem}. In models with the Higgs field one
may get such situation multiple times, therefore naturally
obtaining a multiple degenerate vacuum as advocated
in\cite{Nielsen}. If one to choose $s=+1$ then one can  treat the
fermion vacuum  as a model for dark energy in the FRW cosmology of
the late time universe. Assuming that the scalar field
$\phi\rightarrow\infty$ as $t\rightarrow\infty$ we
obtain\cite{G9}-\cite{G12},\cite{G14},\cite{G15} that the
evolution of the late time universe is governed by the sum of the
cosmological constant and the quintessence-like scalar field
$\phi$ with the potential proportional to the integration constant
$M^{4}$ and having the form of a combination of two exponents of
$\phi$. In the more simple model(see \cite{G9},\cite{G12}) where
the potentials for $\phi$ are not included in the original TMT
action at all, the effective $\phi$-potential is generated due to
spontaneous symmetry breaking by Eq.(5) and it has the form of the
exponential potential studied in quintessence
models,\cite{quint1},\cite{quint2}.

Due to the constraint, physics of primordial fermions at energy
densities comparable with the dark (scalar sector) energy density
turns out to be very different from what we know in normal
particle physics.  In this case, the non-canonical contribution
(proportional to $g_{\mu\nu}$) of the primordial fermion into the
energy-momentum tensor can be larger and even much larger than the
canonical one. The theory predicts that in this regime the state
of the primordial fermion is totally different from what we know
in normal particle physics conditions.  For instance, in the FRW
universe, the primordial fermion can participate in the expansion
of the universe by means of changing its own parameters. We call
this effect "Cosmo-Particle Phenomenon"  and refer to such states
as Cosmo-Low Energy Physics (CLEP) states. A possible way to
approach and get up a CLEP state might be spreading of the
non-relativistic neutrino wave packet during its free motion (that
may last a very long time). As the first step in exploration of
Cosmo-Particle Phenomena, we have studied a simplified
cosmological model\cite{G12},\cite{G14},\cite{G15} where the
spatially flat FRW universe  is filled with a homogeneous scalar
field $\phi$ and uniformly distributed {\it non-relativistic
(primordial) neutrinos}. Some of the features of the CLEP state in
this toy model are the following: neutrino mass increases as
$a^{3/2}$ ($a=a(t)$ is the scale factor); its energy density
scales as a sort of dark energy and its equation-of-state
approaches $w=-1$ as $a\rightarrow\infty$; the total energy
density of such universe is less than it would be in the universe
free of fermionic matter at all. The described effect of the
neutrino contribution into the dark energy is much stronger than
the one studied in Ref.\cite{Nelson}.

When including terms quadratic in curvature, these types of models
can be applied not only for the late time universe but also for
the early inflationary epoch. As it has been demonstrated in
Ref.\cite{G13}, a smooth transition between these epochs is
possible in these models.

\section{ Strings, Branes, Horizon and K-K modes.}

With the 2-dimensional version of the measure $\Phi$ we can
construct the world-sheet density
\begin{equation}
\Phi =  \frac{1}{2}\varepsilon^{ab}  \varepsilon_{ij}
\partial_{a} \varphi_{i} \partial_{b} \varphi_{j}.
\label{2-Phi}
\end{equation}
However, a problem appears in the naively generated Polyakov-type
string action $S_{0} = - \frac{1}{2}\int d^{2}\sigma \Phi
\gamma^{ab}
\partial_{a} X^{\mu}\partial_{b} X^{\nu} g_{\mu \nu}$ because the equation
 that results from the variation of $\gamma^{ab}$
yields  the unacceptable condition $\Phi \partial_{a}
X^{\mu}\partial_{b} X^{\nu} g_{\mu \nu} = 0$, i.e. vanishing of
the induced metric on the world-sheet. To remedy this situation we
have considered\cite{G16}-\cite{G18} an additional term
$S_{g}=-\int d^{2}\sigma\Phi L$ where  $\sqrt{-\gamma}L$ would be
a total derivative. One can see that without loss of generality,
$L$ may be chosen in the form $\frac
{\varepsilon^{ab}}{\sqrt{-\gamma}} F_{ab}$ where $F_{ab} =
\partial_{a} A_{b} - \partial_{b} A_{a}$, \, $A_{a}(\sigma)$ is an
abelian gauge field on the world sheet of the string.

The action $S_{0}+S_{g}$ is invariant under diffeomorphisms
$\varphi_{i} \rightarrow \varphi_{i}^{'} = \varphi_{i}^{'}
(\varphi_{j})$ in the space of the measure fields (so that $\Phi
\rightarrow \Phi^{'} = J \Phi$) combined with a conformal (Weyl)
transformation of the metric $\gamma_{ab}$: \,  $\gamma_{ab}
\rightarrow \gamma^{'}_{ab} = J \gamma_{ab}$ The combination
$\frac {\varepsilon^{ab}}{\sqrt{-\gamma}} F_{ab} $ is a genuine
scalar. In two dimensions it is proportional to $\sqrt{ F_{ab}
F^{ab}}$.

The equation of motion obtained from the variation of the gauge
field $A_{a}$ is $\varepsilon^{ab}\partial_{a}
(\frac{\Phi}{\sqrt{-\gamma}}) = 0$, which can be integrated to
yield a {\em spontaneously induced string tension}
$T=\frac{\Phi}{\sqrt{-\gamma}}$. The string tension appears here
as an integration constant and does not have to be introduced from
the beginning. The string theory Lagrangian in the modified
measure formalism does not have any fundamental scale associated
with it. The gauge field strength $ F_{ab}$ can be solved from a
fundamental constraint of the theory, which is obtained from the
variation of the action with respect to the measure fields
$\varphi_{j}$ and which requires that $L=M=constant$. Consistency
demands $M=0$ and finally all the equations are the same as those
of standard bosonic string theory.

The described model can be extended\cite{G17},\cite{G18} by
putting point-like charges on the string world-sheet which
interact with the world-sheet gauge field $A_{a}$. Then the
induced tension is not a constant anymore and it suffers
discontinuous jumps at the points where electric charges are
located. The generalization of this model to the non-Abelian gauge
fields is straightforward\cite{G17},\cite{G18} by using $\sqrt{Tr
F_{ab} F^{ab}}$ instead of $\frac
{\varepsilon^{ab}}{\sqrt{-\gamma}} F_{ab} $ (in the non-Abelian
case the latter is not a scalar in the internal space ). In this
case {\em the induced tension is identified as the magnitude of an
effective non-Abelian electric field-strength on the world-sheet
obeying the standard Gauss-low constraint}. As a result, a simple
classical mechanism for confinement via modified-measure "color"
strings has been proposed\cite{G17},\cite{G18} where {\em the
colorlessness of the "hadrons" is an automatic consequence of the
new string dynamics}.

We have studied two types of branes: the first one is similar to
the  kind of branes well known in literature; the branes of the
second type have totally new features and have no analog in the
literature. In order to construct the bosonic $p$-branes of the
first type, a term of the form $\frac
{\varepsilon^{a_{1}a_{2}...a_{p+1}}}{\sqrt{-\gamma}}
\partial_{[a_{1}}A_{a_{2}...a_{p+1}]}$ has to be considered instead of
the $\frac {\varepsilon^{ab}}{\sqrt{-\gamma}} F_{ab} $. The branes
of the second type are constructed\cite{G19},\cite{G20} by using
$\sqrt{F_{ab}F^{ab}}$ in order to provide the conformal (Weyl)
invariance for any $p$. For even $p$ the branes must contain
light-like directions. In particular, for $p=2$ the spherically
symmetric solutions {\em automatically adjust themselves to sit at
the black hole horizon}. This suggests that the second type of
branes can serve as a relevant candidate for realization of the
idea of the black hole membrane paradigm\cite{paradigm} and of the
't Hooft approach\cite{hooft} to description of the degrees of
freedom of the horizon.

In the "Kaluza-Klein" context we have found\cite{G20} solutions
describing \textsl{WILL}-branes wrapped around the internal
(compact) dimensions and moving as a whole with the speed of light
in the non-compact (space-time) dimensions. Although the {\em
WILL}-brane is wrapping the extra dimensions in a topologically
non-trivial way, its modes remain {\em massless} from the
projected $d$-dimensional space-time point of view. This is a
highly non-trivial result since we have here particles (membrane
modes), which acquire in this way non-zero quantum numbers, while
at the same time remaining massless. In contrast, one should
recall that in ordinary Kaluza-Klein theory, non-trivial
dependence on the extra dimensions is possible for point particles
or even standard strings and branes only at a very high energy
cost (either by momentum modes or winding modes), which implies a
very high mass from the projected $D=4$ space-time point of view.

\section{ Braneworld Scenarios.}

A six dimensional braneworld scenario based on a model describing
the interaction of gravity, gauge fields and $3+1$ branes in a
conformally invariant way is described by the action
\begin{equation}
S = \int L\Phi_{(6)} d^{6}x ,\qquad L = -\frac{1}{\kappa_{(6)}}
R^{(6)} + \sqrt{|F_{CD}F^{CD}|}, \label{III-1}
\end{equation}
where $\kappa_{(6)}$ and $R^{(6)}$ are 6-D gravitational constant
and scalar curvature. The action of this model is defined using a
6-dimensional version $\Phi_{(6)}$ of the measure $\Phi$. This
allows for theory to be conformal invariant. In this theory the
branes do not need to be postulated separately. They result here
from delta-function configuration of the gauge fields. As it is
known, $\sqrt{|F_{CD}F^{CD}|}$-gauge theory allows for such type
of extended object solutions[25],[26]. It was shown in
Refs.[25],[26] that in such a model there is no need to fine tune
any bulk cosmological constant or the tension of the two parallel
branes to obtain zero 4-D cosmological constant: the only
solutions are those with zero 4-D cosmological constant. In
contrast, the extra dimensions in these solutions are highly
curved.

\title{Randomness in Random Dynamics}
\author{A. Kleppe}
\institute{%
Bjornvn. 52\\
07730 Oslo\\
Norway}
\titlerunning{Randomness in Random Dynamics}
\authorrunning{A. Kleppe}
\maketitle                                       
\begin{abstract}

Unlike the conventional schemes where the fundamental level is assumed to be
simple, the Random Dynamics approach 
is based on the assumption that there is an irreducibly complex, random, 
bottom layer.

The purpose of this paper is to discuss the possible meanings of this
assumption.

\end{abstract}         
\section{Random dynamics}
Symmetry is an essential guiding principle in the development of physical 
theories. The highly 
successful Standard Model is an outstanding example \cite{SM} of this.
According to the Standard Model philosophy, as we go up in energy, we expect a 
larger symmetry group.
In the search of the ultimate fundamental theory, Grand Unification \cite{SM}
is a scenario in this spirit, advocating that at higher energies there are 
forces corresponding to larger symmetry groups which break down at lower 
energies, ultimately giving rise to the symmetry group corresponding to the 
forces we see here at our low energy level.

The striving to reach an ultimate, fundamental force or forces that we find in
the conventional Theory of Everything schemes, is based on the assumption that
all physical phenomena eventually can be brought back onto a finite set of
laws of nature, where a law is characterized by a finite complexity.
There are however some stumbling blocks on this path, such as the inherent 
randomness 
of quantum mechanics. That is hard to reconcile with a
conventional Theory of Everything scheme.

If symmetry is one corner stone in the search for physical laws, another 
guiding principle is simplicity. This is Occam's principle - simplicity is 
``sigillum veri''. The idea is that fundamental principles are simple, which 
however does not necessarily imply that Nature itself is ``simple'' at a 
fundamental scale, on the contrary:
as we climb up the energy scale there are more and more degrees of
freedom, meaning a growing complexity. What goes on at a fundamental scale, 
like the Planck scale, is probably enormously complicated and most simply 
described in terms of randomness. This is the punch line of Random 
Dynamics, the theory developed by Holger Bech Nielsen \cite{hol} and his
collaborators.  Unlike the conventional schemes where fundamentality is 
assumed to be characterized by simplicity, the Random Dynamics approach 
is based on the contrary assumption that there is an irreducibly complex
bottom layer.

The idea is that a sufficiently complex and general model for the
fundamental physics at or above the Planck scale, will in the low energy 
limit where we operate, yield the physics we know. The reason is that as 
we slide down the energy scale, the structure and complexity characteristic 
for the high energy level is shaved away. The features that survive are 
those that are common for the long wavelength limit of any generic model 
of fundamental supra-Planck scale physics. The ambition of Random Dynamics 
is to ``derive'' \cite{holderive} all the known physical laws as an almost unavoidable
consequence of a random fundamental ``world machinery''.

According to this scheme, the fundamental ``world machinery'' is a very general,
random mathematical 
structure ${\cal{M}}$, which contains non-identical elements and some 
set-theoretical notions.
There are also strong exchange forces present. There is as yet no physics.
At some stage ${\cal{M}}$ comes about, and then physics follows. 

That the fundamental structure ${\cal{M}}$ comes without differentiability and
with no concept of distance, that is, no geometry, implies an apriori lack of 
locality in the model. We cannot put in locality by hand, since the lack of 
geometry forbids locality to be properly stated. Thus the principle of 
locality, taken say as a path way integration
$\int {\cal{D}}e^{\int{\cal{L}}d^4x}$ with a Lagrangian density ${\cal{L}}$ 
only locally depending on the fields, cannot be put in before we have space 
and time.

We are so used to the urge for simplicity in science, that the idea that 
something fundamental could be non-simple seems impossible to imagine, the 
Random Dynamics assumption of non-simple fundamental laws thus appears as an 
oxymoron. How do we furthermore recognize a law as fundamental if it is 
non-simple?

Even more alarming is the Random Dynamics assumption that the fundamental 
level is characterized by a high degree of randomness.
The purpose of this paper is to discuss the possible meaning of this
assumption.

\section{A comprehensible Universe}
What does it mean that the world is comprehensible?
According to Gottfried Wilhelm Leibniz \cite{Leib Discours de la metode} it 
means that the Universe is rationally comprehensible: that God used but a few 
principles to create the whole Universe, with all its complex beauty. And that
we can backtrack it all by tracing the whole world back to the laws of nature. 

The act of understanding thus amounts to a reduction of complexity, a kind of
mapping from a set of ideas onto a smaller set of already accepted,
well-defined notions.
That is, we look for a cause, which is a phenomenon which is ``smaller'' and 
more general than the phenomenon we want to explain; the ultimate causes being 
the laws of nature.

When phenomena that we judge to have a
resemblance with each other occur in a way that however cannot be assigned any
regularity or pattern - i.e. no cause can be defined - we perceive them 
as $random$. Something random cannot be ascribed to some 
$simpler$ underlying mechanism or algorithm.
Albeit the concept of simplicity according to Herman Weyl 
\cite{Weyl Philofnature}, `` appears to be inaccessible to objective 
formulation...it has been attempted to reduce it to that of probability'',
we think of something simple as opposed to something complex, as something 
with few elements related in a transparent way.
A simple dress is unadorned, with a straightforward cut, a simple language 
usually means short sentences with common words, and a simple model comprehends
uncomplicted rules and few elements.

There certainly is simplicity out there, in the sense of comprehensible
regularity, otherwise we would not have the insight that we actually have,
otherwise our cars would not be running, our lamps would not be shining..
The question is how deep the simplicity sticks, i.e. whether the Universe is
simple also at the most fundamental level - regular, 
ruled by some few and formulable principles - or complex, maybe even random, 
at heart.
Is the idea that we should look for a simple ultimate law or principle, nothing
but a prejudice? Maybe nature is not fundamentally simple, maybe nature
is fundamentally quite complex.
It may even be fundamentally random, and yet subjected to simple rules. Or does
the randomness we find in the world only exist on the interface between us 
and Nature?
Whatever the answer is, it is clear that we cannot get rid of the point where 
language touches upon Nature, like the nerve of sight touches the eye in the 
blind spot.

According to the Random Dynamics approach, the question is not whether 
the Universe is complex or not, the
question is rather $how$ complex it is, finitely or infinitely, i.e random.

\subsection{Randomness}

There is a story by Honor\'{e} de Balzac \cite{Balzac} called ``The Unknown 
Masterpiece''.
It is about the young Nicolas Poussin who in the early 17th century came to
Paris in the hope of becoming the apprentice of one of the great painters of 
his time.

As the Master's apprentice, the young man every day saw the Master
produce the most exquisite paintings. But the 
painting that most of all preoccupied the apprentice's imagination, was a big 
canvas standing in a corner, covered by a piece of cloth.
He knew the Master considered it his chef d'oeuvre, and that he also considered
it as unfinished. The Master now and then worked on the canvas, but never in 
the presence of the apprentice. 

Then one day when nobody else was around, the apprentice gave in to his
curiosity, tiptoed to the chef d'oeuvre canvas, and lifted the cloth. 
He fixed the canvas in astonishment, but could not see a thing. The surface
was covered with paint, layer on layer, a meningless chaos of colour. In one 
corner a first layer of paint was still visible. There an absolutely 
perfect human foot stuck out, as from under a blanket.

In perfecting the painting, the Master had painted and painted on top of 
the first layer,
hiding every structure under another structure, until all structure was
muddled away. The canvas had become patternless. The young man was
staring at a $random$ pattern.

When we use the word random we think of something disorganized, 
without a plan. We speak of making 
a random choice, and mean picking a choice from a set of possible choices, 
without a plan or reason for that particular choice.
Something organized, patterened, on the other hand, is $ordered$, 
non-random. 
 
Order and randomness can be assigned to a process, to a number, to a 
furnished room...
This is something we know from everyday life - when we tidy up, the room 
becomes tidy, ordered. But how do we produce something random?

Tossing a coin is a classical procedure to obtain a $random$ $number$. 
Each toss is independent and random, in the sense that each toss is a fact
that is what it is, head or tails, for no reason.
When you toss a coin N times, and write down the result by letting
0 and 1 represent heads and tails, you get one of $2^N$ binary series 
which all have the same probability. In the traditional understanding of 
a random sequence of tokens is a sequence where all
the tokens appear with the same probablitity, so the series obtained by 
tossing the coin should be reliably random \cite{randdd}.
Some of the $2^N$ series may however display some recognizable inner 
structure, i.e. non-randomness. So merely to emerge from a probabilistic 
event does not guarantee randomness.  

Something random is by definition not specific, in some sense it is so average 
that there is no way of tagging it. To single out a number as being random 
is thus an oxymoron - since there is no way to pin down something random, 
there can be no consensus about the definition of randomness.

\section{The axioms at the bottom}

A random event is an event that cannot be ascribed a cause, it just happens
``for no reason''. 
The scheme of explaining a phenomenon by relating it to a cause, which 
ultimately is a principle or a law of 
nature, originates from Euclid's idea of mathematical proof: a mathematical
truth is established by reducing it to simpler truths until self-evident
truths, i.e. axioms or postulates, are obtained. At the bottom of the world
there is a firm layer of true axioms.

That the language for describing the Universe is maths implies that the 
Universe is perceived as made out of eternal mathematical truth: like in 
Plato's Tima\-eus 
where the building blocks of the Universe are given, as simple, symmetrical 
geometrical forms. And mathematical truth is most certainly based on a set of
simple axioms. This was at least David Hilbert's \cite{Hilbert} conviction when
he addressed his colleagues at the International Congress of Mathematicians in 
Paris in 1900, where he outlined 23 major mathematical problems to be studied 
in the coming century. One of the problems was the the axiomatization of
mathematics: to formulate the axioms onto which all mathematical truth can 
ultimately be brought back. This led to the opening of the first Pandora box.

At the end of the 19th century, Georg Cantor \cite{Cantor} had investigated 
the properties of infinite sets. His work lead to many worrisome results, so
worrysome in fact, that many mathematicians - including the great 
Poincar\'{e} - turned against Cantor. 
What upset people were the insights brought about by Cantor's work, like when
he proved that ``almost all'' numbers are transcendental by proving that the 
real numbers were not countable, or the status of reality that he bestowed 
on the concept of infinity (Gauss had stated that infinity should only 
be used as "a way of speaking"). Cantor furthermore discovered the set 
theoretical paradoxes that constituted the basis for Bertrand Russell's 
\cite{Russell} work on paradoxes in logic itself (like in ``is the set of all 
set a member of itself?''). 

David Hilbert was an follower of Leibniz,
who thought that one could avoid not only logical paradoxes, but all conflict, 
by formulating all statements in an algebraic form (we may laugh at this, but 
it is noteworthy that the majority of dissidents in former Soviet-Union were 
natural scientists and mathematicians, people for whom retouching of facts is 
not that easy to swallow).
Hilbert thus had the idea that one could escape paradoxes like Russell's 
paradox, by creating a completely formal axiomatic system. Once all statements 
were formulated within this system 
there would not be room for paradoxes, Hilbert thought. 

He expected every formal system to be consistent and complete,
and any well-posed mathematical problem to be decideable, in the sense that
there is a mechanical procedure (for example a computer program) for deciding
whether a statement is true or not.
If you can prove both the statement $A$ and the statement
$\neg A$ within a given formalism, the formalism is inconsistent.
A formal axiomatic system is moreover complete if any statement $A$ formulated
within the system can be settled by proving or disproving it.
That is, from the set of axioms of the system, you should be able to prove 
the whole truth and nothing but the truth implied by these axioms.

Hilbert's ambition was thus to project all of mathematics onto a formal 
complete and consistent system.
In 1931 Kurt G\"{o}del \cite{Godel} however showed that Hil\-bert's idea that 
all paradoxes would evaporate when we decide to exclusively use formal 
language, was wrong.
G\"{o}del's point of departure was (a variant of) the liar's paradox, which
is classically formulated as ``this is a lie'', or ``I lie now''; an equivalent
statement is ``I am unprovable''. G\"{o}del showed that there is no formal 
axiomatic system that can make it clear whether a sentence like this is true 
or not - the implication being that any formal system, any language, is 
either incomplete or inconsistent.
Hilbert's first two demands were thus shown to be mutually exclusive.
Moreover, in 1936, Hilbert's third demand was abolished as Alan Turing 
\cite{Tur} discovered $uncomputability$. 

A number is $computable$ if there is an algorithm for computing its digits 
one by one,  approximating it to arbitrary precision.
${\pi}$ is a computable number, even though its decimals may look
totally patternless, without redundancy, as its digits all seem
to have the same probability.
$\pi$ however has but finite complexity:, since there are algorithms \cite{Plouffe}
for calculating it, like the Bailey-Borwein-Plouffe algorithm 
\begin{equation}
\pi=\sum_{n=0}^\infty \left(\frac{4}{8n+1}-\frac{2}{8n+4}-\frac{1}{8n+5}
-\frac{1}{8n+6}\right)\left(\frac{1}{16}\right)^n
\end{equation}
Most real numbers are however not computable. 
In physics there is a large occurrence of measurable albeit non-computable
numbers, and likewise in maths itself. "A rather startling result" writes Roger
Penrose \cite{Pen} about the fact that the wave equation with computable 
initial conditons can have non-computable solutions.

Turing addressed the halting problem, which concerns whether a computer 
program can tell in advance if another program will eventually halt or not.
He concluded that no such program exists, since if you can find a mechanical 
procedure for deciding whether a computer program will halt, you end up being 
able to compute a real number which is not computable, i.e. in
self-contradiction.

Do the logical limitations in the formal system constitute limitations 
to scientific knowledge? 
Is there moreover any physical G\"{o}del's theorem, implying unanswerable 
questions and limits to scientific knowledge \cite{Traub}? 

One answer is that the limitations discovered by
G\"{o}del and Turing imply is that certain mathematical ``observables'',
i.e. deduced results that scientists can understand, cannot be obtained.
According to this view, since this is information that is excluded from the 
tentative mathematical model, it cannot contribute to the scientific method 
and should therefore be discarded. To pay a more serious attention to 
the limitations within the formal system would namely be to consecrate
mathematical equations with a physical reality that they do not have.
(Einstein: ``as far as the propositions of mathematics refer to reality,
they are not certain; and as far as they are certain, they do not refer
to reality'').

Physical observation consists of finite detectable information corresponding 
to a finite set of domains of real numbers (precision determined by the 
precision of the instruments), while a mathematical model is associated with 
infinitely precise real numbers, and computer algorithms correspond
to a finite set of integers.
These forms of information are thus inherently distinct, to the point that 
there is no one-to-one mapping between them.
The correlations emerging from the physical observations are
not necessarily the same as those emerging from the mathematical models or
from the computer algorithms, the scientific method is however expected to
supply a relationship between the correlations from physical observation,
and mathematical or computer modelling.

The G\"{o}delian trap is that we are able to intuit the truth of a sentence
without being able to formally prove it. If you can prove the sentence 
``this is an unprovable sentence'', it is a false sentence. Thus you have
formally proven something false - a terrible oxymoron. On the other hand,
if you $cannot$ prove it, albeit you perceive its truth, you sit in the trap.
The crux is that to possess information is not the same as having
knowledge or understanding. Insight and understanding come with 
the correlations between observables. Since understanding per se cannot be 
logically formalized, it cannot be logically proved or disproved either. 


\subsection{Information Theory}
In information theory, one studies how to measure the rate at 
which a message source generates information. It tells us how to represent or 
encode messages from a particular source over a given channel, 
avoiding errors of transmission.
The question is basically how messages are conveyed from a message source, 
such as a writer or speaker, to a recipient. The amount of information in 
a message is measured in $bits$, one bit being the answer to a yes/no-question.
Bits are thought of as abstract zeros and ones, but information is 
always encoded in real physical objects. A string of bits can thus be 
regarded as a physical resource.
The essential elements of information science, classical or quantum, can 
be summarized in a three step procedure.

1. Identify a physical resource - e.g. A string of bits.

2. Identify an information-compressing task, like gunzipping.

3. Identify a criterion for successful completion of 2. (like 
controling that the output from the compression stage perfectly matches 
the input of the compression stage).

The fundamental question of information science is then ``what is the 
minimal quantity of physical resource 1. needed to perform the 
information processing task 2. in compliance with the success criterion 
3''.

In classical information science it can be stated as ``what is the minimum 
number of bits needed to store the information produced by some source''?
This was solved by Shannon in 1948 \cite{Sha}.
Shannon quantified the information content produced by an information 
source, defining it to be the minimum number of bits needed to reliably 
store the output of the source.


\subsection{Algorithmic Information Theory}
While classical information theory (Shannon\cite{Sha}, Wiener\cite{Win}) 
uses concepts like ensembles and probability distributions, Algorithmic 
Information Theory \cite{AIT} instead focuses on individual objects, by posing
questions like: what is the size of the smallest program for calculating a 
given object, how many bits are needed to compute it?
In algorithmic information theory the complexity of an object is measured 
by the size (say in bits) of the smallest algorithm generating it, i.e. 
the amount of information needed to give to a computer in order to have 
it perform a given task. 
The size of a computer program is analogous to the degree of disorder of a 
physical system, algorithmic information theory in this way supplies a 
definition of what it means for a string to be unstructured, i.e. random.
According to algorithmic information theory the simplest theory is defined 
as one corresponding to the smallest algorithm, that is, the scheme using the 
fewest bits.

As algorithmic complexity of a sequence is measured as the length of
the smallest algorithm that reproduces the sequence in question, randomness 
refers to something that informationwise cannot be compressed at all.
A completely random sequence requires an algorithm as long as the sequence
itself, i.e. something random is algorithmically incompressible or irreducible.
Most strings are moreover algorithmically irreducible and therefore random. 

Consider a string of the length of N bits. The algorithmic information content
of such a string is (generically) 
$H(S) \leq N + H(N)$, where $H(N)$ is the algorithmic information content
for $N$ in the base-2, $H(N)$ given as $H(N) \sim N + \log N$.

For strings that have some pattern or regularity, $H(S) < N + H(N)$, while for
strings that have no regularities that could diminish their information
content, $H(S) = N + H(N)$. Such strings are random or algorithmically 
incompressible,
the border between random and non-random occurring at $H(S) \approx N$. 
The probability that an infinite sequence obtained by tossing a (fair) coin 
is (algorithmically) random is 1, while the calculation of the first $N$ bits of
$\pi$ takes $H(N) \approx \log N$ bits, implying that $\pi$ is definitely a
non-random number.

\subsection{Attempts to define randomness}
Order and randomness are both characteristic of the set of relations between 
parts of a system, not of the parts themselves.
It is a whole system of entities that is random or ordered. 

While ``order'' is a very special way of organizing the parts of a system (think
of a tidy room), a random state is ``typical'', non-specific, and thus has no 
distinguishing features (think of all the different ways a room can be untidy).
According to the traditional understanding of randomness, the parts of a 
random system are all on the same footing - complete democracy, complete 
symmetry rule.
In that sense it is patternless and very hard to define, indeed impossible.
To see this, assume that you actually can single out a random number. The 
property of being random is then a feature that makes random numbers stick
out. But since randomness is by generic, non-specific, this is a 
self-contradiction, implying
that there cannot be any definitive definition of randomness. The 
more one tries to pin it down, the more evasive it becomes. Every definition 
of randomness therefore has a taint of something preliminary.

In algorithmic information theory, the degree of randomness of a given sequence
is however established in terms of the smallest algorithm that reproduces the 
sequence in question. If this algorithm is as long as the sequence 
itself, the sequence is considered as random. This still does not 
ensure that something is random in any absolute sense, it merely establishes
a randomness hierarchy.
It is thus impossible to prove randomness,
but non-randomness can be stated exactly.
We may establish that a number is not random by finding an algorithm that
generates it, but we cannot by the same means establish that a 
number $is$ random: that I have not succeeded to find an algorithm generating 
it does not prove that the number is random. How can I know that I will not
find such an algorithm tomorrow - I cannot prove that such an algorithm will 
never be found. In this sense the property of randomness is unprovable.   
The definition of randomness is
by necessity heuristic and preliminary - we can never say if there
is some algorithm that is yet to be discovered.
According to the algorithmic information theory scheme, the Random Dynamics 
philosophy thus means that fundamental algorithms and laws are not
necessarily short. 

The program-size randomness also implies that something random is 
undistinguished and typical. Incompressibility is characteristic of 
something typical or generic - i.e. there is no structure
that singles out this object, that makes it non-random \cite{rand}.
The only way to describe a random thing is by 
stipulative definition, to point at it: here it is!

So, something is 
random if we cannot find an algorithm, i.e. a shorter description, from which 
we can derive the structure in question: no redundancies remain to be removed.
This clearly is true for the elementary parts of a system, like the axioms 
in an axiomatic system. 
In this sense axioms, and indeed any elementary entity is also - trivially - 
random. 
(A G\"{o}delian statement, impossible to prove or disprove, is thus also a 
sort of axiom). 
When we speak of randomness we however usually do not mean the trivial 
randomness of axioms. When we speak of randomness and order, we mean 
properties of systems above the elementary, axiomatic level.

In the causal explanatory scheme, it is assumed that $if$ it were possible to 
control all the influences 
over a physical experiment, the outcome would always be the same. Therefore,
since a random event cannot be ascribed a well-defined cause, 
randomness is often perceived as a lack of knowledge, an ignorance on the
part of the observer,
In quantum mechanics it assumed to be possible to set up an experiment 
with perfect control of all relevant parameters. Even in such an experiment 
the outcome can still be totally random. There have been unsuccessful attempts
to save the situation by means of ``hidden variables''; the randomness of the 
outcome however remains, indicating that the world might be irreducibly random.

A large body of non-random physics and maths has been formulated, but quantum 
physics places randomness at the fundamental level of physics, and even in 
classical mechanics we find unpredictability and randomness.
But this does not prove that there is an innermost random core - maybe 
what we perceive as random today, will one day be punctuated, maybe it is our
ignorance that speaks, not nature's fundamental properties!


\section{The emergence of order}

In the Random Dynamics scheme, the world is assumed to be fundamentally
random. If one takes this view, the question is how order emerges, the order
we see, in the form of symmetry, laws, all the organized forms - in 
short, the world we live in!
Can there be an inherent randomness in a world with such a high degree of
organization? 
To explain the emergence of the observed order of the world, is one of the 
challenges of the Random Dynamics scheme, while the ontological state of 
randomness remains a matter of discussion.

Some proponents claim that the world contains real randomness, others that 
what we believe is randomness is really pseudo-randomness.

A third approach is that randomness may exist in the realm of mathematics, 
while not in the physical world. An proponent of this view is physicist Karl 
Svozil \cite{Svozil}, 
who claims that the randomness displayed in quantum mechanics is a matter of
ignorance. He believes that a new, deeper hidden-variable quantum theory 
will eventually emerge, which amends the present quantum theory and gives us 
a randomness-free, deterministic theory. Svozil however accepts that 
mathematics contains real randomness.


\subsection{Apparent randomness}
According to the rationalist view of the physical world, everything
happens for a reason, implying that the Universe is logical and comprehensible.
Supporters of this view claim that there is no real randomness,
only pseudo-randomness exixts. This is the kind of randomness produced by
random-number generators, which are deterministic sequences of numbers 
generated by algorithms (not by quantum mechanical processes!). The Universe 
is accordingly deterministic, governed by deterministic physical laws: 
- the world has finite complexity. According to this view, albeit the world
looks so complex, it is really quite simple, like in the case of $\pi$. 
We just don't know the underlying law, or algorithm, therefore this overwhelming 
impression of complexity and even randomness.  
This resembles the classical picture, where we formulate a story based on 
observations, and then deal with the information content of that story by 
using mathematics, with the intention of pinning down the substructure of laws
and principles.

$\pi$ really constitutes a very interesting example. The $number$ $\pi$ 
looks impossibly convoluted, while the $geometrical$ $relationship$
defining $\pi$ is quite simple. And indeed, the $number$ $\pi$ turns out to be
non-random. 
The simple, transparent geometrical description of $\pi$ constitutes an
algorithm with finite complexity, indicating that there should also exist
numerical algorithms for generating the number $\pi$.

In his book ``A New Kind of Science'' \cite{Wol}, Stephen Wolfram reports on 
a systematic computer search for simple rules with very complicated 
consequences. 
In traditional physics one studies systems that satisfy certain constraints.
Wolfram instead studies systems that develop according to given algorithms,
and subsequently asks what pattern or algorithm 
corresponds to a given constraint, if any. 
He looks for fundamental algorithms rather than for fundamental laws,
by investigating the emergence of organized patterns in a system of
cellular automata, using simple rules \`{a} la the Game of Life.
Challenging the Pythagorianian ``Number rules the Universe'', 
Wolfram asserts that it is not number that rules the Universe, but Algorithm. 
Namely discrete algorithm.

Wolfram's claim is moreover on the line with that of digital philosophy 
\cite{Fredkin}, which advocates that everything fundamentally comes in discrete
bits. That implies a picture of the world as a giant digital information 
processor, a computer, in agreement with algorithmic information theory.
This challenges traditional physics which is based on continuous mathematics.
Digital philosophy is really a scheme for describing the world in terms of 
automata theory. The assumption is that everything fundamental is atomic
or discrete, and the continuous flow of time is replaced by a sequence of
time steps. 
So the basis of digital philosophy is:

$\bullet$ All information can ultimately be digitally represented. 

$\bullet$ All change in information is due to digital information processes.
\\
According to digital philosophy a physical state is represented by a pattern 
of bits, like in a computer.
The digital philosophy bits exist in a digital spacetime, and each point contains 1 bit of
information. Digital spacetime thus consists of bits that all have integer 
coordinates.
A digital philosophy model is always specific, unlike a mathematical model that is generic.
Every digital mechanics model can be put into a computer where it runs and evolves. 
In physics, if space and time are discrete, all other physical quantities 
must also be discrete. In digital philosophy, the dynamics of a system is 
therefore described by difference
equations, which can be transformed into computational algorithms.

A simple example of cellular automata is a fixed array of cells together with 
the rule that a cell becomes black if one of its neighbours is black:
\begin{center}\begin{picture}(340,140)(0,0)
\GBox(15,60)(35,80){0}
\Line(0,60)(50,60)
\Line(0,80)(50,80)
\Line(15,45)(15,95)
\Line(35,45)(35,95)
\LongArrow(55,70)(70,70)
\Line(80,40)(170,40)
\Line(80,60)(170,60)
\Line(80,80)(170,80)
\Line(80,100)(170,100)
\GBox(95,60)(115,80){0}
\GBox(115,60)(135,80){0}
\GBox(135,60)(155,80){0}
\GBox(115,80)(135,100){0}
\GBox(115,40)(135,60){0}
\Line(95,20)(95,115)
\Line(115,20)(115,115)
\Line(135,20)(135,115)
\Line(155,20)(155,115)
\LongArrow(180,70)(195,70)
\Line(210,20)(340,20)
\Line(210,40)(340,40)
\Line(210,60)(340,60)
\Line(210,80)(340,80)
\Line(210,100)(340,100)
\Line(210,120)(340,120)
\Line(225,0)(225,135)
\Line(245,0)(245,135)
\Line(265,0)(265,135)
\Line(285,0)(285,135)
\Line(305,0)(305,135)
\Line(325,0)(325,135)
\GBox(265,20)(285,40){0}
\GBox(265,40)(285,60){0}
\GBox(265,60)(285,80){0}
\GBox(265,80)(285,100){0}
\GBox(265,100)(285,120){0}
\GBox(245,40)(265,60){0}
\GBox(245,60)(265,80){0}
\GBox(245,80)(265,100){0}
\GBox(225,60)(245,80){0}
\GBox(285,40)(305,60){0}
\GBox(285,60)(305,80){0}
\GBox(285,80)(305,100){0}
\GBox(305,60)(325,80){0}
\end{picture}\end{center}
{\vspace{1cm}}
One of the features that cellular automata and Turing machines have in 
common is that they
consist of a fixed array of cells. The colours and arrangement of colours
vary, but the number and organization of cells remains constant. 
With a substitutions system one may however also change the number of
elements, if the substitution law e.g. implies to replace each element by a
block of new elements.
To complicate the matter further, one can combine the substitution of elements
with some form of interaction between the elements. The changes then depend
both on the ``internal properties'' of each element, as well as on its 
environment.

By studying cellular automata that develop according to given algorithms, 
Wolfram found that behaviour of great complexity can be produced by certain 
rules starting from certain initial states. Other systems than cellular 
automata can also develop complexity, Wolfram indeed concludes that complexity
is a universal phenomenon that is quite independent of the details of the given
system. 
If the rules governing a system are simple enough, the system will display
very simple, repetetive behaviour. With rules somewhat less simple, nesting
will appear, namely self-similar, fractal patterns. As the complexity in the 
underlying rules goes beyond a certain critical value, the system's overall 
behaviour will
be complex. The criticality threshold is typically rather low, meaning that 
the deviation from simplicity is not very large. Quite simple rules may lead to
a complex behaviour. Moreover, once the critical value has been passed, adding 
complexity to the underlying rules does not increase the complexity of the 
system's behaviour. 

So the stages from simplicity to complexity can be described by repetition, 
nesting and complexity.
The typical types of behaviour are quite universal and practically independent
of the underlying rules. Behaviour with non-random features can develop even 
from completely random initial conditions, there are many instances of
cellular automata that start from random initial conditions and quickly settle
down in a stable state. The end result is 
not necessarily ``simple'' and transparent, it may be quite complicated, almost
random. But also in a random final state, there is almost always some small 
non-random structures that emerge in the evolution of the system.
The most complex results lie between the extremes where the
end result is a completely trivial, uniform state; or a 
seemingly random state. In the complex final state the cellular automata 
are organized in a set of definite localized structures that do not remain 
fixed but move around and interact with each other.

The patterns that emerge, quite different among themselves, fall into four 
fundamental types of
patterns or classes, numbered in growing order of complexity:

{\bf{Class 1.}} displays very simple behaviour: almost all initial conditions 
lead to the same uniform final state. The information about the initial 
conditions is simply wiped out 
and the same final state is reached, redardless of the initial conditions.   
 
{\bf{Class 2.}} emcompasses many different final states, but they all display
a certain set of simple structures that remain the same or repeat every
few steps.  Changes may persist, but are always localized in some small region 
of the system. That is, the information about the initial conditions always
remains, but localized and not communicated from one part of the system to 
the others. 

{\bf{Class 3.}} more complicated, more random; but at some level some 
small-scale structures exist. Any change that is made typically spreads with a
uniform rate throughout the system. That is, there is a long-range
communication of information.

{\bf{Class 4.}} mixture of order and randomness. Changes also spread here,
but in a more sporadic way than in the class 3.-systems.

Class 1. And 2. rapidly settle down to states in which there is essentially
no further activity, while class 3.-systems continue to change at each step;
class 4. systems are somewhere between class 3 and the two first classes.
The differences between the classes reflect that systems from different classes
handle information in different ways. 
Also continuous cellular automata - where the underlying rules involve 
parameters that vary smoothly between 0 and 1 - can be classified in 
accordance with this scheme.

\subsection{Randomness at the heart of mathematics}

Unlike Stephen Wolfram, 
Gregory Chaitin \cite{Chaitin} is convinced that true randomness exists.
The crux is to pin it down. 

According to Chaitin G\"{o}del's and Turing's discoveries indicate that there 
is an inherent randomness in mathematics, Chaitin furthermore formulated the 
source of randomness at the heart of mathematics \cite{Chaitin2}. 
Chaitin's reasoning goes as follows: 
equations can be classified according 
to their numbers of solutions. 
For example, some equations have no solutions, like $x = x + 4$, or one 
solution, like $4x = 12$, or two solutions, like $x^2 = 3x - 5$, or 
an infinite number of solutions, like $(1+x)^2 = x^2 + 2x + 1$.
Now, Chaitin has found an equation for which it is mathematically undecideable 
whether it has a finite or an infinite number of solutions, leading to the 
definition of the number $\Omega$ = the probability that a probabilistically 
generated program will ever terminate. This number is so random that not one 
sole of its digits can be predicted!

The prescription for the random number $\Omega$ starts with running
a program on a computer. Each time the computer requests the next bit
of the program, insert the result obtained by flipping a fair coin. The computer
must decide by itself when to stop reading the program, this turns the program 
into self-delimiting binary information. 
For each program $p$ that halts, sum the probailities of getting precisely
that program by chance:

\begin{equation}
\Omega = \Sigma_{(program{\hspace{1mm}}p{\hspace{1mm}}halts)}  \frac{1}{2^P} 
\end{equation}
where $P$ is the size in bits of the program $p$.
Each n-bit self-delimiting program p that halts contributes 1/$2^n$ to the 
value of $\Omega$.

$\Omega$ is an algorithmically random or irreducible number, thus a formal 
algebraic system can determine only finitely many bits of such a number; really
only as many bits of $\Omega$ as its own complexity.
The only way to determine the bits of $\Omega$ is thus to put the information 
directly into the axioms of the formal algebraic system. There are thus no 
short-cuts, no algorithms to compress the information needed to generate
$\Omega$. The information needed to generate $\Omega$ is $\Omega$ itself. The
bits of $\Omega$ are logically irreducible, i.e. cannot be obtained from axioms
simpler than they are.

Mathematics thus contains randomness - the bits of $\Omega$.
That doesn't mean that mathematics is random in the sense of being arbitrary. 
It means that mathematics contains irreducible information, like $\Omega$.
$\Omega$ is, both algorithmically or computationally and logically, i.e. by 
means of proofs, irreducible. No shift of perspective, no reparametrization
can make $\Omega$ more transparent, like for example in the case of $\pi$. 
This means that $\Omega$ has many of the 
characteristics of the typical outcome of a random process.     

$\Omega$ is a random real with lots of meaning, since it contains lots of
information about the halting problem. This information is stored in
$\Omega$ in an irreducible fashion, with no redundancy. Once redundancy is
squeezed out of something meaningful, it may look meaninglesseven though
it is in reality dense with meaning. Just like the Master's chef d'oeuvre.
A random pattern may be meaningless or extremely meaningful, there is no
way to distinguish. That is the crux of the matter.

This implies that the mathematical Universe is 
infinitely complex, and thus the whole of mathematical ideas cannot be 
comprehended in it entirety. This is Chaitin's radical refutation of 
Hilbert's project of summing all of mathematics up by formulating the set 
of mathematical axioms.

\section{Conclusion}
It is by definition impossible to formulate an exact definition of randomness,
but degrees of randomness can nevertheless be established, and most precisely
so in algorithmic information theory.
There are divergeing opinions about the ontological state of randomness. 
While some claim that what we take for randomness is only an effective
pseudo-randomness, others state that randomness exists at the heart of
mathematics, and most probably also at the heart of physics. 

G\"{o}del's, Turing's and 
Chaitin's work certainly gives strong support to a fundamental 
randomness, as advocated by Random Dynamics, which on the one hand
concerns the choice of a primary set ${\cal{M}}$ from the set ${\bf{M}}$ 
of all generic sets or proto-models.
The choice is random in the sense that any set of such general nature will do. 
Nature picked ${\cal{M}}$, but could just as well have picked ${\cal{M'}}$. 
The randomness of choice of ${\cal{M}}$ can be interpreted as the type
of randomness that occurs in coin tossing.

The randomness inherent in the structure of the set ${\cal{M}}$ corresponds to
a lack of an organizing principle governing the elements
of ${\cal{M}}$, as energy and information are evenly distributed
over all the degrees of freedom.
Only by going down in energy is redundancy added to the system, allowing the
information that is inherent in ${\cal{M}}$ to be displayed.
The elements of ${\cal{M}}$ are different, yet undistinguishible, because 
there is not enough
structure at hand to enable categorization of the elements.
This randomness inherent in the composition of ${\cal{M}}$ implies that 
``algorithms'' or recipes for generating a set like ${\cal{M}}$ can by no 
means be short.

\title{An Example of  Kaluza-Klein-like Theories Leading After 
Compactification to Massless Spinors Coupled
to a Gauge Field---Derivations and Proofs}
\author{N. Manko\v c Bor\v stnik${}^1$, H. B. Nielsen${}^2$ and %
D. Lukman${}^1$}
\institute{%
${}^1$ Department of Physics, University of
Ljubljana, Jadranska 19, SI-1111 Ljubljana\\
${}^2$ Department of Physics, Niels Bohr Institute,
Blegdamsvej 17, Copenhagen, DK-2100}
\titlerunning{An Example of  Kaluza-Klein-like Theories Leading \ldots}
\authorrunning{N. Manko\v c Bor\v stnik, H. B. Nielsen and D. Lukman}
\maketitle

\begin{abstract}
The genuine Kaluza-Klein-like theories (with no fields in addition to gravity with torsion) have difficulties 
with the existence of massless spinors after the compactification
of some of dimensions of space\cite{witten}. We wrote a letter\cite{HN2004} in which we demonstrate 
on an example of a torus - as 
a compactified part
of an (1+5)-dimensional space - that there exists a Kaluza-Klein charge - as a consequence of boundary 
conditions leading to allowance of torsion - which couples chirally a spinor to the corresponding
$U(1)$ field and splits a real spinor representation in the $(1+3)$  space   
into  two parts, each of different handedness and different  Kaluza-Klein charge, making the mass protection mechanism work.
We also showed on a very special example how the procedure goes. In this contribution we present in addition in details most
of proofs, needed in this letter.
\end{abstract}



\section{Introduction}

{\it Genuine Kaluza-Klein-like theories}, assuming nothing but  a gravitational field 
in $d$-dimensional space (no additional gauge or scalar fields), 
which after the spontaneous compactification of a $(d-4)$-dimensional part of space manifest in 
four dimensions as all the known gauge fields including gravity, have difficulties\cite{witten} with 
masslessness of fermionic fields at low energies. It looks  
namely very difficult to avoid after the compactification of a part of space  the appearance of  
representations of both handedness in this part of space and consequently also in the 
(1+3)-dimensional space. Accordingly, the gauge fields can hardly couple chirally in the (1+3) - dimensional 
space.

In an approach by one of us\cite{pikanorma,norma}
it has long been the wish to obtain the gauge fields from only gravity, so that ''everything'' would become gravity.
This approach has taken the inspiration from looking for unifying all the internal degrees of freedom, that is 
the spin and the charges into only spins. This approach is also a kind of the genuine Kaluza-Klein theory, suffering
the same problems, with the problem of getting chiral fermions included, unless we can solve them.

It is the purpose of this  contribution to present (most of) detailed derivations and proofs for statements and lemmas, 
used in the letter\cite{HN2004}, where we pointed out on an example of a torus-shaped compactifying space (a flat $M^{1+5}$
space with the $SO(1,5)$ symmetry compactifying into a flat $M^{1+3}$ part of space with the $SO(1,3)$
and $S^1 \times S^1$ symmetry)
that using a gravitational field with a vielbein and a spin connection and a torsion (comming from the boundary condition)
on a torus, we
indeed achieve to get a chirally coupled genuine Kaluza-Klein originating gauge fields!

We assume an action for a free field, which is linear in the Riemann scalar and in first order formulation.
In a two dimensional 
compactified part of space, the Euler-Lagrange equations of motion for a free 
gravitational field with zweibein and spin connection fields give no conditions on  the fields. It is the requirement 
that a covariant derivative of vielbeins is zero, which relates  spin connections,  vielbeins and a torsion.
There are also boundary conditions, which  put limitations on these fields.  

By applying a special (physical) choice of boundary conditions on a meeting place of  patching regions of a two-dimensional
manifold  we make the ''curvature'' integral $\int  
\omega_{56 [\sigma, \tau]} \; dx^{\sigma} dx^{\tau}$  to be proportional to an integer, and accordingly
quantized - as we can quantize a magnetic flux through a two dimensional surface\footnote{Here and in what follows
$[,]$ denotes the 
anti symmetrization with respect to $\mu$ and $\nu$.}.

We then {\it prove}   that there exists in this case  a conserved  {\it  Kaluza-Klein gauge charge}, 
which {\it by marking a representation, enables the choice of the representation of only one handedness}.
Accordingly a ground state solution of the Weyl equation in (1+3)-dimensional ''realistic'' space  
is  massless and mass protected as well as coupled chirally to the $U(1)$ gauge field represented by 
the corresponding spin connection and vielbein fields.

We first study solutions in the two-dimensional compactified space as essencially decoupled from
 (1+3)-dimensional space. We found that equations of motion allowed torsion. We are able to
prove that the Kaluza-Klein gauge field  induced 
couples chirally to the Weyl-spinor  from the (1 + 3)- dimensional point of view. 




This contribution is, except for the proofs, the repetition of the letter, sent to Phys. Lett.\cite{HN2004}.

\section{Weyl spinor in gravitational fields with spin connections and vielbeins} 
\label{Weylspinor}

We let a spinor interact with a gravitational field through vielbeins $f^{\alpha}{}_{a}$ (inverted vielbeins to 
$e^{a}{}_{\alpha}$ with the properties $e^a{}_{\alpha} f^{\alpha}{}_b = \delta^a{}_b,\; 
e^a{}_{\alpha} f^{\beta}{}_a = \delta^{\beta}_{\alpha} $ ) and spin connections
$\omega_{ab\alpha}$
\begin{eqnarray}
(\gamma^a p_{0a} =0)\psi, \quad p_{0a} = f^{\alpha}{}_a p_{0\alpha}, \quad p_{0\alpha} = p_{\alpha} - \frac{1}{2}
S^{ab} \omega_{ab \alpha}.
\label{weylgravity}
\end{eqnarray}
The operators $\gamma^a$ fulfill the Clifford algebra $\{\gamma^a, \gamma^b \}_+ = 2 \eta^{ab}$.
Latin indices  $a$, $b$, $m$, $n$,\ldots $s,t,.$ denote a tangent space (a flat index),
while Greek indices $\alpha$, $\beta$, \ldots,$\mu$, $\nu$, \ldots, $\sigma$,
$\tau$ denote an Einstein 
index (a curved index). Letters  from the beginning of both the alphabets
indicate a general index ($a,b,c,..$   and $\alpha, \beta, \gamma $ ), 
 from the middle of both the alphabets   
 the observed dimensions $0,1,2,3$ ($m,n,..$ and $\mu,\nu,..$), indices from the bottom of the alphabets
 the compactified dimensions ($s,t,..$ and $\sigma,\tau,..$)
Taking into account that $\gamma^a \gamma^b = \eta^{ab}- 2i S^{ab}$,  $\{\gamma^a, S^{bc}\}_- = i
(\eta^{ab} \gamma^c - \eta^{ac} \gamma^b)$,  one easily finds
\begin{eqnarray}
(\gamma^a p_{0a} )^2 &=& p_{0}{}^a p_{0a} + \frac{1}{2} S^{ab} S^{cd} {\cal R}_{abcd} + 
S^{ab} {\cal T}^{\beta}{}_{ab} p_{0 \beta}, \nonumber\\
{\cal R}_{\alpha \beta cd} &=&  \omega_{cd[\alpha,\beta]} + 
\omega_{ce[\alpha}\omega^e{}_{d\beta]}, \nonumber\\
{\cal T}^a{}_{\alpha \beta} &=&  
e^a{}_{[\alpha,\beta]} + \omega^a{}_{b[\alpha} e^b{}_{\beta]}.
\label{RandTdet}
\end{eqnarray}

We require\cite{milutin2002} that the total covariant derivative of a vielbein $e^a{}_{\alpha}$ is equal to
zero 
\begin{eqnarray}
e^a{}_{\beta ; \alpha} = 0 = e^a{}_{\beta , \alpha} + \omega^a{}_{b \alpha} e^b{}_{\beta} - 
\Gamma^{\gamma}{}_{\beta \alpha} e^a{}_{\gamma}.
\label{covderviel}
\end{eqnarray}
From Eq.(\ref{covderviel}) it then follows that the covariant derivative of a metric tensor $g_{\alpha \beta}= 
e^a{}_{\alpha} e_{a \beta}$  is also equal to zero $g_{\alpha \beta; \gamma}= 0 $.
Eq.(\ref{covderviel}) connects the spin connections $\omega^a{}_{b \alpha}$, vielbeins and the quantity
$\Gamma^{\alpha}{}_{\beta \gamma} $, whose antisymmetric part is a torsion $
{\cal T}^{a}{}_{\alpha \beta} = \Gamma^{a}{}_{\alpha \beta} - \Gamma^{a}{}_{\beta \alpha},$
in agreement with the definition of  the torsion in Eq.(\ref{RandTdet}). The symmetric part
is the Christoffel symbol $\{{}^{\alpha}_{\beta \gamma}\}$ $
\Gamma^{\alpha}{}_{\beta\gamma} = \{{}^{\alpha}_{\beta \gamma}\} +{\cal K}^{\alpha}{}_{\beta \gamma},$
with $\{{}^{\alpha}_{\beta \gamma}\}= \frac{1}{2} g^{\alpha \delta} (g_{\delta \gamma ,\beta} + 
g_{\delta \beta ,\gamma} - g_{\beta \gamma ,\delta} )$, $ g_{\alpha \beta} = e^a{}_{\alpha} e_{a \beta},$ 
and $g^{\alpha \beta}
= f^{\alpha}{}_{a} f^{\beta a}$.
${\cal K}^{\alpha}{}_{\beta\gamma}$ is a tensor of a contorsion $
{\cal K}^{\alpha}{}_{\beta \gamma} = \frac{1}{2} ({\cal T}^{\alpha}{}_{\beta \gamma} - 
{\cal T}{}_{\gamma}{}^{\alpha}{}_{\beta}+ {\cal T}{}_{\beta \gamma}{}^{\alpha} )$.
%

We assume the Einstein action for a free gravitational field, which is linear in the curvature
\begin{eqnarray}
S&=& \int \; d^d{} x \; E \; R, \nonumber\\  
R &=& f^{\alpha [a} f^{\beta b]} \;(\omega_{a b \alpha,\beta} - \omega_{c a \alpha}
\omega^{c}{}_{b \beta}), 
\label{Riemannaction}
\end{eqnarray}
with $ E = \det(e^a{}_{\alpha}) $.
Varying this action with respect to spin connections and vielbeins, for a general $d>2$,
equations of motion follow accordingly for spin connections and vielbeins, which allow in general case
to express spin connections in terms of vielbeins.

{ \it Statement 1}: Spin connections are for $d>2$ and the action of Eq.(\ref{Riemannaction}) expressible 
in terms of vielbeins as follows
\begin{eqnarray}
 \omega_{a b \alpha} &=& - \frac{1}{2E}\left\{ e^d{}_{\alpha} [e_{[a \gamma} \partial_{\beta} 
 (E f^{\gamma}_{[b]} f^{\beta}{}_{d]}) - e^d{}_{\gamma} \partial_{\beta} 
 (E f^{\gamma}_{[a} f^{\beta}{}_{b]})]\right\} \nonumber\\
 & & \qquad {} + \frac{1}{(d-2)E}\left\{ e_{[b\alpha} e^{d}{}_{\gamma} \partial_{\beta} 
 (E f^{\gamma}_{[d} f^{\beta}{}_{a]]})\right\},
\label{omegaoff}
\end{eqnarray}
where $[a  ...b]$ means, that the expression must be antisymmetrized with respect to $a,b$. 
The proof is presented in subsection \ref{proof1}.

This is not, however, the case for $d=2$. 

{\it Statement 2:}
{\it For $d=2$,   variations of the action} (\ref{Riemannaction})with no sources 
{\it with  
respect to $\omega_{ab \alpha}$ and $f^{\alpha}{}_{a}$ bring no conditions on either of these two types of fields,
so that any zweibein and any spin connection can be assumed.}
The proof for this statement is presented in subsection \ref{proof2}.

For this particular case of $d=2$, the only limitation
on zweibeins and spin connections might come from boundary conditions.
{\it Accordingly } (\ref{covderviel}) {\it can be understood as the defining equation 
for $\Gamma^{\alpha}_{\beta \gamma}$,}
for any spin connection and any zweibein, which fulfill desired boundary conditions.
 Because of these facts, the Euler index
\begin{eqnarray}
{\rm Euler}\;{\rm index} &=& \int \; d^d{} x \; E \; {\cal R}, \quad {\rm with }\nonumber\\  
\mathcal {R} &=& {\cal R}^{\alpha}{}_{\beta \delta \gamma} \delta^{\delta}{}_{\alpha} g^{\beta \gamma} \nonumber\\
 &=& (\{{}^{\alpha}_{\beta \gamma}\}_{, \delta}
- \{{}^{\alpha}_{\beta\delta}\}_{,\gamma} + \{{}^{\alpha}_{\sigma \delta}\} \{{}^{\sigma}_{\beta \gamma
}\} - \{{}^{\alpha}_{\sigma \gamma}\}
\{{}^{\sigma}_{\beta \delta}\} ) \delta^{\delta}{}_{\alpha} g^{\beta \gamma}, 
\label{Eulerindex}
\end{eqnarray}
{\it is not in general equal to $R$, with 
$R = (f^{\alpha} f^{\beta b} - f^{\alpha b} f^{\beta a})\;\omega_{a b \alpha,\beta}$}. (the qudratic term in 
$\omega_{ab \alpha}$ 
from Eq.(\ref{Riemannaction}) contributes nothing
in the case of $d=2.$). These two ${\cal R}$ and $R$ are independent\cite{Deser}.

\subsection{Proof of  {\it Statement 1}}
\label{proof1}

{\it Proof for Statement 1:} 
We vary the action for vielbeins and spin connections, linear in the curvature, in the presence of a
spinor field
\begin{equation}\label{actionSources}
S = \int d^d x\, E\mathcal{R} 
    + \int d^d x\, E \bar{\Psi}
        \gamma^d \left(-\frac{1}{2}S^{ab}\omega_{ab\alpha} \right)
        f^\alpha{}_d \Psi.
\end{equation}
first  with respect to $\omega_{a b \alpha}$. It then follows
\begin{eqnarray}
  0 &=&
\int d^d x\,E\delta\omega_{ab\alpha}\cdot
      \left(\frac{\partial\mathcal{R}}{\partial\omega_{ab\alpha}}
            -\frac{1}{2}\bar{\Psi}S^{ab} f^\alpha{}_d \Psi\right)  
  \nonumber\\ 
 &=& \int d^d x\, \partial_\beta\left(Ef^\alpha{}_{[a}f^\beta{}_{b]}
            \delta\omega_{cd\alpha}\eta^{ac}\eta^{bd} \right) 
      \label{varAction}\\
 &+& \int d^d x\, \delta\omega^{ab}{}_{\alpha} 
       \left\{ -\partial_\beta\left(Ef^\alpha{}_{[a}f^\beta{}_{b]} \right) 
               +E \omega^c{}_{a\beta} f^\beta{}_{[c}f^\alpha{}_{b]} 
               -\frac{1}{2} E\bar{\Psi}f^\alpha{}_d\gamma^d S^{ab}\Psi 
       \right\} \nonumber
\end{eqnarray}
The above relation can be rewritten as
\begin{equation}\label{Srel}
\frac{1}{E} \partial_\beta\left(Ef^\alpha{}_{[a}f^\beta{}_{b]} \right)
+ \frac{1}{2} \bar{\Psi}f^\alpha{}_d\gamma^d S^{ab}\Psi
= f^\alpha{}_{[a}\,\omega_{b]}{}^c{}_c 
  - f^\alpha{}_c \omega_{[b}{}^c{}_{a]}.
\end{equation}
Multiplying this equation with the vielbein $e^a{}_\alpha$ and summing over the two indices, which appear
twice, we end up with the relation
\begin{equation}\label{omegabcc}
(d-2)\,\omega_b{}^c{}_c = 
\frac{1}{E} \partial_\beta\left(Ef^\alpha{}_{[a}f^\beta{}_{b]} \right)
+ \frac{1}{2} \bar{\Psi}f^\alpha{}_d\gamma^d S^{ab}\Psi.
\end{equation}
To express $\omega_a{}^e{}_b$ in terms of vielbeins, we first multiply Eq.(\ref{Srel}) with $e^e{}_\alpha$ and sum
over $\alpha$
\begin{equation}\label{omegaaeb}
\omega_a{}^e{}_b - \omega_b{}^e{}_a 
= \frac{1}{E} e^e{}_\alpha 
    \partial_\beta\left(Ef^\alpha{}_{[a}f^\beta{}_{b]} \right)
  + \frac{1}{2} \bar{\Psi}\gamma^e S^{ab}\Psi
  - \delta^e_a \omega_b{}^c{}_c + \delta^e_b \omega_a{}^c{}_c.  
\end{equation} 
Then we denote the left hand side of Eq.(\ref{omegaaeb}) by $ -A_b{}^e{}_a$
\begin{eqnarray}
A_b{}^e{}_a &=& 
- \omega_a{}^e{}_b + \omega_b{}^e{}_a \nonumber\\ 
 &=& -\left(\frac{1}{E} e^e{}_\alpha 
    \partial_\beta\left(Ef^\alpha{}_{[a}f^\beta{}_{b]} \right)
  + \frac{1}{2} \bar{\Psi}\gamma^e S^{ab}\Psi
  - \delta^e_a \omega_b{}^c{}_c + \delta^e_b \omega_a{}^c{}_c\right).  
  \label{omegaA}
\end{eqnarray} 
The quantity $-A_b{}^e{}_a$ is
 antisymmetric in the first two indices: 
$A_b{}^e{}_a=-A{}^e{}_b{}_a$. The linear combination 
$A_{ab}{}^e+A^e{}_{ab}-A_b{}^e{}_a$ is  proportional 
to $\omega_{ab}{}^e$
\begin{equation}\label{omegaabe}
\omega_{ab}{}^e = \frac{1}{2}\left(A_{ab}{}^e+A^e{}_{ab}-A_b{}^e{}_a\right).
\end{equation}
Taking into account the relation(\ref{omegaaeb})
with suitably permuted indices we obtain finally
\begin{eqnarray*}
\omega_{ab}{}^e &=& 
 -\frac{1}{2E}\biggl\{
   e_{b\alpha}\,\partial_\beta(Ef^{\alpha[e}f^\beta{}_{a]} )
   + e_{a\alpha}\,\partial_\beta(Ef^{\alpha}{}_{[b}f^{\beta e]})\\
  & &  \qquad\qquad  {} - e^e{}_\alpha\,
     \partial_\beta\bigl(Ef^\alpha{}_{[a}f^\beta{}_{b]} \bigr)
   \biggr\} \\
  &-& \frac{1}{4}
   \biggl\{\bar{\Psi}\left(\gamma_{[b} S^e{}_{a]}  
      - \gamma^e S_{ab}\right) \Psi \biggr\} \\
  &-& \frac{1}{d-2}  
   \biggl\{ \delta^e_a \left[
            \frac{1}{E} e^d{}_\alpha \partial_\beta
             \left(Ef^\alpha{}_{[d}f^\beta{}_{b]}\right)
            + \frac{1}{2}\bar{\Psi} \gamma^d S_{db}\Psi 
            \right] \\
  & & \qquad {} - \delta^e_b \left[
            \frac{1}{E} e^d{}_\alpha \partial_\beta
             \left(Ef^\alpha{}_{[d}f^\beta{}_{a]}\right)
            + \frac{1}{2}\bar{\Psi} \gamma^d S_{da}\Psi 
            \right] 
   \biggr\}.
\end{eqnarray*}
To obtain the exression for $\omega_{ab\alpha}$ in terms of vielbeins, we multiply the above equation with $e_{e\alpha}$
and sum over $e$
\begin{eqnarray}
\omega_{ab\alpha} &=& 
 -\frac{1}{2E}\biggl\{
   e_{e\alpha}e_{b\gamma}\,\partial_\beta(Ef^{\gamma[e}f^\beta{}_{a]} )
   + e_{e\alpha}e_{a\gamma}\,\partial_\beta(Ef^{\gamma}{}_{[b}f^{\beta e]})
  \nonumber\\
  & &  \qquad\qquad  {} - e_{e\alpha}e^e{}_\gamma\,
     \partial_\beta\bigl(Ef^\gamma{}_{[a}f^\beta{}_{b]} \bigr)
   \biggr\} \nonumber\\
  &-& \frac{e_{e\alpha}}{4}
   \biggl\{\bar{\Psi}\left(\gamma_e S_{ab} 
      + \frac{3i}{2} \delta^e_{[b}\gamma_{a]} 
          \right) \Psi \biggr\} \nonumber\\
  &-& \frac{1}{d-2}  
   \biggl\{ e_{a\alpha} \left[
            \frac{1}{E} e^d{}_\gamma \partial_\beta
             \left(Ef^\gamma{}_{[d}f^\beta{}_{b]}\right)
            + \frac{1}{2}\bar{\Psi} \gamma^d S_{db}\Psi 
            \right] \nonumber\\
  & & \qquad {} - e_{b\alpha} \left[
            \frac{1}{E} e^d{}_\gamma \partial_\beta
             \left(Ef^\gamma{}_{[d}f^\beta{}_{a]}\right)
            + \frac{1}{2}\bar{\Psi} \gamma^d S_{da}\Psi 
            \right] 
   \biggr\}.\label{omegaabalpha} 
\end{eqnarray}

This complets the proof.

\subsection{Proof of  {\it Statement 2}}
\label{proof2}

For $d=2$ the relation follows
\begin{eqnarray}
E f^{\sigma}{}_{[s} f^{\tau}{}_{t]} &=& \frac{1}{\varepsilon_{\sigma' \tau'} f^{\sigma'}{}_{s'} f^{\tau'}{}_{t'}
\varepsilon^{s' t'}} f^{\sigma }{}_{[s} f^{\tau }{}_{t]},\nonumber\\
&=& \frac{1}{4} \varepsilon^{\sigma \tau} \varepsilon_{s t} 
\frac{1}{\varepsilon_{\sigma' \tau'} f^{\sigma'}{}_{s'} f^{\tau'}{}_{t'}
\varepsilon^{s' t'} \varepsilon_{\sigma '' t''}} f^{\sigma '' }{}_{[s''} f^{\tau '' }{}_{t'']} 
\varepsilon^{s'' t''} \nonumber\\
&=& \frac{1}{4} \varepsilon^{\sigma \tau} \varepsilon_{s t}. 
\label{varactd2}
\end{eqnarray}

Since for $d=2$ the variation of the action (\ref{Riemannaction}) without sources with respect to $\omega_{st \sigma}$
gives
\begin{eqnarray}
\partial_{\tau} (E f^{\sigma}{}_{s} F^{\tau}{}_{t}) =0 = \partial_{\tau} 
(\frac{1}{4} \varepsilon^{\sigma \tau} \varepsilon_{s t}),
\label{omegafd2}
\end{eqnarray}
which is obviously zero, we must conclude that this variation gives no relation between spin connections and vielbeins 
and no condition on either of them. 

Variation of the same action (again without sources)  with respect to vielbeins leads to the equation
\begin{eqnarray}
-e^s{}_{\sigma} R + 4 f^{\tau t} \omega_{st \sigma,\tau} = 0, 
\label{fomegad2}
\end{eqnarray}
which is trivially zero for any $R$. This can be seen by multiplying the above equation by $ f^{\sigma}{}_{s}$ and summing
over the two indices $\sigma$ and $s$. It follows then that $(d-2) R =0.$
The proof is completed.

\section{Spinors in $M^{(1+5)}$ space}
\label{spinor1+5}

The operator of handedness $\Gamma^{(d)}$ is one of 
the invariants\cite{norma} of the Lorentz (and the Poincar\' e) group
\begin{eqnarray}
\Gamma^{(d)}:&=& i^{d/2} \;\; \prod_a \quad (\sqrt{\eta^{aa}} \gamma^a), \quad  {\rm if} \quad d=2n,
\label{handedness}
\end{eqnarray}
with the product of $\gamma^a$'s to be understood in the ascending order with respect to the index $a$.
One finds that since in even dimensional spaces $\Gamma^{(d)}$ anticommutes with $\gamma^a$, it also 
anticommutes with the Weyl equations of motion operator for a free spinor $
\{\Gamma^{(d)}, \gamma^a p_a \}_+ = 0,\quad {\rm if} \quad d=2n,$
while $M^{ab}$ ($ = L^{ab} +S^{ab}$) is the constant of motion
$ \{ M^{ab}, \gamma^{c} p_c \}_- = 0.$ 

Making use of the technique for generating spinor 
representations from the Clifford algebra objects\cite{holgernorma,norma,pikanorma}
one can write down a Weyl representation for an even $d$ as a product of $d/2$ nilpotents $\stackrel{ab}{(k)}$
and projectors $\stackrel{ab}{[k]}$, which have the property:
$S^{ab}\stackrel{ab}{(k)} = \frac{k}{2} \stackrel{ab}{(k)}$ and $S^{ab} \stackrel{ab}{[k]}
= \frac{k}{2} \stackrel{ab}{[k]}$, with $S^{ab} = \frac{i}{4} \{ \gamma^a,\gamma^b \}_-$ and $k^2 =\eta^{aa} \eta^{bb}$.
In Table~\ref{NHDtable1}, one Weyl representation of $SO(1,5)$ is presented,
\begin{table}
\begin{center}
\begin{tabular}{|rc|c|ccc|ccc|}
\hline
i&$$&$\psi_i$&$\Gamma^{(1,5)}$&$ \Gamma^{(1,3)}$&$\Gamma^{(2)}$&
$S^{12}$&$S^{03}$&$S^{56}$\\
\hline\hline
1&$\psi_1$&$\stackrel{03}{(+i)}\stackrel{12}{(+)}|\stackrel{56}{(+)}$
&-1&+1&-1&+1/2&+i/2&+1/2\\
\hline 
2&$\psi_2$&$\stackrel{03}{[-i]}\stackrel{12}{[-]}|\stackrel{56}{(+)}$
&-1&+1&-1&-1/2&-i/2&+1/2\\
\hline\hline
3&$\psi_3$&$\stackrel{03}{[-i]}\stackrel{12}{(+)}|\stackrel{56}{[-]}$
&-1&-1&+1&+1/2&-i/2&-1/2\\
\hline 
4&$\psi_4$&$\stackrel{03}{(+i)}\stackrel{12}{[-]}|\stackrel{56}{[-]}$
&-1&-1&+1&-1/2&+i/2&-1/2\\
\hline\hline
\end{tabular}
\end{center}
\caption{\label{NHDtable1}
A left handed Weyl representation of $SO(1,5)$ is presented in terms
of nilpotents $\stackrel{ab}{(k)}$ and projectors $\stackrel{ab}{[k]}$, which are ''eigenvectors''
of the Cartan sub algebra $S^{12}, S^{03}, S^{56}$. The handedness  $\Gamma^{(1,5)} = 
\gamma^0 \gamma^1 \gamma^2 \gamma^3 \gamma^5 \gamma^6 = 8i S^{03} S^{12} S^{56}= \Gamma^{(1,3)}\Gamma^{(2)}$ and
$\Gamma^{(1,3)} = -4i S^{03} S^{56}$,  $\Gamma^{(2)} = -2S^{56}$, together with the
''eigenvalues'' of the Cartan sub algebra operators are also presented.}
\end{table}
with the choice of a starting state: 
$\stackrel{03}{(+i)} \stackrel{12}{(+)} \stackrel{56}{(+)}$. There are four basic states, two right handed spinors 
with respect 
to the subgroup $SO(1,3)$ ($\Gamma^{(1,3)}= 1$) with the ''$S^{56}$ charge''
equal to $+1/2$, and two left handed spinors with respect to the group $SO(1,3)$ ($\Gamma^{(1,3)}= - 1$) 
with the ''$S^{56}$ charge'' 
equal to $-1/2$.

A free spinor in $d(=1+5)$-dimensional space with a momentum $p^a=(p^0,\overrightarrow{p})$, obeying the
Weyl equations of motion $
(\gamma^a p_a=0)\psi,$ is in $M^{1+5}$ a plane wave, with the spinor part, which is  a 
superposition of the vectors, presented in Table~\ref{NHDtable1}  
\begin{eqnarray}
\psi(p) &=& e^{-ip^ax_a} \mathcal{N} \biggl\{\stackrel{03}{(+i)}\stackrel{12}{(+)}\stackrel{56}{(+)} - 
\frac{p^0-p^3}{p^5 -i p^6} 
\stackrel{03}{[-i]}\stackrel{12}{(+)}\stackrel{56}{[-]} \nonumber\\ 
  & & \qquad {}+ \frac{p^1+ip^2}{p^5-ip^6} \stackrel{03}{(+i)}\stackrel{12}{[-]}\stackrel{56}{[-]} \biggr\}, 
\label{weylin6}
\end{eqnarray}
with  $(p^0)^2=(\overrightarrow{p})^2$. 
If $p^5=0=p^6$, the above solution goes into the
right handed one with respect to $SO(1,3)$ ($\Gamma^{(1,3)}=1$) with $S^{56}=1/2$.
 Looking from the point of view of the
four-dimensional space, we call the eigenvalue of $S^{56}$ a charge,
while nonzero
components of the momentum $p^a$ in higher than four dimensions (nonzero  $p^5$, $p^6$) manifest as a mass term,
causing a superposition of  the left and the  right handed components of  $\Gamma^{(1,3)}$ which have ''charges''
of both signs 
(Eq.(\ref{weylin6}) and Table~\ref{NHDtable1}). 

One finds that for a free massless left handed spinor in $M^{1+5}$ the relations  $  
 \{ M^{bc}, \gamma^a p_{a}\}_- =0, \; \{ \Gamma^{(d)}, \gamma^a p_{a}\}_+=0, \;
\{ M^{bc}, \Gamma^{(d)}\}_- = 0,$
are fulfilled.

\section{ Spin connection, vielbein, torsion on $M^{(2)}$ and boundary conditions}
\label{d=2case}

We first treat a compactified part of space alone, namely a general case of $M^{(2)}$ . We assume the $M^{(2)}$ part of space
covered by patches and pay attention on boundary conditions on meeting places among patches.
We use symbols  $5,6$ to denote flat indices and $(5),(6)$ to denote Einstein indices 
of the two compactified dimensions. At the end we apply the result on a torus, with the choice of only one patch.

The most general vielbein for $d=2$ can be written by an appropriate parame\-trization choice
\begin{eqnarray}
e^s{}_{\sigma} = e^{\varphi/2}
\begin{pmatrix}
\cos \phi   & \sin \phi \\
- \sin \phi & \cos \phi 
\end{pmatrix}
f^{\sigma}{}_{s} = e^{-\varphi/2} \;
\begin{pmatrix}
\cos \phi   & -\sin \phi \\
\sin \phi & \cos \phi
\end{pmatrix}
\label{f1and1}
\end{eqnarray}
with $s=5,6$ and $\sigma =(5),(6)$ and 
$g_{\sigma \tau} = e^{\varphi} \eta_{\sigma \tau}$, 
$g^{\sigma \tau} = e^{- \varphi}\eta^{\sigma \tau}$, 
and $\eta_{\sigma \tau} = diag(-1,-1) = \eta^{\sigma \tau}$.
Taking into account Eqs.(\ref{RandTdet},\ref{covderviel},\ref{f1and1}), the second determining the relation
among a spin-connection, 
a vielbein and a torsion, we find for the torsion ${\cal T}_{\sigma}: = {\cal T}^{\sigma'}{}_{\tau \tau'}
\varepsilon^{\tau \tau'}
g_{\sigma' \sigma}$ the expression
\begin{equation}
\mathcal{T}_{\sigma} = \phi_{; \sigma} + \varepsilon_{\sigma}{}^{\tau} \frac{1}{2} \varphi_{,\tau}, 
\label{Tsigma}
\end{equation}
with $\phi_{; \sigma} = \phi_{, \sigma} - \omega^{5}{}_{6 \sigma}$,
and $\mathcal{T}_{[\sigma, \tau]} = - \omega ^{5}{}_{6 [\sigma, \tau]}$. 
The detailed derivations of the above expressions, as well of
all further expressions, can be found in the subsection \ref{cauchy}.

Any choice of patching regions makes that two types of expressions meet on the overlapping parts of regions. 
Let $x^{[1]\sigma}$ and $x^{[2]\sigma}$, $\sigma =(5),(6)$,
stays for the two types of coordinates on any overlap region of two chosen patches, while
a conformal map connects these two types of coordinates. There are two types of transformations, which have
to be taken into account on such an overlapping  region: 
i) {\it a zweibein rotation gauge transformation} and ii) {\it a conformal map}, 
connecting $\omega^{[1]5}{}_{6 \sigma}, \phi^{[1]}$ and $\varphi^{[1]}$ with
$\omega^{[2]5}{}_{6 \sigma}, \phi^{[2]}$ and $\varphi^{[2]}$. Accordingly (Eq.(\ref{Tsigma})) 
also ${\cal T}^{[1]}{}_{\sigma}$ is connected with ${\cal T}^{[2]}{}_{\sigma}$.

i) A zweibein rotation gauge transformation, determined by $\lambda$ 
(where $\psi^{[1]} = e^{i \lambda S^{56}} \psi^{[2]}$),
requires  
\begin{eqnarray}
\frac{\partial x^{[1]\tau}}{\partial x^{[2] \sigma}} \omega^{[1] 5}{}_{6 \tau} (x^{[1]}) &=&
\omega^{[2] 5}{}_{6 \sigma} (x^{[2]}) + \partial^{[2]}{}_{\sigma} \lambda, 
 \nonumber\\ 
 \phi^{[1]} &=& \phi^{[2]} + \lambda, \label{sabpatch}\\
 \varphi^{[1]} &=& \varphi^{[2]}.\nonumber 
\end{eqnarray}
The transition boundary condition $\lambda$ can locally be replaced by choosing an appropriate gauge of
one of the meeting patches. But if we make a ''physical'' choice of $\lambda$ (with putting, for example, on 
a particular patch a Dirac string) so that $\lambda$
 steps up  by $2\pi n$, with $n$ an integer, when going around a non simply connected patch-meeting-place, 
 such a $\lambda$ can not be gauged away.
 
 ii) Any coordinate transformation manifests for $g_{\sigma \tau} = \eta_{\sigma \tau} e^{\varphi}$ 
 (Eq.(\ref{f1and1}))  the following boundary condition on two meeting patches 
\begin{equation*}
 \frac{\partial x^{[1]\sigma}}{\partial x^{[2] \sigma'}} \frac{\partial x^{[1]\tau}}{\partial x^{[2] \tau'}} 
g^{[1]}{}_{ \sigma \tau} (x^{[1]}) =
g^{[2]}{}_{\sigma' \tau'} (x^{[2]})
\end{equation*} 
leading to
 \begin{eqnarray}
 det(\frac{\partial x^{[1]}}{\partial x^{[2]}} )&=&  e^{\varphi^{[2]} - \varphi^{[1]}},\nonumber \\
 \frac{\partial x^{[1](5)}}{\partial x^{[2] (5)}} &=&  \frac{\partial x^{[1](6)}}{\partial x^{[2] (6)}}, \quad
 \frac{\partial x^{[1](5)}}{\partial x^{[2] (6)}} = - \frac{\partial x^{[1](6)}}{\partial x^{[2] (5)}}. 
 \label{conformalpatch}
 \end{eqnarray}
The last two equations are just the Cauchy-Riemann equations, which guarantee the analyticity of 
$z^{[1]} = x^{[1](5)} + i x^{[1] (6)}$ in terms of  $z^{[2]} = x^{[2](5)} + i x^{[2] (6)}$.
From the zweibein transition relation $e^{[1]s}{}_{\sigma } \frac{\partial x^{[1]\sigma}}{\partial x^{[2] \tau}}
= e^{[2]s}{}_{\tau}$ and by taking into account the Cauchy-Riemann equations (Eq.(\ref{conformalpatch})) we
find $\phi^{[1]} + \chi= \phi^{[2]}$, with $\chi = - arg(\frac{dz^{[1]}}{dz^{[2]}}). $

If we introduce the notation $\omega^{[i]5}{}_{6} = \omega^{[i]5}{}_{6(5)} + i \omega^{[i]5}{}_{6 (6)}$,
${\cal T}^{[i]} = {\cal T}^{[i]}{}_{(5)} + i {\cal T}^{[i]}{}_{(6)}$, $i=1,2,$ and recognize that since $g_{\sigma \tau}
=\eta_{\sigma \tau} e^{\varphi}$ the antisymmetric tensor $\varepsilon^{\sigma}{}_{\tau } = 
\varepsilon ^{\sigma \rho} e^{\varphi} \eta_{\rho \tau}$ is an invariant tensor,
we may summarize the boundary conditions on a meeting place of two patches as follows
\begin{eqnarray}
\phi^{[1]} &=& \phi^{[2]} + \lambda + arg \frac{dz^{[2]}}{dz^{[1]}}, \quad 
\varphi^{[1]} = \varphi^{[2]} - 2 log|\frac{dz^{[2]}}{dz^{[1]}} |, \nonumber\\
\bar{\omega}^{[1]5}{}_{6} &=& \frac{dz^{[2]}}{dz^{[1]}} \bar{\omega}^{[2]5}{}_{6}  + 2 \frac{d\lambda}{dz^{[1]}}, \quad
\bar{{\cal T}}^{[1]} = \frac{dz^{[2]}}{dz^{[1]}} \bar{{\cal T}}^{[2]}.
\label{genboundarypatch}
\end{eqnarray}

All written up to now is valid for a general two dimensional surface, for which we may require some conditions
 on $\lambda$ and $\varphi$. On a torus ($S^1 \times S^1$), however, we may choose $\varphi=0$ and then it follows
$g_{\sigma \tau}= \eta_{\sigma \tau}$. If our patch is the whole torus, then the overlap region is defined
by going around the rectangular with the two sides $2\pi r^{\sigma}$, where $ r^{\sigma}$ are the two radii of a torus.
Since $z^{[2]} = z^{[1]} + 2 \pi (r^{(5)} +i r^{(6)})$, it follows that $\frac{dz^{[2]}}{dz^{[1]}} =1$ and $arg
\frac{dz^{[2]}}{dz^{[1]}} =0.$
The boundary conditions (Eq.(\ref{genboundarypatch})) then simplify a lot. 

For the ''perpendicular'' boundary conditions,
when $\lambda_{\sigma} $ is allowed to step up by $2\pi n^{\sigma}$, with $n^{\sigma}$ an integer, so that 
$\lambda_{r^{(5)}} (z + 2\pi i r^{(6)}) = \lambda_{r^{(5)}}(z) + 2\pi n^{(5)} $ and 
$\lambda_{r^{(6)}} (z + 2\pi r^{(5)})=
2 \pi n^{6} + \lambda_{r^{(6)}}(z)$,
it then follows that the integral of $\int \omega^{5}{}_{6\sigma} dx^{\sigma}$ along the overlap region
is equal to
\begin{eqnarray}
\int \partial_{[\sigma} \omega^{5}{}_{6 \tau]} dx^{\sigma} dx^{\tau}=  2\pi n^{(5)} + 2\pi n^{(6)}. 
\label{quantizationw}
\end{eqnarray}
It is nonzero and, since it is proportional to an integer, it is quantized. The Euler index of a torus 
(Eq.(\ref{Eulerindex}))  is, of course, zero. Due to the relation 
${\cal T}_{[\sigma \tau]} = - \omega^{5}{}_{6 [\sigma, \tau]}$ (Eq.(\ref{Tsigma})), the torsion part 
is accordingly quantized as well. We could
interpret that the  stepping up process of  $\lambda$ is due to an appearance of something like 
a Dirac string within the patch.

\subsection{Derivations for section~\ref{d=2case}}
\label{cauchy}

We assume a $2$-dimensional manifold covered by patches. 
We label the two coordinates on a $2$-dimensional manifold as $x^{(5)}$
and $x^{(6)}$ and use superscripts $[1]$ and $[2]$ to label variables 
belonging to the two overlap regions of the two patches.  Then for a metric tensor with a general 
transformation rule
\begin{equation}\label{g2dtrule}
g_{\rho\lambda}=\eta_{\rho\lambda}\cdot e^{\phi},
\end{equation}
where $\eta_{\rho\lambda}$ denotes the diagonal metric tensor $diag(-1,-1)$,
 a conformal mapping from $x^{[1](5)}, x^{[1](6)}$ to $x^{[2](5)}, 
x^{[2](6)}$  follows
\begin{equation}\label{g2transrule}
\frac{\partial x^{[1]\sigma}}{\partial x^{[1]\tau}}
\frac{\partial x^{[2]\rho}}{\partial x^{[2]\lambda}}
\delta_{\sigma\tau} 
= \delta_{\rho\lambda} e^{\phi^{[2]}-\phi^{[1]}}.
\end{equation}
This can further be written as
\begin{equation}\label{g2trans55}
\frac{\partial x^{[1](5)}}{\partial x^{[2](5)}} 
\frac{\partial x^{[1](5)}}{\partial x^{[2](5)}}
+ \frac{\partial x^{[1](6)}}{\partial x^{[2](5)}} 
  \frac{\partial x^{[1](6)}}{\partial x^{[2](5)}}
=  e^{\phi^{[2]}-\phi^{[1]}},
\end{equation}
\begin{equation}\label{g2trans66}
\frac{\partial x^{[1](5)}}{\partial x^{[2](6)}} 
\frac{\partial x^{[1](5)}}{\partial x^{[2](6)}}
+ \frac{\partial x^{[1](6)}}{\partial x^{[2](6)}} 
  \frac{\partial x^{[1](6)}}{\partial x^{[2](6)}}
=  e^{\phi^{[2]}-\phi^{[1]}},
\end{equation}
and
\begin{equation}\label{g2trans56}
\frac{\partial x^{[1](5)}}{\partial x^{[2](5)}} 
\frac{\partial x^{[1](5)}}{\partial x^{[2](6)}}
+ \frac{\partial x^{[1](6)}}{\partial x^{[2](5)}} 
  \frac{\partial x^{[1](6)}}{\partial x^{[2](6)}}
= 0.
\end{equation}

To calculate the square of 
$\det\left(\displaystyle\frac{\partial x^{[1]\sigma}}{\partial x^{[2]\rho}}\right)$ we
explicitly  write it down and use the above 
equations~\ref{g2trans55}--\ref{g2trans56}. We get
\begin{equation}\label{detsqval}
\left[\det\left(
  \displaystyle\frac{\partial x^{[1]}}{\partial x^{[2]\rho}}\right)  
\right]^2 = e^{2(\phi^{[2]}-\phi^{[1]})}.
\end{equation}
It then follows 
\begin{equation}\label{detval}
\det\left(\displaystyle\frac{\partial x^{[1]\sigma}}{\partial x^{[2]\rho}}\right)
= \pm e^{\phi^{[2]}-\phi^{[1]}}.
\end{equation}
This enables us to conclude 
\begin{equation}\label{CR1}
\frac{\partial x^{[1](5)}}{\partial x^{[2](5)}} 
= \pm \frac{\partial x^{[1](6)}}{\partial x^{[2](6)}},\nonumber\\
\frac{\partial x^{[1](5)}}{\partial x^{[2](6)}} 
= \mp \frac{\partial x^{[1](6)}}{\partial x^{[2](5)}}
\end{equation}
These equations are just the familiar Cauchy-Riemann equations, guaranteeing  
 the analiticity of a complex function 
$z^{[1]}=x^{[1](5)}+ i x^{[1](6)}$,  expressed in terms of variable
$z^{[2]}=x^{[2](5)}+ i x^{[2](6)}$. 

Accordingly we find the transformation rules for the vielbeins
\begin{equation}\label{logdetsq}
\phi^{[2]}-\phi^{[1]} 
= 2\,\log \left|
         \frac{dz^{[1]}}{dz^{[2]}} 
       \right|.
\end{equation}
Transformation rule for zweibeins $e^s{}_\tau$ for these transformations is
\begin{equation}\label{zweibeintrule}
e^{[1]h}{}_\sigma \frac{\partial x^{[1]\sigma}}{\partial x^{[2]\rho}}
= e^{[2]h}{}_\rho.
\end{equation}
For the zweibein parametrized as
\begin{equation}\label{zweiparam}
e^s{}_\sigma = e^{\phi/2}
  \begin{pmatrix}
     \cos\varphi & \sin\varphi \\
     -\sin\varphi & \cos\varphi
  \end{pmatrix}.    
\end{equation}
it follows
\begin{eqnarray}
 & &
e^{\phi^{[1]}/2}\,
 \begin{pmatrix}
     \cos\varphi^{[1]} & \sin\varphi^{[1]} \\
     -\sin\varphi^{[1]} & \cos\varphi^{[1]}
  \end{pmatrix}
\,\begin{pmatrix}
    \displaystyle
    \frac{\partial x^{[1](5)}}{\partial x^{[2](5)}} &
    \displaystyle
    \frac{\partial x^{[1](5)}}{\partial x^{[2](6)}} \\
    \displaystyle
    \frac{\partial x^{[1](6)}}{\partial x^{[2](5)}} &
    \displaystyle
    \frac{\partial x^{[1](6)}}{\partial x^{[2](6)}}
  \end{pmatrix}\label{matrtrans}\\
&=&
e^{\phi^{[1]}/2}\,
 \begin{pmatrix}
     \cos\varphi^{[2]} & \sin\varphi^{[2]} \\
     -\sin\varphi^{[2]} & \cos\varphi^{[2]}
  \end{pmatrix}.\nonumber
\end{eqnarray}
Then it follows for the logarithmic derivative  $\frac{dz^{[1]}}{dz^{[2]}}$ is
\begin{equation}\label{logderdz}
\frac{\frac{dz^{[1]}}{dz^{[2]}}}{\left|\frac{dz^{[1]}}{dz^{[2]}} \right|}
= e^{i\chi} = e^{i\cdot\arg(\frac{dz^{[1]}}{dz^{[2]}})}
\end{equation} 
or
\begin{equation}\label{ichi}
\log \left\{  
\frac{\frac{dz^{[1]}}{dz^{[2]}}}{\left|\frac{dz^{[1]}}{dz^{[2]}} \right|}
\right\}
= i\chi.
\end{equation}
Recalling Eq.(~\ref{detsqval}) and rewriting Cauchy-Riemann's 
equations~(\ref{CR1}) as
\begin{equation}\label{CR1and2alt}
\frac{\partial x^{[1](5)}}{\partial x^{[2](5)}} 
= \Re \frac{dz^{[1]}}{dz^{[2]}}
\qquad
\mathrm{and} 
\qquad
\frac{\partial x^{[1](6)}}{\partial x^{[2](5)}} 
= \Im \frac{dz^{[1]}}{dz^{[2]}}
\end{equation}
it follows for the zweibein transformation equation (~\ref{matrtrans})
\begin{eqnarray}
 & &
e^{\phi^{[1]}/2}\,
 \begin{pmatrix}
     \cos\varphi^{[1]} & \sin\varphi^{[1]} \\
     -\sin\varphi^{[1]} & \cos\varphi^{[1]}
  \end{pmatrix}
\,\frac{1}{e^{(\phi^{[2]}-\phi^{[1]})/2}}
\,\begin{pmatrix}
    \displaystyle
    \Re \frac{dz^{[1]}}{dz^{[2]}} &
    \displaystyle
    -\Im \frac{dz^{[1]}}{dz^{[2]}} \\
    \displaystyle
    \Im \frac{dz^{[1]}}{dz^{[2]}} &
    \displaystyle
    \Re \frac{dz^{[1]}}{dz^{[2]}}
  \end{pmatrix}\nonumber\\
&=&
 \begin{pmatrix}
     \cos\varphi^{[2]} & \sin\varphi^{[2]} \\
     -\sin\varphi^{[2]} & \cos\varphi^{[2]}
  \end{pmatrix},\label{matrtrans2intermed}
\end{eqnarray}
that is
\begin{eqnarray}
 & &
e^{\phi^{[1]}/2}\,
 \begin{pmatrix}
     \cos\varphi^{[1]} & \sin\varphi^{[1]} \\
     -\sin\varphi^{[1]} & \cos\varphi^{[1]}
  \end{pmatrix}
\,\frac{1}{e^{(\phi^{[2]}-\phi^{[1]})/2}}
\,\begin{pmatrix}
    \cos\chi & -\sin\chi \\
    \sin\chi & \cos\chi
  \end{pmatrix}\nonumber\\
&=&
 \begin{pmatrix}
     \cos\varphi^{[2]} & \sin\varphi^{[2]} \\
     -\sin\varphi^{[2]} & \cos\varphi^{[2]}
  \end{pmatrix}.\label{matrtrans2final}
\end{eqnarray}

This can be rewritten as
\begin{equation*}
 \begin{pmatrix}
     \cos(\varphi^{[1]}+\chi) & \sin(\varphi^{[1]}+\chi) \\
     -\sin(\varphi^{[1]}+\chi) & \cos(\varphi^{[1]}+\chi)
  \end{pmatrix}
=
 \begin{pmatrix}
     \cos\varphi^{[2]} & \sin\varphi^{[2]} \\
     -\sin\varphi^{[2]} & \cos\varphi^{[2]}
  \end{pmatrix},
\end{equation*}
with
\begin{equation}\label{phi1pluschi}
\varphi^{[1]}+\chi = \varphi^{[2]}.
\end{equation}

We take into account now also the gauge transformation condition for a chosen coordinate patch
\begin{equation}\label{patchgaugetr}
\frac{\partial x^{[1]\rho}}{\partial x^{[2]\mu}}
\omega^{[1]5}{}_{6\rho}(x^{[1]})
=
\omega^{[2]5}{}_{6\rho}(x^{[2]}) + \lambda_{,\mu},
\end{equation} 
and
\begin{equation}\label{pathcgaugevarphi}
\varphi^{[1]}=\varphi^{[2]} + \lambda, \qquad
\phi^{[1]} = \phi^{[2]}.
\end{equation}
Then we have
\begin{equation}\label{phi1gaugecond}
\varphi^{[1]}=\varphi^{[2]} + \lambda -\chi.
\end{equation}
From Eq.(~\ref{patchgaugetr}) it follows
\begin{eqnarray}
\omega^{[2]5}{}_{6(5)} + \lambda_{,(5)} 
&=& \Re \frac{dz^{[1]}}{dz^{[2]}}\cdot\omega^{[1]5}{}_{6(5)}
   + \Im \frac{dz^{[1]}}{dz^{[2]}}\cdot\omega^{[1]5}{}_{6(6)}
\label{omegalambda5}\\ 
\omega^{[2]5}{}_{6(6)} + \lambda_{,(6)} 
&=& \Re \frac{dz^{[1]}}{dz^{[2]}}\cdot\omega^{[1]5}{}_{6(6)}
   - \Im \frac{dz^{[1]}}{dz^{[2]}}\cdot\omega^{[1]5}{}_{6(5)}.
\label{omegalambda6} 
\end{eqnarray}

One can  prove that the  $\varepsilon^{\mu}{}_{\nu}$ is for  the $2$-dimensional case 
(and not the tensor $\varepsilon_{\mu\nu}$) equal  in every coordinate patch:
\begin{equation}\label{eps2inv}
\varepsilon^{[2]\mu}{}_{\nu}
= \varepsilon^{[1]\mu}{}_{\nu},
\end{equation}
where the relation
\begin{equation}\label{epsrelation}
\varepsilon^{\mu}{}_{\nu}
= \varepsilon^{\mu\rho}\eta_{\rho\nu} e^\phi
\end{equation}
holds.
Defining the quantity $T^\sigma$ as
\begin{equation}\label{defT}
T^\sigma=T^\sigma{}_{\mu\nu}\varepsilon^{\mu\nu}\, e^{-\phi}
\end{equation}
we get
\begin{equation*}
T_\sigma=g_{\mu\nu}T^\nu=T^\mu{}_{\rho\lambda}\varepsilon^{\rho\lambda}.
\end{equation*}
This quantity is related to the torsion tensor and can be written using
familiar definitions for a covariant derivative and our choice of
the zweibein parametrization (Eq.(\ref{zweiparam})) as 
\begin{equation}\label{Tcovar}
T_\sigma=\varphi_{;\sigma} + \varepsilon_\sigma{}^\mu\frac{1}{2}\phi_{,\mu},
\end{equation}
where the following holds:
\begin{equation}\label{varphicov}
\varphi_{;\sigma}=\varphi_{,\sigma}-\omega^5{}_{6\sigma}. 
\end{equation}

\section{Weyl spinor in $M^2$}
\label{WeylM2}

The Weyl spinor wave functions $\psi^{[i]}$, $i =1,2,$ in a space 
$M^2$ is influenced
by the boundary conditions of Eq.(\ref{genboundarypatch}) 
\begin{eqnarray}
\psi^{[1]} = e^{- i \lambda S^{56}} \psi^{[2]} 
\label{NHDboundarypsi}
\end{eqnarray}
and must obey the Weyl equations of motion
\begin{eqnarray}
\gamma^s f^{[i]\sigma}{}_{s} p^{[i]}{}_{0\sigma} \psi^{[i]} =0, \quad {\rm with}\; p^{[i]}{}_{0 \sigma} = 
p^{[i]}{}_{\sigma} - 
\frac{1}{2} S^{s't} \omega^{[i]}{}_{s't \sigma}
\label{boundarypsiweyl}
\end{eqnarray}
with $i=1,2$.

We  find for the zweibeins from Eq.(\ref{f1and1}) the Weyl equation
\begin{eqnarray}
- e^{-\varphi^{[i]}/2}
\begin{pmatrix}
0  & -e^{i\phi^{[i]}} (p^{[i]} + \frac{1}{2} \bar{\omega}^{[i]5}{}_6) \\
e^{-i\phi^{[i]}} (\bar{p}^{[i]} - \frac{1}{2} \omega^{[i]5}{}_6 )& 0 
\end{pmatrix} \;\;\psi^{[i]} = 0,\label{boundarypsi}\\
\quad i=1,2, \nonumber
\end{eqnarray}
with
\begin{eqnarray}
p^{[i]} &=& p^{[i]}{}_{(5)} - i p^{[i]}{}_{(6)},\quad \bar{p}^{[i]} = p^{[i]}{}_{(5)} +i p^{[i]}{}_{(6)} ,\nonumber\\
\omega^{[i]5}{}_{6} &=& \omega^{[i]5}{}_{6(5)} + i \omega^{[i]5}{}_{6(6)},
\quad \bar{\omega}^{[i]5}{}_{6} 
= \omega^{[i]5}{}_{6(5)} -i \omega^{[i]5}{}_{6(6)}, 
\label{ppbar}
\end{eqnarray}
while $\omega^{[2]5}{}_{6\sigma} = - \partial^{[2]}{}_{\sigma} \lambda +
\omega^{[1]5}{}_{6\sigma}, \phi^{[2]} = \phi^{[1]} +\lambda$.
For  details see subsection \ref{weylder}.
One easily sees that for massless spinors neither
$\varphi$ nor $\phi$ contributes in the equations of motion for the left and the right handed solution 
($\Gamma^{(2)} =\pm 1$, respectively).

\subsection{Derivation of Weyl equations}
\label{weylder}

We may write the Weyl equations in $d=2$ -dimensional space as follows
\begin{equation}\label{Diractheta}
(\gamma^{5}p_{05}+\gamma^{6}p_{06}) |\psi>=0 = (\gamma^{5} f^{\sigma}{}_{5}p_{0\sigma}+\gamma^{6} f^{\sigma}{}_{6} 
p_{0\sigma}) |\psi>.
\end{equation}
Multiplying this equation with $\gamma^{5}$ and noticing that 
\begin{equation}\label{defp0}
p_{0\sigma}=p_\sigma - \frac{1}{2} S^{ab}\omega_{ab\sigma},
\end{equation}
where, as shown elsewhere in this article, 
$S^{ab}=\frac{i}{4}[\gamma^a,\gamma^b]$, our Weyl equation takes  (for the most general form of zweibein of
Eq.(\ref{f1and1})) the form
\begin{eqnarray}
& & 
- e^{\varphi}\biggl[
-e^(\phi) \left(-p_{(5)}+ (\pm\frac{1}{2})\omega_{(5)}\right)
\begin{pmatrix}
  1 & 0 \\
  0 & 1
\end{pmatrix} \nonumber\\
& & \qquad {} + 
(-1)(\pm)
\begin{pmatrix}
  i & 0 \\
  0 & i
\end{pmatrix}
-e^{\phi} \left( p_{(6)}- (\pm\frac{1}{2})\omega_{(6)}\right) 
\biggr] |\psi_\pm> =0, \label{Dirac1}
\end{eqnarray}
for the two solutions $|\psi_{\pm}>$, which correspond to the eigenvectors of $S^{56}$ with the eigenvalues
$\pm \frac{1}{2}$, correspondingly.  
It then follows, independently of the two angles $\varphi$ an $\phi$
\begin{equation}\label{Dirac2}
\left[p_{(5)} \pm ip_{(6)} - 
   (\pm)\frac{1}{2}(\omega_{(5)}+ (\pm) i\omega_{(6)}) \right]|\psi_\pm> =0.
\end{equation}

If we introduce the dimensionless variables
\begin{eqnarray}
v= \hat{x}^{(5)} +i \hat{x}^{(6)},\quad v^* = \hat{x}^{(5)} -i \hat{x}^{(6)}, \nonumber\\
\frac{\partial}{\partial v} = \frac{1}{2} ( \frac{\partial}{\partial \hat{x}^{(5)}} -i \frac{\partial}{\partial \hat{x}^{(6)}}),
\quad \frac{\partial}{\partial v} = 
\frac{1}{2} ( \frac{\partial}{\partial \hat{x}^{(5)}} -i \frac{\partial}{\partial \hat{x}^{(6)}}), \nonumber\\
{\rm with} \quad \hat{x}^{\sigma} = x^{\sigma}/(2\pi r),
\label{vvstar}
\end{eqnarray}
where $r$ is the radius of each of the compactified dimensions, the relations
follow
\begin{equation}\label{dbarz}
\bar{p}: = p_{(5)} + ip_{(6)}  
=  \frac{i}{\pi r} \frac{\partial}{\partial v*} ,\nonumber\\
p=: p_{(5)} - ip_{(6)}= \frac{i}{\pi r}\frac{\partial}{\partial v}.
\end{equation}

We  define in addition
\begin{equation}\label{defomombar}
\omega=\omega_{(5)}+i\omega_{(6)}, \qquad
\bar{\omega}=\omega_{(5)}-i\omega_{(6)}.
\end{equation}
With these definitions it follows for Eq.(\ref{Dirac2}) 
\begin{eqnarray}
\left(\bar{p} - S^{56}\omega \right) |\psi_+> =0, \nonumber\\
\left(p - S^{56}\bar{\omega} \right) |\psi_-> =0, \label{s56w}
\end{eqnarray}
and also 
\begin{eqnarray}
\left(\frac{\partial}{\partial v^*} 
  - \frac{i\pi r}{2}\omega \right) \psi_+ &=& 0, \label{Dirac3plus}\\
\left(\frac{\partial}{\partial v} 
  + \frac{i\pi r}{2}\bar{\omega} \right) \psi_- &=& 0.
  \label{Dirac3minus}
\end{eqnarray}

\section{ An example of massless spinors coupled to $U(1)$ field}
\label{theta}

We shall demonstrate in this section one example of spinors living on a torus and obeying  the Weyl equations
as well as  all the properties, required
in Eqs.(\ref{genboundarypatch}) -
 (\ref{boundarypsi}). The wavefunctions for these spinors can be  written in terms of well known 
  $\vartheta_{\alpha},\; \alpha =1,2,$ functions\cite{theta} 
\begin{eqnarray}
\psi^{(2)}{}_{-} = {\cal N} \; \vartheta_{1} (v) \vartheta_{2} (v^*)  |\downarrow>, \quad
\psi^{(2)}{}_{+} = {\cal N} \; \vartheta_{1} (v^*) \vartheta_{2} (v)  |\uparrow> , \nonumber\\
{\rm with} \; v= {\hat x}^{(5)} + i {\hat x}^{(6)}\; {\rm and}\; x^{\sigma} = (2\pi r) {\hat x}^{\sigma}, 
\label{psisoltheta}
\end{eqnarray}
for the spin connection fields $\omega^{5}{}_{6}$
and $\bar{\omega}^{5}{}_{6}$, which can be expressed as  follows 
\begin{eqnarray}
\omega^{5}{}_{6} = \frac{2i}{\pi r}\; \frac{\partial \psi^{(2)}_{+}(v^*)}{\partial v^*}/\psi^{(2)}_{+}(v^*) 
=  \frac{2i}{\pi r} 
\frac{\vartheta_{1}(v^*)_{,v^*}}{\vartheta_{1}(v^*)}, \nonumber\\
\bar{\omega}^{5}{}_{6} = \frac{-2i}{\pi r} \;\frac{\partial \psi^{(2)}_{-}(v)}{\partial v}/\psi^{(2)}_{-}(v)  =
\frac{-2i}{\pi r} \; \frac{ \vartheta_{1}(v)_{,v}}{\vartheta_{1}(v)}. 
\label{omegatheta}
\end{eqnarray}
The second equation is the complex conjugate of the first.
We see that the two solutions have the desired periodic properties
\begin{eqnarray}
\psi^{(2)}{}_{\pm}({\hat x}^{(5)} +1, {\hat x}^{(6)} + 1 ) 
&=& - e^{-i 4\pi {\hat x}^{(5)}}\; \psi^{(2)}{}_{\pm}({\hat x}^{(5)}, {\hat x}^{(6)}) \nonumber\\
 &=& e^{- iS^{56} \lambda} 
\psi^{(2)}{}_{\pm}({\hat x}^{(5)}, {\hat x}^{(6)} ). 
\label{checktheta}
\end{eqnarray}
From here it follows that
\begin{eqnarray}
\pm \frac{1}{2} \lambda = 4 \pi {\hat x}^{(5)} +  \pi. 
\label{lambdatheta}
\end{eqnarray}
Using the spin connection fields from Eq.(\ref{omegatheta}) (or the expression for $\lambda$ from Eq.(\ref{lambdatheta})),
the integral $\int ER dx^{(5)} dx^{(6)}$ can be calculated. For a simple choice of a zweibein 
$f^{\sigma}{}_{s} = \delta^{\sigma}{}_{s}$
we found for the integral of the curvature
\begin{eqnarray}
\int \; E \delta^{\sigma}{}_{[s} \delta^{\tau}{}_{t]} \; \omega^{st}{}_{\sigma,\tau}\; dx^{(5)} dx^{(6)}
= -16 \pi. 
\label{intR}
\end{eqnarray}

\subsection{Spinors, products of $\vartheta_{\alpha} functions$}
\label{dertheta}

We shall demonstrate that theta functions (see for example Bateman \& Erd\'elyi~\cite{theta}) 
 are solutions of the Weyl equations for a very particular spin connection field. We shall take
$\psi_+=\mathcal{N} \vartheta_1(v^*) \vartheta_2(v)|\uparrow>$ and $\psi_-= {\cal N} \vartheta_1(v)
\vartheta_2 (v^*) |\downarrow>$, with $|\uparrow>,|\downarrow>$ for the two eigenstates of $S^{56}$, with
eigenvalues $1/2$ for $|\uparrow>$ and $-1/2$ for $|\downarrow>$ and with 
$v$ and $v^*$ defined in Eq.(\ref{vvstar}).

Theta functions have well known  series expansions
\begin{eqnarray}
\vartheta_1(v) = i \sum^{+\inf}_{-\infty}\; (-)^n q^{(n-1/2)^2} e^{(2n-1)\pi v i},\nonumber\\
\vartheta_2(v) = i \sum^{+\inf}_{-\infty}\;  q^{(n-1/2)^2} e^{(2n-1)\pi v i},
\label{series}
\end{eqnarray}
with $q=e^{i\pi \tau}$ and we take $\tau =i$. In $\hat{x}^{(5)}$ direction 
$\vartheta_\alpha$ are normal periodic functions, while in
the $\hat{x}^{(6)}$ direction they are periodic in the complex sense.

One easily finds that 
\begin{eqnarray}
\vartheta_1(v \pm 1) = - \vartheta_1(v), \quad \vartheta_2(v \pm 1) = \vartheta_2(v),\nonumber\\
\vartheta_1(v \pm i) = - q^{-1} e^{\mp\pi i(2 v \pm \tau} \vartheta_1(v), \quad \vartheta_2(v \pm 1) = 
e^{\mp\pi i(2 v \pm \tau}\vartheta_2(v).
\label{series2}
\end{eqnarray}

One also can derive
\begin{equation}\label{Theta1series}
\frac{\frac{\partial}{\partial v}\vartheta_1(v)}{\vartheta_1(v)}
= \pi\cot(\pi v) 
+ 4\pi\sum_{m=1}^\infty \frac{q^{2m}}{1-q^{2m}}\sin2m\pi v,
\end{equation}
with the property
\begin{equation}\label{Thetaperiod}
\frac{\frac{\partial}{\partial v}\vartheta_1(v +m + n i)}{\vartheta_1(v+m +n i)} - \frac{\frac{\partial}{\partial v}\vartheta_1(v )}{\vartheta_1(v)} = -2 \pi n i,
\end{equation}
for any integers $n,m$, since the period of $\vartheta_1$ is $1+i.$
Taking into account that $\omega = \omega_{(5)} + i \omega_{(6)} = \frac{2i}{\pi r} 
\frac{\vartheta_1(v^*)_{,v^8}}{\vartheta_1(v^*)}  $, we end up with the relations
\begin{eqnarray}
\omega^{5}{}_{6(5)} &=& 
   -\frac{1}{r} \biggl\{- \frac{\sinh 2\pi \hat{x}^{(6)}}{(\cosh 2\pi \hat{x}^{(6)}- \cos 2\pi \hat{x}^{(5)} )/2} \nonumber\\
 & & \qquad {}+ 8 
   \sum_{m=1}^\infty \frac{q^{2m}}{1-q^{2m}} \cos 2m \pi \hat{x}^{(5)} \sinh 2m \pi \hat{x}^{(6)}\biggr\},\nonumber\\
\omega^{5}{}_{6(6)} &=& 
-\frac{1}{ r} \biggl\{ \frac{\sin 2\pi \hat{x}^{(5)}}{(\cosh 2\pi \hat{x}^{(6)}- \cos 2\pi \hat{x}^{(5)} )/2} \nonumber\\
 & & \qquad {}+ 8 
   \sum_{m=1}^\infty \frac{q^{2m}}{1-q^{2m}} \sin 2m \pi \hat{x}^{(5)} \cosh 2m \pi \hat{x}^{(6)}\biggr\}.
   \label{omega56}
\end{eqnarray}

Accordingly we find that 
\begin{eqnarray}
\omega^{5}{}_{6(5)}(x^{(5)} +1, -(x^{(6)} +1)) - \omega_{(5)}(x^{(5)}, -x^{(6)}) &=& 
   -\frac{4}{ r},\nonumber\\
 \omega^{5}{}_{6(6)}(x^{(5)} +1, -(x^{(6)} +1)) - \omega_{(6)}(x^{(5)}, -x^{(6)}) &=& 
   0,  
   \label{omega56per}
\end{eqnarray}
One also notices that
\begin{eqnarray}
\omega^{5}{}_{6(5),(6)} &=& 
   \frac{1}{r^2} \biggl\{ 
       \frac{(- 1 + \cos 2\pi \hat{x}^{(5)} \cosh 2\pi \hat{x}^{(6)})}{A^2} 
 \nonumber\\ 
  & &\qquad\qquad {}+ 8 \sum_{m=1}^\infty \frac{q^{2m}}{1-q^{2m}} 
              \cos 2m \pi \hat{x}^{(5)} \cosh 2m \pi \hat{x}^{(6)}
                  \biggr\} \nonumber\\ 
 &=& \omega^{5}{}_{6(6),(5)}, \label{omega56a} 
\end{eqnarray}
with  
\begin{equation*}
A = \frac{1}{2} (\cosh 2 \pi \hat{x}^{(6)}- \cos 2 \pi \hat{x}^{(5)}),
\end{equation*}
so that we immediately see that the curvature $R$, which is proportional to $\omega_{[(5),(6)]}$ is equal to 
zero all over, except at the point when $A=0,$ which is happening for $x^{(5)} = 0 = x^{(6)}$, where both
$\omega_{(5),(6)}$ and $\omega_{(6),(5)}$ have a pole.

Besause of the pole, which $\omega_{{(5),(6)}}$ has for $x^{(5)} = 0 = x^{(6)}$, the integral of the curvature is nonzero.
We obtain
\begin{eqnarray*}
 & & \int_{x^{(5)}=a}^{a+2\pi r}\,\int_{x^{(6)}=b}^{b+2\pi r}\,\,
 E \omega^{5}{}_{6[(5),(6)]}\,dx^{(5)}\,dx^{(6)} =\\
 &=& \int_{x^{(5)}=a}^{a+2\pi r}\,dx^{(5)}\,
    \omega^{5}{}_{6(5)}\big|_{x^{(6)}=b}^{b+2\pi r}
 - \int_{x^{(6)}=b}^{b+2\pi r}\,dx^{(6)}\,
    \omega^{5}{}_{6(6)}\big|_{x^{(5)}=b}^{b+2\pi r},
\end{eqnarray*}
which when taking into account Eq.(\ref{omega56per})  leads to
\begin{equation}\label{intnon0}
\iint\,E \omega^{5}{}_{6[(5),(6)]}\,dx^{(5)}\,dx^{(6)} =
\int_a^{a+2\pi r}\,dx^{(5)}\,\left( -\frac{4}{r}\right)
=-8\pi.
\end{equation}
in agreement with Eq.(\ref{intR}), since it is just twice the above value, as it  should be.

\section{ Spinor coupled to gauge fields in $M^{(1+3)} \times (S^1 \times S^1)$}
\label{spinorcoupled}

To study how do spinors couple to the Kaluza-Klein gauge fields in the case of $M^{(1+5)}$, compactified
to $M^{(1+3)} \times (S^1 \times S^1)$, we first look for the appearance of 
pure gauge fields,  when coordinate transformations  of the type $x^{' \mu}= x^{\mu}, x^{' \sigma}= x^{\sigma}
+ \vartheta^{\sigma}(x^{\mu})$ are performed. We start with  vielbeins  $e^m{}_{\mu}=\delta^m{}_{\mu}$ ($f^{\mu}{}_{m}
=\delta^{\mu}{}_{m}$) and 
 $e^s{}_{\sigma} (f^{\sigma}{}_{s})$ from Eq.(\ref{f1and1}). Since
$f^{\alpha}{}_{a}$ transform as  vectors under general  coordinate transformations ($\delta f^{\alpha}{}_{a} = 
f^{\alpha}{}_{a,\beta} \vartheta^{\beta} + f^{\gamma}{}_{a}\vartheta^{\alpha}{}_{,\gamma}$) 
and since we are interested in only the transformations 
which keep $f^{\sigma}{}_{s}$ unchanged), we end up with $\delta f^{\sigma}{}_{m} = 
\delta^{\mu}{}_{m}\vartheta^{\sigma}{}_{,\mu}, \delta f^{\mu}{}_{s}=0, \delta f^{\sigma}{}_{s} = 0. $ 
Replacing pure gauge fields with
true fields to which these pure 
gauge transformations should belong, we find for new vielbeins
\begin{eqnarray}
e^a{}_{\alpha} = 
\begin{pmatrix}
\delta^{m}{}_{\mu}  & e^{m}{}_{\sigma} \\
0= e^{s}{}_{\mu} & e^s{}_{\sigma} 
\end{pmatrix},
f^{\alpha}{}_{a} =
\begin{pmatrix}
\delta^{\mu}{}_{m}  & f^{\sigma}{}_{m} \\
0= f^{\mu}{}_{s} & f^{\sigma}{}_{s} 
\end{pmatrix},  
\label{f6}
\end{eqnarray}
with $f^{\sigma}{}_{m} = A^{\sigma}{}_{\mu} \delta^{\mu}{}_{m}$. In the last equation we define the $U(1)$ gauge fields $A^{\sigma}{}_{\mu}$.

Taking into account that also $\omega^{a}{}_{b \alpha}$ transform as  vectors and that before ''switching'' 
on the $U(1)$ field $A^{\sigma}{}_{\mu}$, only  $\omega^{s}{}_{t \sigma}$ were non zero (which do not change
under these transformations), we end up with new fields
\begin{eqnarray}
\omega^{s}{}_{t \sigma},\quad \omega^{s}{}_{t \mu}  = -A^{\sigma}{}_{\mu} \omega^{s}{}_{t \sigma},
\label{omega6}
\end{eqnarray}
while all the other components of $ \omega^{m}{}_{b \alpha} =0$, 
since for simplicity we allow no gravity in
$(1+3)$ dimensional space. 
We can obtain the corresponding ${\cal T}^{\alpha}{}_{\beta \gamma}$
from Eqs.(\ref{RandTdet},\ref{Tsigma}) or by taking into account that ${\cal T}^{\alpha}{}_{\beta \gamma}$ transforms as a 
third rank tensor.

To find out the current, coupled to the Kaluza-Klein gauge fields $A^{\sigma}{}_{\mu}$, we
analyze the spinor action 
\begin{eqnarray}
{\cal S} = \int \; d^dx E \bar{\psi} \gamma^a p_{0a} \psi.
\label{spinoractionwhole}
\end{eqnarray}

{\it Statement 3:} The above action reduces  for the case that 
$\omega^{s}{}_{t \mu}  = -A^{\sigma}{}_{\mu} \omega^{s}{}_{t \sigma}$, for
the vielbeins from Eq.(\ref{f6}) and for the solutions of the Weyl  
Eq.(\ref{s56w}) to
\begin{eqnarray}
S = \int \; 
d^dx E \bar{\psi} \gamma^m \delta^{\mu}{}_{m} p_{\mu} \psi + \int \; d^dx E \bar{\psi} \gamma^s f^{\sigma}{}_{s} p_{0\sigma} \psi
\nonumber\\ 
+ \int \;\bar{\psi} \gamma^m \delta^{\mu}{}_{m} A^{\sigma}{}_{\mu} p_{\sigma} \psi.
\label{spinoraction}
\end{eqnarray}
$\psi$ is now defined in $d=(1+ 5)$ dimensional space and has the spin part defined in the Table~ref{NHDtable1}.
The proof is in subsection \ref{statement3}.
 The first term on the right hand side is the kinetic term, while the second
gives zero for the massless solutions from Eq.(\ref{boundarypsi},\ref{psisoltheta}).

To evaluate the charge and correspondingly the current in $(1+3)$ - dimensional space, one must evaluate 
\begin{eqnarray}
 \int \; dx^{(5)} dx^{(6)} \; \psi^+\; \gamma^0 \gamma^{m} \delta^{\mu}{}_{m} p_{\sigma} A^{\sigma}{}_{\mu}
  \psi_, 
\label{KKsolcharge}
\end{eqnarray}
with $E=1$ in our particular case. We expect that for massless spinors from Eq.(\ref{boundarypsi}) the relation that
$\int dx^{(5)} dx^{(6)} \; \psi^+{}_{\pm} A^{\sigma}{}_{\mu} p_{0\sigma} \psi_{\pm} =0$ has to be fullfilled.
In such a case  accordingly also the current, which massless spinors with $\psi_{-}$ manifest, is just minus the 
current determined by spinors with $\psi_{+}$. All this happens indeed if spinors are described in terms of the
$\vartheta_{\alpha}$ functions of Eq.(\ref{psisoltheta}).
%
For such cases the current of massless spinors  
\begin{eqnarray}
j^{\mu}{}_{\sigma} = \int \; dx^{(5)} dx^{(6)} \; \bar{\psi} \gamma^m \delta^{\mu}{}_{m}  \; Q_{\sigma} \psi
\label{kkcurrent}
\end{eqnarray}
originates from
the Kaluza-Klein charge 
\begin{eqnarray}
<Q_{\sigma}>= < p_{0\sigma} + S^{56} \omega_{56 \sigma}> = <p_{\sigma}> = <S^{56} \omega_{56 \sigma}>,
\label{kkchargemassless}
\end{eqnarray}
where $<{}> $ means the integration of the current over the two coordinates $x^{(\sigma)}, \sigma = (5),(6)$.
The spin connection $\omega_{56 \sigma}$ can not be gauged away due to the boundary conditions on the patching 
meeting place, where
$\lambda_{\sigma} $ is allowed to step up by an integer times $\pi$ (Eq.(\ref{genboundarypatch},\ref{lambdatheta})).
The Kaluza-Klein gauge charge  is for massles spinors, which are desribed by $\psi_{\pm}$, proportional to $ S^{56}$, 
which is on $\psi_{\pm}$ equal to $\pm 1/2$. The handedness in the fifth and the sixth dimension is proportional to $S^{56}$
($\Gamma^{(2)}=-2S^{56}$), which means that  {\em the Kaluza-Klein gauge charge and the 
operator of handedness are proportional to each other}. This also means that massless spinors chirally couple to the
corresponding $U(1)$ field and that the masslessness is protected.

\subsection{Current in d=1+5}
\label{statement3}

We prove that the action $S= \int d^d x \; \psi^+ \gamma^0 \gamma^a p_{0a}$ reduces to 
\begin{eqnarray}
 S &=& \int \; d^dx (\psi^+\; \gamma^0 \gamma^{m} \delta^{\mu}{}_{m} p_{\mu}
  \psi) \nonumber\\
 & & \qquad {}+ \int \; d^dx (\psi^+\; \gamma^0 \gamma^{m} \delta^{\mu}{}_{m} 
     - S^{56}\omega_{56\mu} A^{\sigma}{}_{\mu} \psi) , 
\label{Ssimlpe}
\end{eqnarray}
$\omega^{s}{}_{t \mu}  = -A^{\sigma}{}_{\mu} \omega^{s}{}_{t \sigma}$ (Eq.(\ref{omega6})) and for vielbein
from Eq.(\ref{f6}).

To see this we first recognize that $\gamma^a p_{0a} = \gamma^m p_{0m} + \gamma^s p_{0s} = 
\gamma^m f^{\mu}{}_{m} p_{0 \mu} + \gamma^m f^{\sigma}{}_{m} p_{0 \sigma} + \gamma^s f^{\sigma}{}_{s} p_{0\sigma}$.
Since for spinors which obey the Weyl equations of motion the term $\int d^dx \psi^+ \gamma^0 
\gamma^s f^{\sigma}{}_{s} p_{0\sigma} \psi $ is zero and since 
$\gamma^m f^{\mu}{}_{m} p_{0 \mu} + \gamma^m f^{\sigma}{}_{m} p_{0 \sigma} = 
\gamma^m (\delta^{\mu}{}_{m} p_{ \mu} + (-)S^{56} \omega_{56\mu}) + \gamma^m \delta^{\mu}{}_{m} A^{\sigma}{}_{\mu} (p_{ \sigma}
-S^{56} \omega_{56 \sigma})$, with $\omega_{56\mu} = - A^{\sigma}{}_{\mu} \omega_{56 \sigma}$, the above statement
is proved.

Next we recognize that
\begin{eqnarray}
 p_{\sigma} A^{\sigma}{}_{\mu} &=& \frac{1}{2} (\bar{p} A^+{}_{\mu} + p A^{-}{}_{\mu}),\nonumber\\
 A^{\pm} &=& A^{(5)}{}_{\mu} \mp i A^{(6)}{}_{\mu}, \;\bar{p} = p_{(5)} + i p_{(6)},\; p = p_{(5)} - i p_{(6)}.  
\label{currentpbar}
\end{eqnarray}
Next we prove that  
\begin{eqnarray}
 & & \qquad \int dx^{(5)} dx^{(6)} \psi^+_+ \gamma^0 \gamma^m \delta^{\mu}{}_{\mu } p_{\sigma} A^{\sigma}{}_{\mu} \psi_+ \nonumber \\
&=& -\int dx^{(5)} dx^{(6)} \psi^+_- \gamma^0 \gamma^m \delta^{\mu}{}_{\mu } p_{\sigma} A^{\sigma}{}_{\mu} \psi_-. 
\label{opposite}
\end{eqnarray}
To prove this we must see that the two integrals
\begin{eqnarray}
I_+ &=& <+|\int\,dx^{(5)}\,dx^{(6)}\,
    \biggl\{ 
      \vartheta_1(v^*)\vartheta_2(v)
      \frac{d}{dv^*}\left(\vartheta_1(v^)\vartheta_2(v^*)\right) A^+{}_{\mu} 
      \nonumber\\
 & & \qquad\qquad {} +
   \vartheta_1(v^*)\vartheta_2(v)
      \frac{d}{dv^*}\left(\vartheta_1(v)\vartheta_2(v^*)\right) A^-{}_{\mu}
    \biggr\} |+> ,\\
I_- &=& <-| \int\,dx^{(5)}\,dx^{(6)}\,
   \biggl\{
      \vartheta_1(v)\vartheta_2(v^*)
      \frac{d}{dv^*}\left(\vartheta_1(v^*)\vartheta_2(v)\right) A^+{}_{\mu} 
      \nonumber\\
 & & \qquad\qquad {} +
   \vartheta_1(v)\vartheta_2(v^*)
      \frac{d}{dv}\left(\vartheta_1(v^*)\vartheta_2(v)\right) A^-{}_{\mu}
   \biggr\} |->.
\end{eqnarray}
have opposite values $I_+=-I_-$.
This can easily be seen by  recognizing that 
\begin{eqnarray*}
 & & \qquad
\int\,dx^{(5)}\,dx^{(6)}\,
   \{ \vartheta_1(v^*)\vartheta_2(v)
      \frac{d}{dv^*}\left(\vartheta_1(v^)\vartheta_2(v^*)\right) \\
 &=& 
      \int\,dx^{(5)}\,dx^{(6)}\,
   \left\{ \frac{d}{dv^*} \left(\vartheta_1(v^*)\vartheta_2(v)
      \vartheta_1(v)\vartheta_2(v^*)\right) - \frac{d}{dv^*} \left( \vartheta_1(v^*)\vartheta_2(v)\right)
      \vartheta_1(v)\vartheta_2(v^*)\right\} 
\end{eqnarray*}
and similarly for the other part. We find then 
\begin{eqnarray}
I_- &=& -I_+ + \int_{\alpha}^{\alpha+1}\,dx^{(5)} \, \int_{\beta}^{\beta+1}\; dx^{(6)}\, \nonumber\\
 & & \qquad \qquad
  \biggl\{\frac{d}{dv^*}\left(\vartheta_1(v)\vartheta_1(v^*)\vartheta_2(v)\vartheta_2(v^*)\right)\nonumber\\
 & & \qquad \qquad {} +
   \frac{d}{dv}\left(\vartheta_1(v)\vartheta_1(v^*) \vartheta_2(v)\vartheta_2(v^*)\right)\biggr\}.
\end{eqnarray}
Using periodicity relations  of $\vartheta_{\alpha}$ functions, presented in subsection~\ref{theta},
we find out that both surface terms are zero and that accordingly
\begin{equation*}
I_-  =- I_+.
\end{equation*}
We assume that $A^{\sigma}{}_{\mu}$ do not depend on $x^{(5)}$ and $x^{(6)}$. 
It has to be proved next that for our choice of massless spinors it must be 
\begin{equation}
 S= \int \; d^dx (\psi^+\; \gamma^0 \gamma^{m} \delta^{\mu}{}_{m} p_{0\sigma} A^{\sigma}{}_{\mu}
  \psi) + \int \; d^dx (\psi^+\; \gamma^0 \gamma^{m} \delta^{\mu}{}_{m} 
    p_{\mu}\psi),
  \label{posigma}
\end{equation}
with $p_{0\mu}=p_\mu+S^{56}\omega_{56\sigma}A^\sigma{}_\mu$.
To make the proof we first recognize that
\begin{equation}
 p_{0\sigma} A^{\sigma}{}_{\mu}
= \frac{1}{2} (\bar{p} -S^{56} \omega^5{}_{6}) A^+{}_{\mu} +  \frac{1}{2} (p -S^{56} \bar{\omega^5{}_{6}}) A^-{}_{\mu}.
  \label{posigmaa}
\end{equation}
Taking into account the Weyl equations we recognize 
$ <\psi_{\pm}|p_{0\sigma} A^{\sigma}{}_{\mu}|\psi_{\pm}> =0$, if
$<\psi_{+}|\frac{1}{2} (p -S^{56} \bar{\omega}^5{}_{6}) A^{-}{}_{\mu}|\psi_{+}> =0$ and 
$<\psi_{-}|\frac{1}{2} (\bar{p} -S^{56} \omega^5{}_{6}) A^{+}{}_{\mu}|\psi_{-}> =0$.
Next we recognize that $\omega^5{}_{6}(v^*) = \frac{2i}{\pi r} \frac{\vartheta_1(v^*)_{'v*}}{\vartheta_1(v^*)}$ and
$\bar{\omega}^5{}_{6}(v) = -\frac{2i}{\pi r} \frac{\vartheta_1(v)_{'v}}{\vartheta_1(v)}$.
We recognize that Eq.(~\ref{posigmaa})
leads to the two integrals
\begin{eqnarray}
 & & 
\frac{i}{2\pi r} \int_{\alpha}^{\alpha + 1}\,dx^{(5)} \, \int_{\beta}^{\beta+1}\; dx^{(6)}\,
  \biggl\{\left(\vartheta_1(v)\vartheta_2(v^*)\frac{\partial}{\partial v}\vartheta_2(v)\vartheta_1(v^*)\right) \nonumber\\
 & & \qquad {} +
   \left(\vartheta_1(v^*)\vartheta_2(v^*) \frac{\partial}{\partial v}\vartheta_1(v)\vartheta_2(v)\right)\biggr\} A^{-}{}_{\mu},\nonumber\\
 & & 
   \frac{i}{2\pi r} \int_{\alpha}^{\alpha + 1}\,dx^{(5)} \, \int_{\beta}^{\beta+1}\; dx^{(6)}\,
     \biggl\{\left(\vartheta_1(v^*)\vartheta_2(v)\frac{\partial}{\partial v^*}\vartheta_2(v^*)\vartheta_1(v)\right)\nonumber\\
 & & \qquad {} +
   \left(\vartheta_2(v)\frac{\partial}{\partial v^*}\vartheta_1(v^*) \vartheta_1(v)\vartheta_2(v^*)\right)\biggr\} A^{+}{}_{\mu}.
 \label{posigma0}  
\end{eqnarray}
Again, if transforming one of the two terms of each integral into minus the other plus the surface term, we end
up with only a surface term, which is, according to the periodic properties of $\vartheta_{\alpha}$ functions, equal to
zero.
This completes the proof.

\section{Conclusions}
Starting from a Weyl spinor of only one handedness in a flat Riemann space $M^{1+5}$, we were able to find an example,
for which the compactified space of $S^1 \times S^1$ flat Riemann space torus with the vielbeins and the spin 
connections (which in $d=2$ are not related by the equations of motion but rather by the requirement that the 
covariant derivative of the viel(zwei) bein are equal to zero), allows massless spinors to be chirally coupled
to the Kaluza-Klein $U(1)$ field. The handedness in the non physical dimensions is proportional to the Kaluza-Klein charge.
The change of the handedness, and accordingly of the spinor wave function, causes the change of the sign of the 
Kaluza-Klein charge.  This assures that massless spinors chirally couple to the Corresponding gauge field.

The spin connection field can not be gauged away, due to the particular choice of the boundary conditions.
We found as an example for such a spin connection field  a flat torus, the field, for which the
two solutions of the corresponding Weyl equation, are expressed in terms 
of the well known functions $\vartheta_{\alpha}$ (see Eq.(\ref{psisoltheta})).

The integral of the spin connection over the meeting place of the patch of  the torus $\int \partial_{[\sigma} 
\omega^{5}{}_{6 \tau]} dx^{\sigma} dx^{\tau}$ - which is the integral of the curvature $\int ER dx^{(5)} dx^{(6)}$ -   
is  proportional to an integer
$n$ and accordingly quantized (Eq.(\ref{intR})). The chosen spin connection has a pole in the origin, but the 
curvature is nonzero and accordingly also the integral of it oer the torus is nonzero. 
 
The Kaluza-Klein charges $p_{ \sigma}, \sigma= (5),(6)$, which couple spinors  to the 
Kalu\-za-Klein gauge fields $A^{\sigma}{}_{m}$ are on the Hilbert space
of massless spinors ($\psi_{sol}= \psi_{\pm}$) equal to $S^{56} \omega_{56\sigma}$, that is proportional to 
the generator of the Lorentz transformations in the internal space of spin of the two compactified dimensions.  
By marking  spinors with the Kaluza-Klein charges the choice of states of the desired handedness  
in also the ''physical'' ($M^{1+3}$) part of space is enabled.
Consequently our ''physical'' spinors are massless and are also chirally coupled to the corresponding Kaluza-Klein
gauge fields. 


\section*{Acknowledgments}
It is our pleasure to thank the participants of the series of Bled workshops, requiring one of the authors 
(SNMB\cite{norma,pikanorma}) to prove
that marking spinor representations in a way to select one of the two handedness is indeed possible in 
the Kaluza-Klein like theories, in particular C. Froggatt. SNMB would 
like to thank M. Blagojevi\'c
for stimulating discussions on teleparellel theories.

\title{Geometry Decides Gravity, Demanding General Relativity --- %
it is Thus the Quantum Theory of Gravity} 
\author{R. Mirman\thanks{sssbb@cunyvm.cuny.edu}}
\institute{%
14U\\
155 E 34 Street\\
New York, NY  10016}
\titlerunning{Geometry Decides Gravity}
\authorrunning{R. Mirman}
\maketitle

\begin{abstract}
What decides the laws of physics? Geometry, at least largely. Its
transformation groups (which may not be symmetry groups) greatly
limit physical laws. For massless objects, electromagnetism and
gravitation that can couple to massive matter, these are completely 
determinative~\cite{ml}. Here we only outline reasons and
derivations. Details, and discussions of related subjects, are
elsewhere.
\end{abstract}

\section{Transformation groups}\label{s:trg}

A fundamental transformation group of our geometry is the Poincar\'e 
group (\cite{ia}, sec.~II.3.h, p.~45), the rotations in
3+1 space (the Lorentz group) and the translations (given by the
momentum operators). Whether space is invariant under it is
irrelevant. It is a transformation group, a subgroup of the complete 
one: the conformal group~\cite{cnf} of a 3+1 (locally) flat
real space. This is true even if directions (simulated by the
vertical) were different. Neither points nor directions need be
identical. (With the earth the vertical and its center appear
different --- because there is a material body.)

An example of a transformation group is the rotation group (for
any dimension), which we consider in a space with a direction
different, simulated by a magnetic field. Functions of angles can
be written as sums of basis vectors of the rotation group ---
spherical harmonics for our space --- these forming sets called
representations of the group (for 3-space labeled by the total
angular momentum). A rotation changes each function --- basis
vector --- in a sum replacing it by a sum of basis vectors. Each
basis vector is replaced only by a sum of vectors of the same
representation. States of a representation are mixed with states
of the same representation --- of the transformation group ---
but not with states of another. This cannot be done using states
of unitary groups. Rotations are fundamental properties of real
spaces, ones whose coordinates are real numbers, so more
intrinsic than symmetries. It is however provocative that they
are symmetries also.

The Poincar\'e group is an inhomogeneous group~(\cite{ia},
sec.~II.3.h, p.~46, one with a semisimple part --- the simple
Lorentz group~(\cite{ia}, sec.~II.3.e, p.~44) --- and an Abelian
invariant subgroup, the translations~(\cite{ia}, sec.~II.3.f,
p.~44), this transforming under the regular representation of the
semisimple part~(\cite{ia}, chap.~VI, p.~170). Inhomogeneous
groups are far richer than semisimple ones (like the rotation
group) with which we are more familiar. Prejudices from the latter 
may be completely wrong for richer groups.

\section{Labeling representations and states}\label{s:lb}

States of a group representation are labeled by eigenvalues of a
set of operators invariant under all group operations --- these
giving representations --- plus eigenvalues labeling states. For
the rotation group these are the total angular momentum, and its
component along some axis. These operators are completely
determined (up to isomorphism) but for inhomogeneous groups there
are choices (thus richness). Which operators shall we take
diagonal: semisimple ones, Abelian ones or combinations? These
give representations of different forms. Here we consider only
(as is usual, but not usually explicit) representations with all
momentum operators diagonal (thus no others, which are all semisimple, 
can be).

Rotation representations have one label, those of SU(3) two, and
so on. The Poincar\'e group requires two labels. For an object at
rest these --- its mass and total spin --- are needed to specify
the object. There can be no more (internal labels are not
relevant to these transformations of geometry). For free objects
there is nothing more to say.

Representations with all momentum operators diagonal break into
four sets, those with real mass, $m2 > 0$ (to which we belong);
imaginary mass, $m2 < 0$; zero mass $m = 0$ so $m2 = 0$; and
momentum 0 representations, to which coordinates and momenta
belong. Momentum has no momentum.

Here we consider just massless representations; for these we can
say the most. There is then one more label, the helicity.
Representations with helicity 1 give electromagnetism, with
helicity 2 gravitation. (Neutrinos cannot be massless~(\cite{ml},
sec.~4.4.4, p.~70]).)

There is one further condition --- obvious although its mathematical 
importance may not be. These objects must couple to massive 
matter else we could not know of them --- they would not
exist. This is very difficult, so very determinative. Massive and
massless objects are really quite different.

\section{Little groups}\label{s:lg}

Representations are found using a little group~(\cite{ml},
sec.~1.1.3, p.~4; sec.~2.2, p.~12), a subgroup whose representations 
are known. We need only the action of the remaining
operators on its states. Then we know all states of the full
group, and the action of all operators on them.

For a massive object, which we can take at rest, the little group
is the (simple) rotation group. Its representations, including
explicit expressions for its states, are known. On these we calculate 
the action of the boosts giving the (pseudo-orthogonal)
Lorentz group. On its states we find the effect of the momentum
generators, and then have all representations (of this type) of
the (inhomogeneous) Poincar\'e group.

Massless objects, like the photon, cannot be at rest. Their
little groups are the subgroups leaving invariant a momentum component. 
Such little groups are not semisimple~(\cite{ia},
sec.~IV.9.a, p.~144), but solvable~(\cite{ia}, sec.~XIII.3.a,
p.~376). These types of groups are quite different. Hence it is
almost impossible to construct interactions between (massive)
semisimple objects and (massless) solvable ones. Thus electromagnetism 
and gravitation are fully determined. Restrictions are so
great that there is no choice. We might expect that coupling two
such different objects is impossible. Fortunately it is in two
cases, helicity 1 and 2 (perhaps 0).

\section{All terms must transform the same}\label{s:at}

In an equation all terms (in the sum) must transform as (perhaps
different realizations~(\cite{ia}, sec.~V.3.c, p.~157) of the
same state of the same representation else it would be different
in different systems --- inconsistent. Dirac's equation is a sum
of terms one the mass (a scalar) times the solution, the
statefunction. Hence all terms must transform as the solution (a
bispinor). These include interactions between the massive object
and electromagnetic and gravitational fields. For coupling such
interactions have to transform as the solution. They are products
of semisimple terms and solvable ones (actually functions found
from these by the remaining Poincar\'e transformations, these
different for different types of objects).

This is actually not difficult for electromagnetism. It requires
minimal coupling, the reason that the photon couples this
way~(\cite{ml}, sec.~5.3, p.~81). For helicity-2 gravitation it
is much harder, almost impossible, to couple.

Helicity-2 has five states. Products of it with (massive) Lorentz
group states must transform properly under all groups. We need
scalars formed from products of interaction terms with a Lorentz
basis vector (which solutions of Dirac's equation, massive
statefunctions, are). However there is no irreducible Lorentz
representation with five states~(\cite{ml}, sec.~4.4, p.~67).
There can be no such scalar, just as there cannot be one constructed 
from angular momentum 1 and 2 representations.

The number of components must be reduced requiring relations
between them, nonlinear ones. Fortunately the helicity-2
representation has such: the Bian\-chi identities. Hence massive
objects and gravitation can interact.

A gravitational field is produced by energy, and has energy. Thus
a gravitational field produces a gravitational field --- it is
nonlinear. This argument is correct but it hides the underlying
mathematics. Gravitation must be nonlinear since only that non-
linear representation can couple to matter. Fortunately both
arguments give the same condition.

Also the gravitational field is attractive~(\cite{ml},
sec.~4.2.5, p.~60) while the electromagnetic charge can have
either sign.

\section{Objects for massless representations}\label{s:mo}

Before outlining the derivation of Einstein's equation we need
certain aspects of massless representations: 
connections~(\cite{ml}, sec.~3.2 p.~33), why they are the basis
states of massless representations (and only these), what gauge
transformations are~(\cite{ml}, sec.~3.4 p.~43), why massless
representations (only) have them, and what the fundamental fields
are~(\cite{ml}, sec.~3.3 p.~37).

That gauge transformations are Poincar\'e transformations for
massless objects, and only these, is clear. Take an electron and
photon with momentum and spin parallel (so both spins are parallel 
to the momentum). Lorentz transform to a system in which both
momenta remain the same but the electron's spin is changed. Its
spin and momentum are no longer along the same line, but those of
the photon must be --- it is transverse. Thus there are transformations 
that act on the electron but not on the photon (?).
This cannot be. Poincar\'e transformations do not depend on the
object acted on. What are these extra transformations? Of course
gauge transformations. Their properties are given by the Poincar\'e group.

Gauge transformations are neither rotations nor boosts but
products. Go to the electron's rest frame, rotate its spin, then
reversing the first transformation go the frame with the original
momenta. The momenta remain the same but the spin direction of
the electron is different. Gauge transformations for the photon
are given by this product acting on it.

It is important to understand that the electromagnetic field is
not transverse because of gauge invariance. The form of group
operators is determined by its structure. Generators of the rotation 
group are fixed by their commutation relations, not by the
hydrogen atom for example. Thus the commutation relations of the
Poincar\'e group require that the basis states of its massless
representations be connections and undergo gauge transformations.
And clearly these are possible only for massless representations.

What is the difference between a connection and a tensor? Transformations 
of a tensor are homogeneous (with an $a$ for each
index), schematically,
\begin{equation}
T_{i\ldots }' = a_{ij}\ldots T_{j\ldots}
\end{equation}
Transformations of a connection are inhomogeneous,
\begin{equation}
\Gamma_{i\ldots }' = a_{ij}\ldots \Gamma_{j\ldots} 
 + \Lambda_{i\ldots };
\end{equation}
$\Lambda$ does not depend on $\Gamma$ but is a function of the
transformation (the $a$'s). A tensor transformation changes two
components simultaneously, for a connection only one need be. An
example of a tensor transformation is
\begin{equation}x' =  x\cos\theta  + y\sin\theta, 
 ~~  y' = - x\sin\theta  + y\cos\theta, 
\end{equation}
while for a connection (here the electromagnetic vector potential),
\begin{equation}A_{x}' = A_{x} + \Lambda_{x},  \end{equation}
 where $\Lambda$ is arbitrary, and similarly for the 
gravitational $\Gamma$.

The finite-dimensional representations of the Euclidean group
SE(2), the little group for massless representations with 
rotation operator $M$ diagonal, are not unitary as the algebra
matrices are not hermitian. They consist of blocks of the form,
say, with the $N$'s the other two generators,
\begin{equation}M = \begin{pmatrix}m  &0  \cr 0  &m-1\end{pmatrix}, 
\end{equation}
  and
\begin{equation}N_1 = \begin{pmatrix}0  &u  \cr 0  &0\end{pmatrix}, 
 ~~ N_2 =
\begin{pmatrix}0  &iu  \cr 0  &0\end{pmatrix}, 
\mbox{~~or~~} N_1 = \begin{pmatrix}0  &0
\cr v  &0\end{pmatrix}, 
 ~~ N_2 = \begin{pmatrix}0  &0  \cr -iv  &0\end{pmatrix};
\end{equation}
  $u$, $v$ arbitrary ($N$'s give arbitrary gauge transformations
--- these also depending on group parameters, add arbitrary functions 
to representation basis vectors). There are two representation 
forms, upper and lower.

For semisimple algebras for each off-diagonal entry there is a
corresponding one across the diagonal, but with solvable algebras
this is not true for all entries. Thus, as easily seen from this
simple example, there are transformations of a solvable group
that add terms to basis states, as we are quite familiar for
electromagnetism and gravitation. That is why their states are
connections, and only their states. For other classes of
(momentum-diagonal) representations the little group is semisimple.

We thus see what connections and gauge transformations are, how
they are related, are required by the little group being solvable, 
and why they are properties of massless representations,
and possible only for them.

\section{Fundamental fields}\label{s:ff}

Equations to determine the electromagnetic and gravitational
fields are needed. But which fields? They are connections, the
gravitational connection (not the metric which transforms under a
momen\-tum-zero representation) and the electromagnetic connection,
the potential $A$. These are massless objects, and connections
are the massless states.

Electromagnetic fields $E$ and $B$ are not physical objects, not
gauge invariant~(\cite{ml}, sec.~3.3.1, p.~37) and do not transform 
under a proper representation. They are products of states
of massless and of momentum-zero representations. For rotation
groups a product of representations can be written as a sum.
However this product is of representations of different types
(there is only one type for the rotation group). Such products
have not been characterized, and perhaps cannot be. There may be
nothing further to be said about them. This has to be looked at.

That the electromagnetic field is not a physical object and can-
not be measured is trivial. How do we measure a field? We observe
the behavior of a charged object in it. In elementary physics we
use pith balls. But there are no such things. These are merely
collections of electrons, protons and neutrons (and so on), as
are we. Fields act on these.

What acts on a electron? From Dirac's equation clearly the potential. 
The behavior of the electron then gives that, and that is
what is measured. Other fields are merely functions of it,
unmeasurable thus without basic significance (for this reason
also).

We can now outline derivations of the equations for gravitation
and electromagnetism. These are standard. What is different is
the context. The theories are derived from the Poincar\'e group
thus are unique. Electromagnetism and gravitation are what they
are because that is what geometry wants them to be. They are not
guesses that happened to be correct. And gravitation is not
determined by the equivalence principle --- that is a consequence. 
Too many believe that clues leading to the discovery of
a theory are the reason for it. But how we discover does not
underlie physics. There are reasons for the way physics is. We
generally do not know them (geometry is highly suspect) but for
massless representations we do.

\section{Electromagnetism and what it must be}\label{s:em}

Details for the electromagnetic case can easily be worked
out~(\cite{ml}, sec.~7.2, p.~124) so we just summarize. One Maxwell 
equation is the Bianchi identity, the other is the trace of
the electromagnetic tensor; this is equal to the current. Since
the electromagnetic tensor is not a physical object we need an
expression for the potential. This is given by its covariant
derivative, the momentum operator acting on the electromagnetic
statefunction,
\begin{equation}A_{\mu;\nu} = A_{\mu,\nu}  - {2ie \over
m2}(\psi^{+}\gamma_{\mu }\psi_{,\nu} +
\psi^{+}_{,\nu}\gamma_{\mu }\psi). \end{equation}
The second covariant derivative is zero; the momentum belongs to
the momentum-zero class, so its momentum --- its covariant
derivative --- is zero. The covariant derivative of a spinor
gives minimal coupling.

The equations are for the potential giving it in terms of the
statefunctions of the charged objects; their equations include
it. These equations govern. Maxwell's equations are classical,
and for a nonphysical object: the electromagnetic field. Thus
they are only of calculational use, they are not fundamental.
Since neither the electric nor the magnetic field really exists
that Maxwell's equations seem to distinguish between them is
meaningless. It is purely a matter of notation. Hence for example
a magnetic monopole cannot exist~(\cite{ml}, sec.~7.3, p.~131).
There is no way of putting one in.

\section{Gravitation}\label{s:gv}

These objects are determined by the Poincar\'e group so we need
expressions for its generators, here for the Abelian part, the
momentum operators, the covariant derivatives~(\cite{ml},
sec.~1.2.2, p.~8). Thus we have to find the covariant derivative
of, the momentum acting on, the connection --- the gravitational
statefunction. There are two ways of finding it~(\cite{ml},
sec.~8.2.1, p.~146), using the covariant derivative of the
covariant derivative, and the ordinary derivative of a vector
whose a covariant derivative is given by the usual rules for that
of a tensor. These must be the same. So we get the covariant
derivative of the gravitational statefunction, the connection.
Thus
\begin{equation}ip_{\kappa  }\Gamma^{\lambda}_{\mu \nu } =
\Gamma^{\lambda}_{\mu \nu;\kappa } = \Gamma^{\lambda}_{\mu
\nu,\kappa } + \Gamma^{\lambda}_{\phi \kappa }\Gamma^{\phi}_{\mu
\nu } - \Gamma^{\phi}_{\mu \kappa }\Gamma^{\lambda}_{\phi \nu } -
\Gamma^{\phi}_{\nu \kappa }\Gamma^{\lambda}_{\mu \phi }.
\end{equation}
From this we get the second covariant derivative, then the commutator 
of momentum operators (0 since momenta form an Abelian
subgroup), and the expressions for the Casimir invariants and
curvature tensor. Einstein's equation for free gravitation then
follows using any standard derivation.

With matter present a term has to be added to the covariant
derivative. This gives the energy-momentum tensor. It is here
that there may be some freedom. What is the energy-momentum
tensor? There are expressions for scalars, spin-${1 \over 2}$
objects and the electromagnetic field~(\cite{ml}, sec.~9.2,
p.~153). (There is no such thing as dust.) While these are
reasonable it is not clear there are no other possibilities. This
remains to be looked at. These give the equations for the fields
as determined by the sources.

This raises a problem for scalar objects~(\cite{ml}, sec.~4.2.7,
p.~61).Do they interact with gravity? There is no reason to
believe so. It is an article of faith that gravitation is
universal, interacting with all objects, and in the same way.
Actually it has only been tested in two cases, collections of
spin-${1 \over 2}$ objects and the electromagnetic field. An open
mind can be useful.

\section{Trajectories are geodesics}\label{s:tg}

What determines the behavior of objects in fields? Interaction
terms~(\cite{ml}, sec.~5.3, p.~81). For a scalar, trajectories
are geodesics~(\cite{ml}, sec.~5.2, p.~74). Coordinates,
velocities, momenta and the metric all belong to the momentum-zero 
class of representations. Their momenta --- covariant
derivatives --- are thus 0. Setting the covariant derivative of
the velocity to 0 gives the geodesic. This can also be found
quantum mechanically.

\section{A cosmological constant must be 0}\label{s:cc}

It is traditional to include in Einstein equation a cosmological
constant. Clearly that must be 0~(\cite{ml}, sec.~8.1.4, p.~139).
It sets a constant equal to a function of space and time, and a
real number, the cosmological constant, equal to a complex one
$G$, obviously wrong. Why is a gravitational field complex?
Regarding it as due to curvature of space can be misleading.
Mathematically it is possible to write metric $g$ and $\Gamma$ as
spacetime functions giving the geometry of the entire 3+1-
dimensional space. But fields depends on matter and its behavior.
These are arbitrary and can be varied at will (unless the
statefunction of the entire universe for all time is known). Thus
the field is a function of time. From the field at one time we
get it at all times using (as usual) the equation for the
statefunction (schematically)
\begin{equation}{\it i} {d\Gamma \over dt} = H\Gamma;
\end{equation}
$H$ is the gravitational Hamiltonian which includes arbitrary
matter. Thus $\Gamma$ is a gravitational wave. Solving we get
that $\Gamma$, the gravitational field, is complex. A physical
field, a physical wave, must be complex. More fundamentally $G$
is a function of massless basis vectors, while the cosmological
constant belongs to the momentum-zero representation. Setting
them equal is like equating a vector and a scalar. And with a
cosmological constant gravitational waves would have the fascinating 
property that the metric, thus detectors, react to them,
not only an infinitely long time before they arrive, but even an
infinitely long time before they are emitted. The argument is the
same, but here stronger, as that showing classical physics is
inconsistent and quantum mechanics necessary~(\cite{gf}, chap.~1,
p.~1).

Taking the gravitational field as a purely geometric object is
not fully useful, and likely not fully possible because it is
determined by physics and physical objects have arbitrariness.
This is not surprising since the gravitational field is massless
and that has no meaning for geometrical objects.

Geometry then, through its transformation group --- 
the Poincar\'e group --- determines what gravitation is: knowing that it
is a massless helicity-2 object. What is left? First the functions 
of the matter statefunctions that gives it --- energy-momentum 
tensors might still have some freedom, although perhaps
not. This should be looked at. More mysterious is the value of
the coupling constant, the gravitational constant (perhaps more
than one~(\cite{ml}, sec.~9.3.4, p.~162)). Yet the most fundamental 
question is why gravitation exists. If it does, and
fortunately it does, it is determined. But the Poincar\'e group
does not require its existence. What does?

\section{General relativity is quantum gravity}\label{s:qg}

Can there then be a quantum theory of gravity? There is: general
relativity. It is the first complete, consistent quantum theory.
What is a quantum theory and why do people dislike general
relativity?

A quantum theory~(\cite{gf}, chap.~1, p.~1;~\cite{qm}, chap.~II,
p.~54) is a consistent theory that includes (at least) proper
definitions of the Poincar\'e generators~(\cite{ia},
sec.~XIII.4.b, p.~382), those of the Lorentz transformations and
of translations (thus momentum operators). This is necessary else
it would be impossible to transform to different systems, but
transformations are possible and necessary. Without these physics
cannot be. Beyond that what else is there to require? What else
is possible? General relativity satisfies these requirements,
there are strong reasons to believe it is consistent, and is
unique. Thus it is the quantum theory of gravity. There is nothing 
that can be done to ``quantize'' it.

There is confusion abut quantum mechanics~\cite{qm}. Both this
name and wavefunction are unfortunate. Discreteness is neither
universal nor fundamental. For angular momentum it is a property
of the rotation group (and forms of representations of semisimple
groups in general). For atoms it comes from requirements such as
that there be no infinities. But there is no quantitization for a
free particle, one tunneling through a barrier, or for a huge
number of other cases.

Nor does quantum mechanics lead to objects fluctuating. The
gravitational field does not fluctuate. If an experiment is
redone many times the results are different (usually only a
little) for each repetition. For a box of neutrons the number
decaying in each second varies. But the neutrons do not vary,
they do not oscillate.

There are also fears of infinities. But these occur for intermediate 
steps in a calculational procedure: perturbation theory.
They are regarded as due to point particles. But there are no
such objects as point particles, even classically. There is nothing 
in any fundamental equation of physics that even hints of
them. Overemphasis on infinities is based of a belief that the
universe is carefully designed to make physicists' favorite
approximation method work. That is not likely. What they show is
that perturbation theory has problems, not electromagnetic theory
or gravitation.

There is no reason to think either is inconsistent, but rather
there are strong reasons to believe both are consistent.

If general relativity is a quantum theory does it have
uncertainty relations? As with any field theory it does. But they
are different from those we are more familiar with --- for which
the product of uncertainties is greater than some number. But
gravitation is, necessarily, nonlinear. Thus for it the product
of uncertainties is greater than a function of the statefunction
(the connection). They are far more complicated and depend on the
physical situation. It would be interesting find these for some
cases.

There is much understood but still much to be learned about gravitation. 
Perhaps these comments can stimulate further thought
about such questions.
We see again that geometry imposes its will on
physics~\cite{ia,gf,ml,pt,qm,cnf}. For massless representations
this is particularly clear, for others less so --- but perhaps
only for now.

Some of these topics will be discussed elsewhere~\cite{imp}, for
some in greater depth and in a more elementary manner.

\section*{Acknowledgements}

This discussion could not have existed without Norma Mankoc Bor-
stnik.

\title{Physics Would Be Impossible in Any Dimension But 3+1 ---
There Could Be Only Empty Universes}
\author{R. Mirman\thanks{sssbb@cunyvm.cuny.edu}}
\institute{%
14U\\
155 E 34 Street\\
New York, NY  10016}
\titlerunning{Physics Would Be Impossible in Any Dimension but 3+1}
\authorrunning{R. Mirman}
\maketitle

\begin{abstract}
Our universe has dimension 3+1: three of space, one of time. And
it must --- a universe could not exist if the dimension were different. 
Physics has to be inconsistent --- not possible --- in
any other dimension~(\cite{gf2}, chap.~7, p.~122). Why?
\end{abstract}

\section{There must be observers}\label{s:obs}

A most fundamental fact of nature is that there are different
observers --- physical objects. They must all be able to observe,
their observations must make sense and must be related. And these
are related by geometry. Mathematically all physical laws have to
be expressible --- in a consistent way --- in different coordinate 
systems.

Thus we can use coordinates $x,y$ or
\begin{equation}x = x'\cos\theta  + y'\sin\theta, \end{equation}
\begin{equation}y=y'\cos\theta-x'\sin\theta.\end{equation}
This --- rotation --- is merely a change of symbols with no
physics involved. If it gave an inconsistent set of laws then
there could be no laws, physics would not be possible, thus nor
would a universe.

These transformations need not be symmetries (although it is
quite provocative that they are symmetries also). They form
transformation groups (sets of transformations)~(\cite{gf2},
sec.~A.2, p.~178;~\cite{qm2}, sec.~I.7.b, p.~37), but need not be
symmetry groups (sets of transformations leaving space or other
systems invariant).

Were a direction of space different (simulated by the vertical)
it would not matter. Arguments would not be affected. Since that
is all these are, transformations not symmetries, requirements on
space and nature are very weak thus quite strong. They are weak
because very little (actually it seems nothing) is put in, is
needed. So they are quite strong as it is (almost?) impossible to
avoid them, to have anything else thus avoid what a universe must
be like. And that is not only quite strong but quite disturbing.

\section{So a homomorphism is required}\label{s:hom}

But why are transformations --- just rotations --- so restrictive, 
so difficult? Why is it impossible to have them for physical 
objects in any space but one, that of dimension 3+1, and that
just barely? Coordinates are real numbers thus transformed by
orthogonal (rotation) groups. Physical objects are given by
statefunctions (a better term than wavefunction since nothing
waves). And these are complex numbers, thus transformed by
unitary groups.

Unitary and orthogonal transformations have to be related
(essentially the same), else there could be no physical objects.

Consider an electron with spin up along $z$. It is not up along
$z'$. Its statefunction expressed in terms of $z'$ is different
from that expressed in terms of $z$. One gives spin up, the other
gives it at an angle. They are different, but related. Knowing
that along $z$ and the angle of rotation we know it along $z'$.
The statefunction is transformed --- and by a unitary transformation. 
For each rotation there are corresponding unitary transformations.

Any rotation can be written in an infinite number of ways as a
product of rotations. This is also true of unitary transformations. 
Rotations and unitary transformations both form groups.

A fundamental requirement is that the products coincide (up to a
possible sign) --- the groups are homomorphic. Consider a rotation 
and then the inverse, but one written as a product. This can
be done in an infinite number of ways going from a state back to
the original. The final state is the same as that from which we
started. These transformations are purely mathematical, just
changes of variables, as above.

Suppose that for each product the direction of spin of the electron 
were different --- even though the orientation of the coordinate 
system (the observer) is the same. That is observers,
originally and finally identical, carrying out these (mathematical) 
transformations would see different spins even though these
observers are the same. Obviously physics would be inconsistent
--- not possible, nor would a universe.

\section{Dirac's equation shows this}\label{s:dir}

This can be seen, perhaps more rigorously, using Dirac's equation. 
Writing equations for two objects and an electromagnetic
potential
\begin{equation}i\gamma_{\mu}{\partial \psi (x)_j\over \partial
x_{\mu}} - m_j\psi (x)_j + I(\psi (x)_j,A) = 0; ~~~j = 1,2,
\end{equation}
with statefunction $\psi (x)$ giving the probability of finding
the object at $x$ with some spin direction, $A$ (written
schematically), the electromagnetic potential, for which there is
another equation, and $I$ the interaction term (whose form is
irrelevant --- the argument is very general). We can take the $z$
axis along the spin of one electron or along the spin of the
other (or pick any axes) --- there is no way we can restrict the
direction of arbitrary axes. We must be able to transform (by
changing variables as in the equations above) so such equations
for the statefunctions have to be properly transformable under
(arbitrary) rotations. But coordinates are real and statefunctions 
complex. Therefore transformations on terms in 
the equations are different.

Were conditions not fulfilled then a set of (mathematical) rotations 
returning to the initial coordinates --- so nothing is
changed --- would give different equations. These would depend on
how rotations were (mathematically) carried out. Electrons would
get very confused, they would not know what equation to obey.
Thus there is a set of equations, one for each object (including
the electromagnetic potential). However if conditions were not
met then how we (mathematically) rotate determines these equations, 
clearly ridiculous --- they would be inconsistent. Each
set of rotations, returning to the original orientation, would
give different equations. Statefunctions would have to ---
simultaneously --- satisfy all equations obtained by the infinite
set of possible rotations. Obviously that could not be.

If these equations were inconsistent --- and if the transformations 
were not properly related that proves that they are --- the
only solutions are all zero. Every statefunction would be zero,
thus would the probability of finding any object be --- the
universe is then empty.

What determines the dimension then is that equations of physics
must be invariant under space transformations, here rotations.
But these involve objects that are real (coordinates), and ones
that are complex (statefunctions). Hence these must transform the
same way, which they do not do --- unitary and orthogonal groups
are not homomorphic. Fortunately there is one dimension, 3+1,
with unitary and orthogonal groups that are. And that thus is the
only possible dimension allowing a universe with matter.

\section{Mathematical analysis}\label{s:mtl}

We consider this in more depth.

For mere existence there must be one dimension whose orthogonal
group is homomorphic to a unitary group. Given a space it cannot
contain matter unless there is a unitary group homomorphic to the
(pseudo-)rotation group of that space. The Lie algebras of these
groups have to be isomorphic. Is this possible? Seemingly no. The
number of generators (parameters) and the number of commuting
generators of unitary and orthogonal groups are different. Can
there be a space in which they are the same?
The analysis is essentially trivial, just counting. For
orthogonal groups it requires counting the number of planes, each
giving a generator, and the number of planes that do not share an
axis, each giving commuting generators. Counting need not be done
here since the results are known and the counting is trivial.
(Existence is based on trivialities.)

The number of sets of rotations in a space of dimension $d$ is
\begin{equation}N_{d} = {{(d - 1)(d - 1 + 1)} \over 2} = {{d(d -
1)} \over 2}.\end{equation}

Counting nonintersecting planes gives the number of sets of com-
muting rotations,
\begin{equation}C_{c} = {d \over 2}, \mbox{~~for~} d
\mbox{~even}, \mbox{~~and~~} C_{c} =  {(d - 1) \over 2},
\mbox{~~for~} d \mbox{~odd}. \end{equation}

For a unitary group in a space of dimension $p$ there are $p^{2}
- 1$ transformations, of which $p - 1$ commute.

These quite elementary formulas for the number of rotations and
the number of commuting ones determine whether any universe is
possible.

A space of dimension $d$ cannot contain matter unless there is a
unitary group, in some complex space of dimension $p$, whose Lie
algebra is isomorphic to that of the orthogonal group of the
space. Setting the numbers of generators, and the numbers of 
commuting ones, equal we find that this is possible only for $d$ = 3
or 6. Which does the universe choose?

There is more.

Rotations are defined as those transformations (on real numbers)
leaving angles and lengths unchanged. Angles of rotations are
real numbers but there are also complex parameters for which
these are preserved~(\cite{gf2}, sec.~7.2, p.~124). Both imaginary
and real parts are limited as the formulas show but both can be
nonzero. In general such a transformation in a plane relates
coordinates by
\begin{equation}x' = x\alpha  + y\beta, \end{equation}
with $\alpha, \beta$ complex.

Mathematically if the sets of transformations on real and complex
numbers are the same for real parameters they must be the same
for complex ones. We write
\begin{equation}\alpha = a + {\it i}b, \end{equation}
using two real parameters $a,b$. The equations for the real and
imaginary parts are essentially the same. We then get two 
equivalent sets of transformations, for the real parts of the
parameters and for the imaginary parts. However the number of
sets of commuting transformations does not depend on whether
parameters are real or complex. Transformations in planes sharing
an axis do not commute but do commute if they do not share an
axis. Whether parameters are complex or only real does not change
this.

If transformations on complex numbers (statefunctions) are the
same for real parts of parameters of the generalization of 
rotations they must be the same for imaginary parts. This 
necessitates another condition --- which is fortunate. And fortunately
space gives another transformation: inversion.

A statefunction for an object with spin ${1 \over 2}$ has two
parts, giving spin up and spin down. No matter what the direction
of its spin the statefunction can always be written as a sum of
these two taken along any axis (which does not mean, as sometimes
believed, that it can only point up or down --- it can be in any
direction). Under an inversion the statefunction goes to 
a different one thus it has four components (the Dirac bispinor). The
two two-dimensional spinors behave somewhat differently under all
these transformations with complex parameters (the CO(3,1) group)
as there are differences in minus signs for the
boosts~(\cite{gms}, p.~216),~(\cite{nam}, p.~312),~(\cite{sc},
p.~79).

Thus the bispinor really has 4 components, which is exactly what
is needed. The universe could not exist were this not true.

What condition do these, complex parameters and the inversion,
give? Consider that the statefunction of an object has $p$ parts
(here 4). It transforms under SU($p$). Is there any space whose
orthogonal algebra is isomorphic to that of SU($p$), for some
$p$? We know the number of generators for these algebras. We take
the statefunction in such a space to have $j$ (here 2) blocks,
each then of size ${p \over j}$.

The number of parameters, now complex, for the transformations
are thus twice those of the set of rotations. The rotations are
the same so the number of commuting ones does not depend on
whether the parameters are real or complex. On this complex
statefunction there are transformations going with the rotations
plus others mixing the $j$ blocks. The number of these is the sum
of those of the two sets. The total number of parameters must be
equal to the total number of transformations on this 
$p$-dimensional complex space. So
\begin{equation}(d^{2} - d) + j^{2} = p^{2},  \end{equation}
and for the commuting ones
\begin{equation}{d \over 2} + j = p \mbox{~~or~ } {(d - 1) \over
2} + j = p, \end{equation}
for $d$ even and $d$ odd.

It can easily be checked that the only solution is
\begin{equation}
d=4,~~p=4,~~j=2.
\end{equation}
Neither 3 nor 6 satisfies. This gives that the dimension is 4.
Fortunately $j$ equals 2 (the Dirac bispinor) as it must since it
is the inversion --- which can only interchange one block with
another --- that requires blocks. There can thus be only two.

For larger spin, statefunctions transform as reduced products of
this fundamental representation --- for the same $p$.

The dimension then must be 3 or 6, but can only be 4. That seems
to imply that any universe is impossible. Fortunately the choices
are 3 and 4, thus can be both. Uniquely 3 + 3 = 6; the three
generators of SO(3) plus the three of SO(3) gives the same number
of generators as the six of SO(4). That group is not simple but
only semisimple. It is easy to prove that this is the only
orthogonal group that is not simple --- only for dimension 4 can
this argument work. And that is just the dimension the argument
gives.

Each of the SO(3) groups has the correct number of transformations 
to satisfy the condition from real angles, the two together
satisfy the condition for complex ones. One group acts on each of
the two-dimensional spinors, the other group acts on the pair
treating it as a two-dimensional spinor (each of which has two
components). The six generators of SO(4) break into two sets,
each of the three generators of SO(3). Only for dimension 4 is
this splitting possible.

Why does this work? The (complex) statefunction must split into
two corresponding parts (since space also allows an inversion),
here each themselves of two parts giving four (components)
altogether. Rotations act (identically) on each of the two pieces
but boosts (changes of speed) do not, differing in minus signs.
However as there are two parts an additional set of 
transformations mixes them. Hence the group of transformations of real
space, CO(4), has twelve parameters. But there are an additional
three parameters because the inversion gives a pair of spinors
and this group mixes them. Coordinate systems obtained by these
transformations on space are possible --- it must be possible to
write physical laws using any. Equations have to be 
form-invariant under them. Thus space allows a 15-parameter set of
transformations, the same number as that of SU(4).

The electron statefunction illustrates this. Rotations mix up and
down states. But we have pairs of these. So another set of rotations 
mixes the pairs (each a pair). Two sets of rotations are
thus needed giving two sets of transformations on complex
statefunctions. Each of the two sets of (three) rotations has a
set of (homomorphic) transformations on complex variables ---
each acts in a three-dimensional space which is one of the two
possible spaces allowed by the first condition. Thus the set (of
the two sets together), with six parameters (three complex
parameters so six real numbers) plus the set of unitary 
transformations mixing the blocks (spinors) have a (homomorphic) set
of unitary transformations going with it.

In summary what transformations are there for this 4-dimensional
complex statefunction? Clearly those of SU(4). However if we
rotate (with complex parameters) we induce transformations on
each of the two-dimensional spinors. There are two 3-dimensional
real spaces whose transformations act independently on these
spinors (because SO(4), being uniquely semisimple, splits into
two parts). This gives 12 transformations. Also the pair of
spinors is a 2-dimensional complex number, giving three more
transformations. The total number is thus 15, equal to that of
SU(4).

Space must have dimension 4. Yet that is not unique. Is it 4 of
space, 3 of space and 1 of time, or 2 of each? The signature is
irrelevant to these arguments; only the number of parameters mat-
ters. Both SO(4) and SO(2,2) are only semisimple~(\cite{c},
p.~868;~\cite{z}, p.~274), which is necessary for this
argument~(\cite{gl}, p.~52;~\cite{he}, p.~340). Each consists of
two parts these acting on 3-dimensional spaces. But 3-dimensional
spaces do not satisfy as we see. However SO(3,1) fortunately is
simple. Physics, a universe, is only possible in a space of
dimension 3+1, three of space plus one of time.

These conditions are necessary --- only dimension 3+1 is possible
--- but that does not mean that the groups are homomorphic, only
that they cannot be in any other dimension. It is well-known that
they are~(\cite{c}, p.~65),~(\cite{st}, p.~6), so this need not
be discussed here.

The universe is possible, but just barely. The equations must
have integer solutions. There is no reason to expect that any
have integer solutions, that there is any integer that satisfies
any one --- and certainly not all. Change only one number in any
of the equations even by 1. Then its solution would not be an
integer. That mere change would make universes hopeless. And
there must be two blocks, which is just what the inversion gives.
But the inversion has nothing to do with the counting arguments
that give the requirement of two blocks. As we see there are
several --- independent --- arguments; none should be satisfied
and definitely not all, all together. Yet strangely there is one
dimension and one only --- it is unique --- that does satisfy all
--- and all at once.

The dimension emphasizes again, as seen in so many
ways~\cite{ia2,gf2,ml2,pt2,qm2,cnf2}, that geometry and physics are
deeply intertwined. Geometry limits, perhaps determines, physics.
Physics is possible only in a universe with the proper geometry.
They almost seem, perhaps are, one subject.

Derivation of the dimension was given previously in a somewhat
different manner but with additional analysis~(\cite{gf2}, chap.
7, p.~122), and will be elsewhere~\cite{imp2}. That is on a more
elementary level with greater detail including reasons that the
only possible dimension is just the right dimension --- certainly
for life. What we learn from that more extensive analysis of
nature (going well beyond the dimension) is that the laws of
physics really do love us.

\section*{Acknowledgements}

This could not have existed without the kindness of Norma Mankoc
Borstnik.

\title{Conservation of Energy Prohibits Proton Decay}
\author{R. Mirman\thanks{sssbb@cunyvm.cuny.edu}}
\institute{%
14U\\
155 E 34 Street\\
New York, NY  10016}
\titlerunning{Conservation of Energy Prohibits Proton Decay}
\authorrunning{R. Mirman}
\maketitle

\begin{abstract}
The proton can't decay, baryon number must be conserved. This is
proven rigorously but it can be seen intuitively (using care not
to be misled). The proton, the lightest strongly-interacting
fermion, can be taken (roughly) to have a strong-interaction 
contribution to its mass. If it decays into fermions with no strong
interaction then this contribution disappears, violating 
conservation of energy. This is made rigorous. Why pions are 
different, so can decay, is discussed.
\end{abstract}

\section{Intuitive arguments}\label{s1}
Experimentally the proton is stable; baryon number is conserved.
Must it be so? Actually it is easy to understand why it must.

The proton is the lightest fermion with strong interactions. 
Suppose that the weak interaction caused it to decay to leptons (and
perhaps photons) which are not affected by strong interactions.
Then the weak interaction would turn off the strong one --- a
state with would go to a state without. But one interaction 
cannot turn off another. Anyone who doubts this can write a
Hamiltonian with an interaction that turns off interactions.

This can also be seen with Feynman diagrams. Consider a diagram
(of which there are an infinite number) in which a proton emits a
virtual pion (kaon, or any other possibility) and then reabsorbs
it. Going with that diagram there is another for which the proton
emits the pion and then decays to leptons. The poor pion has to
be reabsorbed. But it does not interact with leptons thus there
is nothing that can absorb it.

Such diagrams violate conservation of energy (thus the entire set
must). While there are various ways of considering this, protons
cannot decay because of conservation of energy as we see formally
below.

\section{But pions can decay}\label{s:pion}
While the weak interaction of baryons cannot turn the strong one
off or on, mesons (like the pion) do decay to leptons. Why? The
photon is analogous; its number is not conserved although charge
is. However it is neutral, it couples to a neutral object, taken
as a particle-antiparticle pair (the current is of this form).
Electron-photon scattering can be regarded as creation of an
electron-positron pair by the photon; the positron annihilates
the electron and that of the pair replaces it. So while 
the electromagnetic interaction (of an electron) cannot be altered (it
cannot be turned off --- charge is conserved), photons can be
taken as not directly interacting. Their creation or annihilation
does not modify an interaction.

It is similar for mesons which can be regarded as having no
strong interactions but rather couple to particle-antiparticle
pairs. We can view meson-baryon scattering as annihilation of the
baryon by its antiparticle with it replaced by the (identical)
baryon of the pair. And decay can be seen as a baryon-antibaryon
pair scattering into leptons or photons. So we take a Feynman
diagram in which the meson goes into a virtual pair of baryons
which then interact (annihilate) turning into 
non-strongly-interacting objects, leptons or photons. This does not violate
conservation of energy. Because of what they are coupled to,
baryon-antibaryon pairs, mesons can decay into objects not
affected by strong interactions.

\section{Mathematical analysis}\label{s:analysis}
With this intuitive understanding of why baryons cannot decay
into only leptons but mesons can, we outline a more formal proof.
It has been given previously with full details~(\cite{cnf3},
sec.~IV-4, p.~212), but an outline clarifies the physical aspects
and shows implications that can be revealing in other ways. To
study the decay we consider the action of the Hamiltonian, $H$,
on a proton. Acting on a state at $t = 0$, time-translation
operator $\exp(iHt)$ gives the state at time $t$. Can this take a
baryon to a state whose fermions are only those not having strong
interactions?

What is a proton? Mathematically (the only way it can be treated)
it is a function obeying Dirac's equation with mass $m_{P}$ and
the (here irrelevant and suppressed electromagnetic, 
gravitational), weak and strong interactions (whose forms are
irrelevant). It is the presence of interactions that determines
what a proton is. (Correctly an object is an eigenstate of the
two Poincar\'e invariants~(\cite{gf3}, sec.~6.3, p.~114). For a
free particle, and one with an electromagnetic interaction,
Dirac's equation is equivalent. Whether this is true with other
interactions seems unknown so consequences of, perhaps important,
differences if any are not clear. And putting interactions in
invariants, which must be done whether Dirac's equation is used
or not, might limit them. Particles are also eigenstates or sums
of the momentum operators~(\cite{ml3}, sec.~5.4, p.~93), of which
the Hamiltonian is one. We ignore these, and refer to Dirac's
equation but discussions should be of invariants which might be
revealing. The statefunction (a better term than wavefunction) of
the proton is a solution of coupled nonlinear equations. We need
information about it but cannot solve so represent it in a way
that allows analysis using an expansion. The arguments though are
exact; we do not calculate so need not truncate.

The physical particle, labeled with a capital, that obeying
Dirac's equation with all interactions, is a sum of states
(schematically):
\begin{equation}\vert P) = c(x,t)\vert p) + \sum c(x,t)_{p\pi
}\vert p)\vert \pi ) + \ldots  + \sum c(x,t)_{K\Lambda }\vert
K)\vert \Lambda ) + \ldots  ,\end{equation}
summing over all states to which the proton is connected by
interactions including any number of pions and so on. 
The summations represent ones over internal labels and integrals over
momenta. All are suppressed; this argument is very general. State
$\vert p)$ is the function satisfying Dirac's equation with the
weak interaction, but not the strong. The effect of that is given
by this sum which is thus an eigenstate with mass $m_{P}$, of the
total Hamiltonian, including all interactions. Individual terms
in the sum differ in energy and momenta; it is the sum that has
the eigenvalues. Thus what this does is to write 
the strongly-interacting object as a sum of terms of objects 
(states satisfying equations) 
that have no strong interactions. Hence we are
putting in the strong interaction explicitly by writing the
physical object as this sum. We then can distinguish between
objects with and without the particular interaction we wish to
consider.

Coefficients are determined by the requirement that this be a
solution of that set of complete equations for all objects to
which the proton is coupled, directly or indirectly, and
normalization $(P\vert P) = 1$. Also the initial state, a
wavepacket, taken as a proton at rest, gives the coefficients at
$t$ = 0.

The effect of the Hamiltonian is seen from that of $H$ which is a
sum of the free particle Hamiltonians for the proton, pion (and
so on) and leptons, plus terms for weak and strong interactions.
The state of the system is a sum of terms, one the state of the
proton, another (if decay were possible) the product of pion and
lepton states and such, each summed over other labels and with
integrals over momenta or space. The free part of $H$ changes the
phase. The weak interaction part, were decay possible, decreases
the coefficient of the proton in the sum, while increasing that
of the (say) pion plus lepton, initially zero --- starting as a
proton, the state becomes a sum of the proton, its contribution
decreasing, plus the pion plus lepton state, with increasing 
contribution (and so on for other states). For the decaying pion the
behavior is similar: starting as a pure pion it becomes a sum of
that plus a state of leptons with the contribution of the first
decreasing, of the second increasing.

What goes wrong? The weak interaction acts on $\vert p)$ 
supposedly causing it to decay, so the final state is
\begin{eqnarray}
\vert fs) = \sum d_{\pi l}c\vert \pi )\vert l) + \sum e_{\pi \pi
l}\vert l)\vert \pi )\vert \pi ) + \ldots
\nonumber  \\ + \sum d_{lll}c\vert l)\vert l)\vert l)  + \sum
d_{\pi lll}\sum c_{p\pi }\vert l)\vert l)\vert l)\vert \pi ) +
\ldots ,
\end{eqnarray}
showing the transition to a pion plus a lepton, and to three 
leptons, and so on, with coefficients of non-occurring terms zero.
The energy of $\vert fs)$ is $m_{P}$, not $m_{p}$, so needs 
contributions from all terms. But $\vert fs)$ is say a lepton plus a
pion so other terms, to which this is orthogonal, cannot contribute.

Hence we can see why the proton cannot decay. State $\vert p)$
satisfies the equation with the weak interaction, thus is caused
to decay by it. But its mass is not the physical mass since the
equation it satisfies is not the equation satisfied by 
the physical object. However the physical object decays (were that 
possible) because of the decay of each of these terms. Each term
however gives a state with energy less than the mass of the mass
of the physical object. It is only the sum of masses that equals
the physical one. But these (smaller) masses cannot be summed
because the final states are different. The physical proton can
decay to only one (for each decay) and the mass of that one is
less than the mass of the initial state.

The decay of the proton cannot conserve energy, thus cannot
occur. An intuitive way (which must be used very carefully as it
can be misleading) of looking at this is that the proton has a
contribution to its mass because of its strong interaction and
this disappears when it decays.

Similarly decays of leptons (the $\tau$) to baryons are ruled
out.

\section{Charge conservation is similar}\label{s:chr}
The argument for electric-charge conservation is the same. Charge
conservation is related to gauge invariance, a partial statement
of Poincar\'e invariance~(\cite{ml3}, sec.~3.4, p.~43) --- this
relates an allowed interaction to the Poincar\'e group. 
An interaction violating charge conservation would not transform under
gauge transformations as other terms in the Hamiltonian, giving
Poincar\'e transformations (on massive objects) that induce gauge
transformations (on massless ones) resulting in physically
identical observers who undergo the different 
gauge transformations --- these cannot be fully specified --- thus physically
identical, but who see different Hamiltonians. The Hamiltonian
would not be well-defined, implying inconsistent physics. It is
fortunate that charge is conserved.

\section{Implications}\label{s:imp}
There are other implications requiring investigation; we mention
a few in hope of stimulating such.
All interactions known are of lowest order. Why? 
For electromagnetism linearity is enforced by gauge (Poincar\'e) 
transformations~(\cite{ml3}, sec.~4.2, p.~57). For strong interactions,
take a particle, a $\Delta$ or $P$, that emits a pion. Higher
order terms would couple it not to a single pion, but to more.
Intuitively we can guess why only lowest order occurs since it
gives diagrams which we interpret (purely heuristically) as two
or more pions emitted sequentially. Higher-order means that these
are emitted together. However this is the limit of the lowest
order in which the time between emissions goes to zero. A higher
order interaction would be this limit which is included in the
lowest order as one case; higher-order terms adding nothing would
be irrelevant. Summing all diagrams, and integrating over time,
would give contributions from terms that have the same effect as
higher-order ones, thus changing only the value of the sum, so
the value of the coupling constant --- an experimental parameter
(at present). Thus we could not distinguish contributions from
terms of different order implying higher order would be
undetectable. This regards particles as virtual. But consider a
decay in two steps, each emitting a pion. If the intermediate
object's life were sufficiently short this would be equivalent to
pions being emitted simultaneously. If a nucleon had 
an interaction of the form $NNN'\pi$, the emission of an $NN'$ pair could
be thought of as due to the decay of a pion, and the interaction
taken as the limit of the emission of a pion and then its decay,
when its lifetime becomes zero, merely changing the sum. These
are purely heuristic and must be investigated in greater depth.

\section{Geometry is destiny(?)}\label{s:geo}

This is part of a long investigation of how geometry through its
transformation groups limits and determines
physics~\cite{ia3,gf3,ml3,pt3,qm3,cnf3}. Geometry is quite powerful,
and quite limiting. And physics is quite limited --- strangely
enough to laws that allow life~\cite{imp3}. It is strange, and as
analysis shows, inexplicable, incomprehensible.

\section*{Acknowledgements}

This discussion could not have existed without Norma Mankoc Borstnik. 
Comments by her and Holger Bech Nielsen improved the
presentation.

\title{Approximate Solutions for the Higgs Masses and Couplings in the
NMSSM}
\author{\underline{R.~Nevzorov}${}^{1,2}$ \and D.J.~Miller${}^{3}$} 
\institute{%
${}^{1}$ School of Physics and Astronomy, University of Southampton, UK \\[0mm]                          
${}^{2}$ Theory Department, ITEP, Moscow, Russia \\[0mm]   
${}^{3}$ Department of Physics and Astronomy, University of Glasgow, UK}
\titlerunning{Approximate Solutions for the Higgs Masses and Couplings in the NMSSM}
\authorrunning{R.~Nevzorov and D.J.~Miller}
\maketitle

\begin{abstract}
We find the approximate solutions for the Higgs masses and couplings in the 
NMSSM with exact and softly broken PQ--symmetry. The obtained solutions indicate 
that there exists a mass hierarchy in the Higgs spectrum which is caused by 
the stability of the physical vacuum.  
\end{abstract}

\section{Introduction}

The minimal SUSY version of the Standard Model (SM) stabilizing the mass hierarchy
does not provide any explanation for its origin.Indeed the Minimal Supersymmetric 
Standard Model (MSSM) being incorporated in the supergravity theories leads to the
$\mbox{$\mu$-problem}$. Within supergravity models the full superpotential
is usually represented as an expansion in powers of observable superfields $\hat{C}_{\alpha}$ 
\begin{equation}
W=\hat{W}_0(h_m)+\mu(h_m)(\hat{H}_1 \epsilon \hat{H}_2)+
h_{\alpha\beta\gamma}\hat{C}_{\alpha}\hat{C}_{\beta}\hat{C}_{\gamma}+\dots~,
\label{1}
\end{equation}
where $h_m$ and $\hat{W}_0(h_m)$ are the ``hidden'' sector fields and
its superpotential respectively.  The ``hidden'' sector fields acquire
vacuum expectation values of the order of Planck scale ($M_{Pl}$)
breaking local supersymmetry and generating a set of soft masses and
couplings in the observable sector. From dimensional considerations
one would naturally expect the parameter $\mu$ to be either zero or
the Planck scale. If $\mu=0$ then the minimum of the Higgs boson
potential occurs for $\langle H_1 \rangle =0$ and down quarks and
charged leptons remain massless.  In the opposite case, when the
values of $\mu\sim M_{Pl}$, there is no spontaneous breakdown of
$SU(2)\times U(1)$ symmetry at all since the Higgs scalars get a huge
positive contribution $\mu^2$ to their squared masses. In order to
provide the correct pattern of electroweak symmetry breaking, $\mu$ is
required to be of the order of the electroweak scale.
 
In the simplest extension of the MSSM, the Next--to--Minimal
Supersymmetric Standard Model (NMSSM) \cite{3,5}, the
superpotential is invariant with respect to the discrete
transformations $\hat{C}_{\alpha}'=e^{2\pi i/3}C_{\alpha}$ of the
$Z_3$ group. The term $\mu (\hat{H}_1 \hat{H}_2)$ does not meet this
requirement. Therefore it is replaced in the superpotential by
\begin{equation}
W_{H}=\lambda \hat{S}(H_1 \epsilon H_2)+\frac{1}{3}\kappa\hat{S}^3\,,
\label{4}
\end{equation} 
where $\hat{S}$ is an additional superfield which is a singlet
with respect to $SU(2)$ and $U(1)$ gauge transformations. A
spontaneous breakdown of the electroweak symmetry leads to the
emergence of the vacuum expectation value of the extra singlet field
$\langle S \rangle =s/\sqrt{2}$ and an effective $\mu$--term is
generated ($\mu=\lambda s/\sqrt{2}$). The $Z_3$ symmetry of the
superpotential naturally arises in string inspired models, where all
observable fields are massless in the limit of unbroken supersymmetry.

In this paper we investigate the Higgs masses and their couplings to the
Z--boson in the NMSSM using approximate solutions.  In Section 2
we specify the Higgs sector of the model. In section 3 the exact
Peccei--Quinn (PQ) symmetry limit in the NMSSM is studied and
approximate solutions for the Higgs masses and couplings are
obtained. The scenario of soft PQ--symmetry breaking is discussed
in section 4. In section 5 we summarize our results .

\section{NMSSM Higgs sector}

The NMSSM Higgs sector involves two Higgs doublets $H_{1,2}$ and one
singlet field $S$.  The interactions of the extra complex scalar $S$ with
other particles is defined by the superpotential (\ref{4}) that leads
to a Higgs boson potential of the following form:
\begin{equation}
\begin{array}{c}
V=\displaystyle\frac{g^2}{8}\left(H_1^+\sigma_a H_1+H_2^+\sigma_a H_2\right)^2+
\frac{{g'}^2}{8}\left(|H_1|^2-|H_2|^2\right)^2+\qquad\qquad\\[3mm]
+\lambda^2|S|^2(|H_1|^2+|H_2|^2)+\lambda\kappa\biggl[S^{*2}(H_1\epsilon
H_2)+h.c.\biggr]+\kappa^2|S|^4+\\[3mm]
+\lambda^2|(H_1\epsilon
H_2)|^2+\biggl[\lambda A_{\lambda}S(H_1\epsilon
H_2)+\displaystyle\frac{\kappa}{3}A_{\kappa}S^3+h.c.\biggr]+\\[3mm]
+m_1^2|H_1|^2+m_2^2|H_2|^2+m_S^2|S|^2+\Delta V\, ,
\end{array}
\label{5}
\end{equation}
where $g$ and $g'$ are $SU(2)$ and $U(1)$ gauge couplings respectively, 
while $\Delta V$ corresponds to the contribution of loop corrections. The couplings $g,\,g',\,
\lambda$ and $\kappa$ do not violate supersymmetry. The set of soft SUSY breaking
parameters includes soft masses $m_1^2,\, m_2^2,\, m_S^2$ and trilinear couplings $A_{\kappa},\, 
A_{\lambda}$. 

At the physical minimum of the potential (\ref{5}) the neutral
components of the Higgs doublets $H_1$ and $H_2$ develop vacuum
expectation values $v_1$ and $v_2$ breaking the electroweak symmetry
down to $U(1)$. Upon the breakdown of $SU(2)\times U(1)$ symmetry
three goldstone modes ($G^{\pm}$ and $G^{0}$) emerge, and are absorbed
by the $W^{\pm}$ and $Z$ bosons.  In the field space basis rotated by an
angle $\beta$ ($\tan\beta=v_2/v_1$) with respect to the initial direction
\begin{equation}
\begin{array}{ll}
Im\, H_1^0= (P \sin \beta + G^0 \cos \beta)/\sqrt{2},&\quad H_1^-=G^- \cos \beta + H^- \sin \beta\, , \\
Im\, H_2^0= (P \cos \beta - G^0 \sin \beta)/\sqrt{2},&\quad H_2^+=H^+ \cos \beta - G^+ \sin \beta\, , \\ 
Im\, S= P_S/\sqrt{2}&
\end{array}
\label{10}
\end{equation}
these unphysical degrees of freedom are removed by a gauge transformation 
and the mass terms in the Higgs boson potential 
can be written as follows
\begin{equation}
V_{mass} = M_{H^{\pm}}^2  H^+ H^- + 
\frac{1}{2} (P \,\, P_S) \tilde{M}^2 
\left( 
\begin{array}{c} 
P \\ 
P_S 
\end{array} \right)+
\frac{1}{2} (H \,\, h \,\, N) M^2 
\left( 
\begin{array}{c} 
H \\ 
h \\ 
N
\end{array} \right)\, ,  
\label{11}
\end{equation}   
where we replace the real parts of the neutral components of the Higgs doublets by their superpositions 
$H\, ,h$ so that
\begin{equation}
\begin{array}{c}
Re \, H_1^0= (h \cos\beta- H \sin\beta+v_1)/\sqrt{2}\,,\qquad Re\, S= (s+N)/\sqrt{2}\,,\\[2mm]
Re \, H_2^0= (h \sin\beta+ H \cos\beta+v_2)/\sqrt{2}\,.
\end{array}
\label{12}
\end{equation}
 
From the conditions for the extrema $\left(\displaystyle\frac{\partial V}{\partial v_1}=\frac{\partial V}{\partial v_2}=\frac{\partial V}{\partial s}=0\right)$ 
of the Higgs effective potential (\ref{5}) one can express $m_S^2$, $m_1^2$, $m_2^2$ via other fundamental parameters, $\mbox{tg}\beta$ and $s$. 
Substituting the obtained relations for the soft masses in the $2\times 2$ CP-odd mass matrix $\tilde{M}^2_{ij}$ we get:
\begin{equation}
\begin{array}{rcl}
\tilde{M}_{11}^2=m_A^2=\displaystyle\frac{4\mu^2}{\sin^2 2\beta}\left(x-\frac{\kappa}{2\lambda}\sin2\beta\right)
+\tilde{\Delta}_{11}\, ,\\
~~~\tilde{M}_{22}^2=\displaystyle\frac{\lambda^2 v^2}{2}x+\frac{\lambda\kappa}{2}v^2\sin2\beta-
3\frac{\kappa}{\lambda}A_{\kappa}\mu+\tilde{\Delta}_{22}\, ,\\
\tilde{M}_{12}^2=\tilde{M}_{21}^2=\displaystyle\sqrt{2}\lambda v \mu\left(\frac{x}{\sin 2\beta}
-2\frac{\kappa}{\lambda}\right)+\tilde{\Delta}_{12}\, ,\\
\end{array}
\label{14}
\end{equation}
where $v=\sqrt{v_1^2+v_2^2}=246\,\mbox{GeV}$, $\mu=\displaystyle\frac{\lambda s}{\sqrt{2}}$, 
$x=\displaystyle\frac{1}{2\mu}\left(A_{\lambda}+2\frac{\kappa}{\lambda}\mu\right)\sin2\beta$ 
and $\tilde{\Delta}_{ij}$ are contributions of the loop corrections to the mass matrix elements. 
The mass matrix (\ref{14}) can be easily diagonalized via a rotation of the fields 
$P$ and $P_S$ by an angle $\theta_A$ ($\tan 2\theta_A=2\tilde{M}^2_{12}/(\tilde{M}^2_{11}-\tilde{M}^2_{22})$)~.  

The charged Higgs fields $H^{\pm}$ are already physical mass eigenstates with
\begin{equation}
M_{H^{\pm}}^2=m_A^2-\frac{\lambda^2 v^2}{2}+M_W^2+\Delta_{\pm}.
\label{16}
\end{equation}
Here $M_W=\displaystyle\frac{g}{2}v$ is the charged W-boson mass and $\Delta_{\pm}$ includes loop 
corrections to the charged Higgs masses. 

In the rotated basis  $H\, ,h\,, N$ the matrix elements of the $3\times 3$ mass matrix of 
the CP--even Higgs sector can be written as \cite{6}--\cite{7}:
\begin{equation}
\begin{array}{rcl}
M_{11}^2&=&\displaystyle m_A^2+\left(\frac{\bar{g}^2}{4}-\frac{\lambda^2}{2}\right)v^2
\sin^2 2\beta+\Delta_{11}\, ,\\[0.2cm]
M_{22}^2&=&\displaystyle M_Z^2\cos^2 2\beta+\frac{\lambda^2}{2}v^2\sin^2 2\beta+
\Delta_{22}\, ,\\[0.2cm]
M_{33}^2&=&\displaystyle 4\frac{\kappa^2}{\lambda^2}\mu^2+\frac{\kappa}{\lambda}A_{\kappa}\mu+
\displaystyle\frac{\lambda^2 v^2}{2}x-\frac{\kappa\lambda}{2}v^2\sin2\beta+\Delta_{33}\, ,\\[0.2cm]
M_{12}^2&=&M_{21}^2=\displaystyle \left(\frac{\lambda^2}{4}-\frac{\bar{g}^2}{8}\right)v^2
\sin 4\beta+\Delta_{12}\, ,\\[0.2cm]
M_{13}^2&=&M_{31}^2=-\sqrt{2}\lambda v \mu x\, \cot  2\beta+\Delta_{13}\, ,
\\[0.2cm]
M_{23}^2&=&M_{32}^2=\sqrt{2}\lambda v \mu (1-x)+\Delta_{23}\, ,
\end{array}
\label{18}
\end{equation}
where $M_Z=\displaystyle\frac{\bar{g}}{2}v$ is the Z--boson mass, $\bar{g}=\sqrt{g^2+g'^2}$, and
$\Delta_{ij}$ can be calculated by differentiating $\Delta V$ \cite{6}. Since the minimal 
eigenvalue of a matrix does not exceed its smallest diagonal element, at least one Higgs scalar in 
the CP--even sector has to be comparatively light: $m_{h_1}\le \sqrt{M_{22}^2}$. At the tree level 
the upper bound on the lightest Higgs mass in the NMSSM was found in \cite{5}. It differs from the 
corresponding theoretical limit in the minimal SUSY model only for moderate values of $\tan\beta$. 
As in the MSSM the loop corrections from the $t$--quark and its superpartners raise the value of 
the upper bound on the lightest Higgs mass resulting in a rather strict restriction on 
$m_{h_1}\le 135\,\mbox{GeV}$ \cite{8}. The Higgs sector of the NMSSM and loop corrections to it
were studied in \cite{9}.  
 
In the field basis $P,\, P_{S},\,H,\,h,\,N$ the trilinear part of the Lagrangian, which is responsible 
for the interaction of the neutral Higgs states with the Z--boson, is simplified:
\begin{equation}
L_{AZH}=\displaystyle\frac{\bar{g}}{2} M_{Z}Z_{\mu}Z_{\mu}h+\frac{\bar{g}}{2}Z_{\mu}
\biggl[H(\partial_{\mu}P)-(\partial_{\mu}H)P\biggr]~.
\label{13}
\end{equation}
Only one CP-even component, $h$, couples to a pair of Z--bosons while another, $H$, interacts with
pseudoscalar $P$ and $Z$. 
The coupling of $h$ to the Z pair is exactly the same as in the SM. In the Yukawa 
interactions with fermions $h$ also manifests itself as the SM like Higgs boson.

The couplings of the physical Higgs scalars to the Z pair ($g_{ZZi}$, $i=1,2,3$) and to the Higgs pseudoscalars and Z boson
($g_{ZA_1i}$ and $g_{ZA_2i}$) appear due to the mixing of $h, H$ and $P$ with other components of
the CP--odd and CP--even Higgs sectors. Following the traditional notation we define the normalized 
$R$--couplings as: $g_{ZZi}=R_{ZZi}\times\displaystyle\frac{\bar{g}}{2}M_Z$ and 
$g_{ZA_{j}i}=\displaystyle\frac{\bar{g}}{2}R_{ZA_{j}i}$. All relative $R$--couplings vary from zero to unity
and are given by
\begin{equation}
R_{ZZi}=U^+_{hi}~,~~~R_{ZA_{1}i}=-U^+_{Hi}\sin\theta_A~,~~~
R_{ZA_2i}=U^+_{Hi}\cos\theta_A~,
\label{20}
\end{equation}
where $U_{ij}$ is unitary matrix relating components of the field basis $H,\,h,$ and $N$ to the physical 
CP-even Higgs eigenstates.

\section{Exact Peccei--Quinn symmetry limit} 

First of all let us discuss the NMSSM with $\kappa=0$. 
At the tree level the Higgs masses and couplings in this model depend on four parameters:
$\lambda$,$\mu$,$\tan\beta$, $m_A$ (or $x$).
When $\lambda$ is small enough (say $\lambda \le 0.1$) the experimental constraints on the SUSY parameters 
obtained in the minimal SUSY model remain valid in the the NMSSM. If $\tan\beta\le 2.5$ the predominant 
part of the NMSSM parameter space is excluded by unsuccessful Higgs searches. Non-observation of 
charginos at LEPII restricts the effective $\mu$-term from below: $|\mu|\ge 90-100\,\mbox{GeV}$. 
Combining these limits one gets a useful lower bound on $m_A$ at the tree level:
\begin{equation}
m_A^2\ge 9M_Z^2 x\,.
\label{25}
\end{equation}
Requirement of the validity of perturbation theory up to the high energy scales constrains the
parameter space further. In order to prevent the appearance of Landau pole during the evolution of the Yukawa couplings 
from the electroweak scale to Grand Unification scale ($M_X$) the value of $\lambda$ has to be always smaller than 0.7 .

In the NMSSM with $\kappa=0$ the mass of the lightest pseudoscalar vanishes. This is a manifestation of the 
enlarged $SU(2)\times [U(1)]^2$ global symmetry of the Lagrangian. The extra $U(1)$ (Peccei--Quinn) symmetry is 
spontaneously broken giving rise to a massless Goldstone boson (axion) \cite{10}. The Peccei--Quinn symmetry 
and the axion allow one to avoid the strong CP problem, eliminating the $\theta$--term in QCD \cite{11}. At low 
energies the axion gains a small mass due to mixing with the pion. The mass of orthogonal superposition of $P$ and $P_S$ is
\begin{equation}
m_{A_2}^2=m_A^2+\frac{\lambda^2 v^2}{2}x+\tilde{\Delta}_{22}\,.
\label{22}
\end{equation}

The lower bound on $m_A$ (\ref{25}) leads to the hierarchical structure of the CP--even Higgs mass matrix.
It can be written as 
\begin{equation} 
M^2= 
\left(
\begin{array}{cc}
A & \varepsilon C^{\dagger} \\[3mm]
\varepsilon C  & \varepsilon^2 B
\end{array} 
\right), 
\label{23}
\end{equation}
where $\varepsilon<<1$. Indeed the top--left entry ($M_{11}^2=A$) of the corresponding $3\times 3$ mass matrix (\ref{18}) 
is the largest one in the dominant part of parameter space. It is proportional to $ m_A^2$ while 
$M_{12}^2\sim M_{22}^2\sim M_{33}^2 \sim M_Z^2$ and $M_{13}^2\sim m_A M_Z$.
Therefore the ratio $M_Z/m_A$ plays the role of a small parameter $\varepsilon$. 

The CP--even Higgs mass matrix can be reduced to block diagonal form:
\begin{equation} 
V M^2 V^{\dagger}\simeq
\left(
\begin{array}{ccc}
\displaystyle M_{11}^2+\frac{M_{13}^4}{M_{11}^2} & O(\varepsilon^3) & O(\varepsilon^3)\\[3mm]
O(\varepsilon^3) & M_{22}^2 & \displaystyle M_{23}^2-\frac{M_{13}^2M_{12}^2}{M_{11}^2}\\[3mm]
O(\varepsilon^3) & \displaystyle M_{23}^2-\frac{M_{13}^2M_{12}^2}{M_{11}^2}& \displaystyle M_{33}^2-\frac{M_{13}^4}{M_{11}^2} 
\end{array} 
\right) 
\label{24}
\end{equation}
by virtue of unitary transformation 
\begin{equation} 
V= 
\left(
\begin{array}{cc}
\displaystyle 1-\frac{\varepsilon^2}{2}\Gamma^{\dagger}\Gamma & \varepsilon \Gamma^{\dagger} \\[3mm]
-\varepsilon \Gamma  &\displaystyle 1-\frac{\varepsilon^2}{2}\Gamma\Gamma^{\dagger}
\end{array} 
\right),\qquad \Gamma=CA^{-1}\,.
\label{241}
\end{equation}
The obtained matrix (\ref{24}) is now easily diagonalized via a rotation by an angle $\theta$ of the 
two lowest states
\begin{equation} 
R=
\left(
\begin{array}{ccc}
1 & 0 & 0\\
0 & \cos\theta & \sin\theta\\
0 & -\sin\theta & \cos\theta 
\end{array} 
\right), \qquad tg\, 2\theta=\frac{\displaystyle 2\left(M_{23}^2-\frac{M_{13}^2M_{12}^2}{M_{11}^2}\right)}
{\displaystyle M_{22}^2-M_{33}^2+\frac{M_{13}^4}{M_{11}^2}} .
\label{243}
\end{equation}
As a result we find the approximate formulae for the masses of the CP--even Higgs bosons
\begin{equation}
\begin{array}{rcl}
m_{h_3}^2&=&\displaystyle M_{11}^2+\frac{M_{13}^4}{M_{11}^2}\, ,\\[3mm]
m_{h_2,h_1}^2&=&\displaystyle\frac{1}{2}\left(M_{22}^2+M_{33}^2-\frac{M_{13}^4}{M_{11}^2} \right. \\[0mm]
&&\left. \displaystyle\pm\sqrt{\left(M_{22}^2-M_{33}^2+\frac{M_{13}^4}{M_{11}^2}\right)^2+
4\left(M_{23}^2-\frac{M_{13}^2M_{12}^2}{M_{11}^2}\right)^2}\right).
\end{array}
\label{244}
\end{equation}
Also using the explicit form of the unitary matrix $U^{\dagger}\approx V^{\dagger}R^{\dagger}$, that
links $H, h$ and $N$ to the mass eigenstates, the approximate expressions for the couplings of the
lightest Higgs particles to the Z--boson can be established:
\begin{equation}
\begin{array}{l}
R_{ZZ2}\approx\cos\theta~,~~~~~~R_{ZZ1}\approx-\sin\theta~,\\[3mm]
R_{ZA_12}\approx\left(\displaystyle\frac{M_{12}^2}{M_{11}^2}\cos\theta+
\frac{M_{13}^2}{M_{11}^2}\sin\theta\right)\sin\theta_A~,\\[3mm]
R_{ZA_11}\approx\left(\displaystyle\frac{M_{13}^2}{M_{11}^2}\cos\theta-
\frac{M_{12}^2}{M_{11}^2}\sin\theta\right)\sin\theta_A~,
\end{array}
\label{42}
\end{equation}

The approximate solutions for the CP-even Higgs boson masses and couplings shed light on their behaviour 
as the NMSSM parameters are varied. As evident from Eq.(\ref{244}), at large values of $\tan\beta$ or $\mu$ 
the mass--squared of the lightest Higgs scalar tends to be negative because $M_{23}^2$ becomes large
while bottom--right entry of 
the matrix (\ref{24}) goes to zero. Due to the vacuum stability requirement, 
which implies the positivity of the mass--squared of all Higgs particles,
the auxiliary variable $x$ is localized near unity. At the tree level we get
\begin{equation}
1-\Delta < x < 1+\Delta\, ,\qquad
\Delta\approx\frac{\sqrt{\displaystyle M_Z^2\cos^22\beta+\frac{\lambda^2 v^2}{2}\sin^22\beta}}{m^0_A}\, ,
\label{31}
\end{equation}
where $m_A^0=2\mu/\sin 2\beta$.
The allowed range of the auxiliary variable $x$ is quite narrow (see \cite{7}). 
According to the definition of $m_A$ (\ref{14}) the tight bounds on $x$  
enforce $m_A$ to be confined in the vicinity of $\mu\, tg\beta$ which is considerably larger than the 
Z-boson mass. As a result the masses of the charged Higgs boson, heaviest CP-odd and 
CP-even Higgs states are rather close to $m_A$. At the tree level the theoretical bounds on the 
masses of the lightest Higgs scalars are
\begin{equation}
\begin{array}{l}
m_{h_1}^2\le \displaystyle\frac{\lambda^2 v^2}{2}x\sin^2 2\beta\,,\\[3mm]
m_{h_2}^2\ge \displaystyle M_Z^2\cos^22\beta+\frac{\lambda^2}{2}v^2\sin^22\beta\, ,\\[3mm]
m_{h_2}^2 \displaystyle\le M_Z^2\cos^22\beta+\frac{\lambda^2}{2}v^2(1+x)\sin^22\beta\, .
\end{array}
\label{32}
\end{equation}
The masses of $h_2$ and $h_1$ are set by the Z-boson mass and $\lambda v$ respectively so that 
$m_{h_1}, m_{h_2}\ll m_{A}$ in the allowed range of the parameter space. 

\section{Soft breaking of the PQ-symmetry}

Unfortunately searches for massless pseudoscalar and light scalar particles exclude any choice of 
the parameters in the NMSSM with $\kappa=0$, unless one allows $\lambda$ to become very small \cite{12}. In order to get a reliable pattern for the Higgs masses and 
couplings the Peccei--Quinn symmetry must be broken. Recently different origins of extra U(1) symmetry breaking 
were discussed \cite{13}. Here we assume that the violation 
of the Peccei--Quinn symmetry is caused by non--zero value of $\kappa$. As follows from the explicit form of the 
mass matrices (\ref{14}), (\ref{16}) and (\ref{18}) in this case the Higgs spectrum depends on six parameters
at the tree level: $\lambda, \kappa, \mu, tg\beta, A_{\kappa}$ and $m_A$ (or $x$). We restrict our consideration
by small values of $\kappa$ when the PQ--symmetry is only slightly broken. To be precise we consider such values
of $\kappa$ that do not change much the vacuum energy density. the last requirement places a strong bound on $\kappa$ when 
$\lambda$ goes to zero:
\begin{equation}
\kappa<\lambda^2\,.
\label{34}
\end{equation} 
If $\kappa \gg \lambda^2$ then the terms $\kappa^2|S|^4$ and $\displaystyle\frac{\kappa}{3}A_{\kappa}S^3$ in the Higgs effective 
potential (\ref{5}) becomes much larger $|\mu|^4\sim M_Z^4$ increasing the absolute value of the vacuum energy
density significantly. A small ratio $\kappa/\lambda$ may naturally arise from the renormalization group flow of $\lambda$ and $\kappa$ from $M_X$ to $M_Z$ \cite{7,14}.

\begin{figure}
\centering
\includegraphics[totalheight=60mm,keepaspectratio=true]{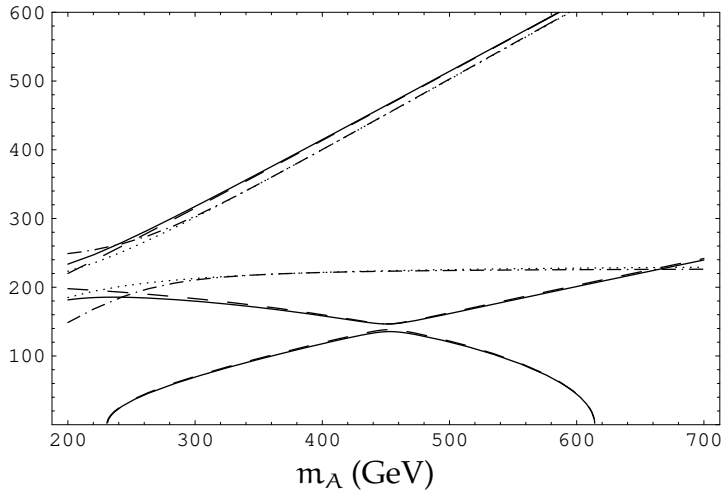}\\
{\large $m_{A}$ (GeV)}\\[2mm]
\caption{\label{rn-fig1}%
 The dependence of the neutral Higgs boson masses on $m_A$ for 
$\lambda=0.6$, $\kappa=0.36$, $\mu=150$\,GeV, $tg \beta=3$ and $A_{\kappa}=135\,\mbox{GeV}$.
The one--loop Higgs masses of scalars (solid curve) and pseudoscalars (dashed--dotted curve) are 
confronted with the approximate solutions (dashed and dotted curves). Masses are in GeV.}
\end{figure}

The soft breaking of the PQ--symmetry does not lead to the realignment of the Higgs spectrum preserving its 
mass hierarchy. Still $M_{11}^2$ is the largest matrix element of the CP-even Higgs mass matrix 
in the admissible part of the NMSSM parameter space. Therefore the approximate formulae (\ref{244})--(\ref{42})
obtained in the previous section remain valid in the considered limit. It is easy to see that
the lightest CP--even Higgs states respect a sum rule
\begin{equation}
m_{h_1}^2+m_{h_2}^2=M_{22}^2+M_{33}^2-\frac{M_{13}^4}{M_{11}^2}~.
\label{37}
\end{equation} 
The right--hand side of Eq.(\ref{37}) is almost insensitive to the choice of $m_A$ and rather weakly varies with
changing $tg\beta$. As a result of the sum rule (\ref{37}) the second lightest Higgs scalar mass is maximized as $m_{h_1}$ 
goes to zero, and vice versa the lightest Higgs scalar gets maximal mass when $m_{h_2}$ attains minimum 
(see also Fig.~\ref{rn-fig1}). According to Eq.(\ref{244}) the mass of the lightest CP--even Higgs 
boson vary within the limits: 
\begin{equation}
\begin{array}{rcccl}
0&\le & m_{h_1}^2 &\le & min\left\{\displaystyle M_{22}^2\, ,\,M_{33}^2-\frac{M_{13}^4}{M_{11}^2}\right\}
\end{array}
\label{38}
\end{equation}

The mass matrix of the CP-odd Higgs sector also exhibits the hierarchical structure. Indeed the  
entry $\tilde{M}^2_{11}$ is determined by $m_A^2$ whereas the off-diagonal element of the matrix (\ref{14}) is of the order
of $\lambda v \cdot m_A$. Since the ratio $\kappa/\lambda$ is small, the other diagonal entry
$\tilde{M}^2_{22} \ll m_A^2$. this again permits one to seek the eigenvalues of the matrix (\ref{14}) as an expansion in 
powers of $\lambda v/m_A$. The perturbation theory being applied for its diagonalization 
results in concise expressions for the squared masses of the Higgs pseudoscalars
\begin{equation}
m^2_{A_1}\approx\tilde{M}^2_{22}+\frac{\tilde{M}^4_{12}}{\tilde{M}^2_{11}}\,,\qquad\qquad
m^2_{A_2}\approx\tilde{M}^2_{11}-\frac{\tilde{M}^4_{12}}{\tilde{M}^2_{11}}\,.
\label{39}
\end{equation} 
Because the PQ--symmetry is now broken the lightest CP--odd Higgs boson also gains non--zero mass.
In compliance with Eq.(\ref{42}) the couplings of $h_1$ and $h_2$ to a Z--pair obey the following sum rule:  
\begin{equation}
R_{ZZ1}^2+R_{ZZ2}^2\simeq 1\, ,
\label{33}
\end{equation}
while $R_{ZA_11}$ and $R_{ZA_12}$ are suppressed by a factor $(\lambda^2 v^2/m_A^2)$.

At the tree level and large values of $tg\beta$ ($tg\beta \gtrsim 10$) the approximate expressions
(\ref{244}) and (\ref{39}) describing the masses of the Higgs scalar and pseudoscalar particles are simplified
\begin{equation}
\begin{array}{rcl}
m^2_{h_3}&=&m_{A_2}^2=\displaystyle m_A^2+\frac{\lambda^2v^2}{2}x\, ,\qquad\qquad 
m^2_{A_1}=-3\frac{\kappa}{\lambda}A_{\kappa}\mu\, ,\\[3mm]
m^2_{h_2,\, h_1}&=&\displaystyle\frac{1}{2}\left[M_Z^2+4\frac{\kappa^2}{\lambda^2}\mu^2+\frac{\kappa}{\lambda}A_{\kappa}\mu
\pm\right.\\[3mm]
&&\left.\pm\sqrt{\left(\displaystyle M_Z^2-4\frac{\kappa^2}{\lambda^2}\mu^2-\frac{\kappa}{\lambda}A_{\kappa}\mu\right)^2+
8\lambda^2v^2\mu^2(1-x)^2}\right]\, 
\end{array}
\label{40}
\end{equation}
making their analysis more transparent. Again the positivity of the mass--squared of the
lightest Higgs scalar restricts the allowed range of $x$
\begin{equation}
1-\left|\frac{\sqrt{2}\kappa M_Z}{\lambda^2 v}\right|<x<1+\left|\frac{\sqrt{2}\kappa M_Z}{\lambda^2 v}\right|\,,
\label{41}
\end{equation}
if the PQ--symmetry is only slightly broken. When $\kappa\ll \lambda^2$ the admissible interval of the auxiliary 
variable $x$ shrinks drastically establishing a very stringent bound on the value of $m_A$ and strong correlation 
between $m_A$, $\mu$ and $tg\, \beta$ that constrains the masses of the heavy Higgs bosons 
$m_{h_3}\approx m_{H^{\pm}}\approx m_{A_2}$ in the vicinity of $\mu\, tg\beta$. 

\begin{figure}
\centering
\includegraphics[height=60mm,keepaspectratio=true]{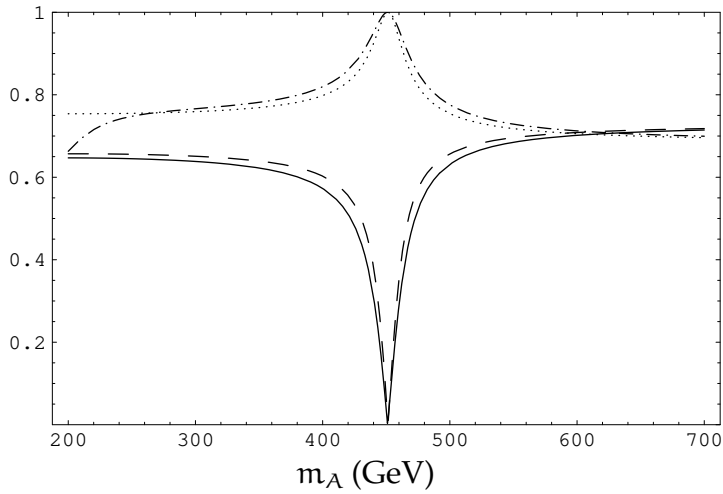}\\
{\large $m_{A}$ (GeV)}\\[2mm]
\caption{\label{rn-fig2}%
The absolute values of $R_{ZZ1}$ and $R_{ZZ2}$,
plotted as a function of $m_A$ for the same values of $\lambda$, $\kappa$, $\mu$, $\tan\beta$ and $A_{\kappa}$
as in Fig.~\ref{rn-fig1}. Solid and dashed--dotted curves reproduce the dependence of $R_{ZZ1}$ and 
$R_{ZZ2}$ on $m_A$ while dashed and dotted curves represent their approximate solutions.}
\end{figure}

\begin{figure}
\centering
\includegraphics[height=60mm,keepaspectratio=true]{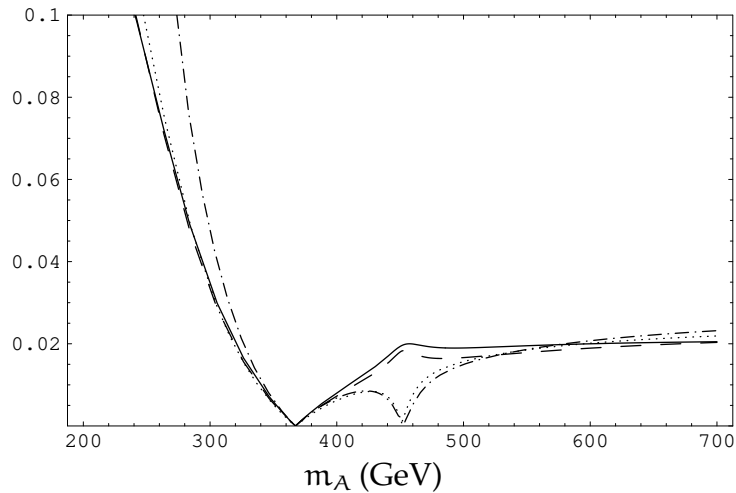}\\
{\large $m_{A}$ (GeV)}\\[2mm]
\caption{\label{rn-fig3}%
The absolute values of $R_{ZA_11}$ and $R_{ZA_12}$ as a function of $m_A$.
The parameters $\lambda$, $\kappa$, $\mu$, $\tan\beta$ and $A_{\kappa}$ are taken to be the same as in 
Fig.~\ref{rn-fig1}. Solid and dashed--dotted curves correspond to $R_{ZA_11}$ and $R_{ZA_12}$ while dashed and dotted curves 
represent their approximate solutions.}
\end{figure}

The results of the numerical studies of the Higgs boson masses and their couplings including leading one--loop 
corrections from the top and stop loops are given in Figs.~\ref{rn-fig1}--\ref{rn-fig3}. As a representative example we fix the Yukawa couplings
at the Grand Unification scale so that $\lambda(M_X)=\kappa(M_X)=2h_t(M_X)=1.6$, that corresponds to $tg\beta\ge 3$,  
$\lambda(M_t)\simeq 0.6$ and $\kappa(M_t)\simeq 0.36$ at the electroweak scale.
We set $\mu=150\,\mbox{GeV}$ which is quite close to the current limit on $\mu$ in the MSSM.
The parameter $A_{\kappa}$ occurs in the right--hand side of the sum rule (\ref{37}) and in the mass of the lightest 
pseudoscalar $m_{A_1}^2$ with opposite sign. As a consequence whereas $m_{h_1}^2$ rises over changing $A_{\kappa}$, $m_{A_1}^2$ 
diminishes and vice versa. Too large positive and negative values of $A_{\kappa}$ pull the mass-squared 
of either lightest scalar or pseudoscalar below zero destabilizing the vacuum that restricts $A_{\kappa}$
from below and above. To represent the results of our numerical analysis the parameter $A_{\kappa}$ is taken to be 
near the center of the admissible interval $A_{\kappa}\simeq 135\,\mbox{GeV}$. 

In Figs.~\ref{rn-fig1}--\ref{rn-fig3} the masses of the neutral Higgs particles and their
couplings to Z are examined as a function of $m_A$.  From the
restrictions (\ref{41}) on the parameter $x$ and numerical results
presented in Fig.~\ref{rn-fig1} it is evident that the requirement of the stability
of the physical vacuum and the experimental constraints on $\mu$ and
$tg\beta$ rules out low values of $m_A$, maintaining mass hierarchy
whilst $\kappa\le\lambda^2$. The lightest Higgs scalar and
pseudoscalar can be heavy enough to escape their production at
LEP. Moreover as one can see from Fig.~\ref{rn-fig2} the lightest Higgs scalar can
be predominantly a singlet field, making its detection more difficult
than in the SM or MSSM.  The lightest Higgs pseudoscalar is also
singlet dominated, making its observation at future colliders quite
problematic; the coupling of the lightest CP--even Higgs boson to a
CP--odd Higgs bosons and a Z is always strongly suppressed (see Fig.~\ref{rn-fig3})
according to (\ref{42}). The hierarchical structure of the mass
matrices ensures that the heaviest CP-even and CP-odd Higgs bosons are
predominantly composed of $H$ and $P$. As a result the coupling 
$R_{ZA_23}$ is rather close to unity while $R_{ZZ3}$ is
almost negligible.  In Figs.~\ref{rn-fig1}--\ref{rn-fig3} the approximate solutions
(\ref{244})--(\ref{42}) are also given. They work remarkably well.
%

\section{Conclusions}

In the present article we have obtained the approximate solutions for the Higgs masses and couplings in the
NMSSM with exact and softly broken PQ-symmetry which describe the numerical solutions with high accuracy. 
The approximate formulae (\ref{244})--(\ref{42}) provide nice insight into mass hierarchies
in the considered model. The vacuum stability requirements and LEP restrictions on the NMSSM parameters 
leads to the splitting in the spectrum of the Higgs bosons. When $\kappa=0$ or $\kappa\le\lambda^2$
the charged Higgs states, the heaviest scalar and pseudoscalar are nearly degenerate around $m_A\sim\mu\, tg\beta$.
The masses of new scalar and pseudoscalar states, which are predominantly singlet fields, are governed by the
combination of parameters $\displaystyle\frac{\kappa}{\lambda}\,\mu$. In the NMSSM with exact and softly broken PQ--symmetry
they are considerably lighter than the heaviest Higgs states. Decreasing $\kappa$ pushes their masses
down so that they can be even the lightest particles in the Higgs boson spectrum. The SM like Higgs boson has a mass around $130\,\mbox{GeV}$. We have established useful sum rules for the masses of the lightest Higgs scalars and 
their couplings to a Z pair. Also we found that the couplings of the lightest CP--even Higgs states to the lightest 
pseudoscalar 
and Z--boson are suppressed. Observing two light scalar and one pseudoscalar Higgs particles but no charged Higgs 
boson, at future colliders would present an opportunity to distinguish the NMSSM with softly broken PQ--symmetry 
from the MSSM even if the heavy states are inaccessible.

\section*{Acknowledgements}
The authors would like to thank P.M.~Zerwas for his continual support and encouragement.
RN is grateful to C.D.~Froggatt, E.I.~Guendelman, N.S.~Mankoc-Borstnik and H.B.~Nielsen for stimulating questions and 
comments, and S.F.~King, S.~Moretti, A.~Pilaftsis, M.~Sher and M.I.~Vysotsky for fruitful discussions and helpful remarks.
The work of RN was supported by the Russian Foundation for Basic Research (projects 00-15-96562 and 02-02-17379) 
and by a Grant of President of Russia for young scientists (MK--3702.2004.2).


\backmatter

\thispagestyle{empty}
\parindent=0pt
\begin{flushleft}
\mbox{}
\vfill
\vrule height 1pt width \textwidth depth 0pt
{\parskip 6pt

{\sc Blejske Delavnice Iz Fizike, \ \ Letnik~5, \v{s}t. 2,} 
\ \ \ \ ISSN 1580--4992

{\sc Bled Workshops in Physics, \ \  Vol.~5, No.~2}

\bigskip

Zbornik 7. delavnice `What Comes Beyond the Standard Models', 
Bled, 19.~-- 31.~julij 2004

Proceedings to the 7th workshop 'What Comes Beyond the Standard Models', 
Bled, July 19.--31.,  2004

\bigskip

Uredili Norma Manko\v c Bor\v stnik, Holger Bech Nielsen, 
Colin D. Froggatt in Dragan Lukman 

Publikacijo sofinancira Ministrstvo za \v solstvo, znanost in \v sport 

Tehni\v{c}ni urednik Vladimir Bensa

\bigskip

Zalo\v{z}ilo: DMFA -- zalo\v{z}ni\v{s}tvo, Jadranska 19,
1000 Ljubljana, Slovenija

Natisnila Tiskarna MIGRAF v nakladi 100 izvodov

\bigskip

Publikacija DMFA \v{s}tevilka 1587

\vrule height 1pt width \textwidth depth 0pt}
\end{flushleft}


\end{document}